\documentclass[draft,tightenlines,nofootinbib,preprint,aps,eqsecnum,amsmath,amssymb]{revtex4}

\newcommand{\beq}{\begin{equation}}
\newcommand{\eeq}{\end{equation}}
\newcommand{\bea}{\begin{eqnarray}}
\newcommand{\eea}{\end{eqnarray}}
\newcommand{\cir}{{\buildrel \circ \over =}}
\newcommand{\sgn}{\mbox{\boldmath $\epsilon$}}

\newcommand{\on}{\stackrel{\circ}{=}}
\newcommand{\byd}{\stackrel{def}{=}}

\newcommand{\h}{h}

\begin{document}

\baselineskip 18pt

\today

\title{Charged Particles and the Electro-Magnetic Field in Non-Inertial Frames
of Minkowski Spacetime: I. Admissible 3+1 Splittings of Minkowski
Spacetime and the Non-Inertial Rest Frames.}

\medskip

\author{David Alba}

\affiliation{
Sezione INFN di Firenze\\Polo Scientifico, via Sansone 1\\
 50019 Sesto Fiorentino, Italy\\
 E-mail ALBA@FI.INFN.IT}

\author{Luca Lusanna}

\affiliation{ Sezione INFN di Firenze\\ Polo Scientifico\\ Via Sansone 1\\
50019 Sesto Fiorentino (FI), Italy\\ Phone: 0039-055-4572334\\
FAX: 0039-055-4572364\\ E-mail: lusanna@fi.infn.it}

\begin{abstract}

By using the 3+1 point of view and parametrized Minkowski theories
we develop the theory of {\it non-inertial} frames in Minkowski
space-time.  The transition from a non-inertial frame to another one
is a gauge transformation connecting the respective notions of
instantaneous 3-space (clock synchronization convention) and of the
3-coordinates inside them. As a particular case we get the extension
of the inertial rest-frame instant form of dynamics to the
non-inertial rest-frame one. We show that every isolated system can
be described as an external decoupled non-covariant canonical center
of mass (described by frozen Jacobi data) carrying a pole-dipole
structure: the invariant mass and an effective spin. Moreover we
identify the constraints eliminating the internal 3-center of mass
inside the instantaneous 3-spaces.

In the case of the isolated system of positive-energy scalar
particles with Grassmann-valued electric charges plus the
electro-magnetic field we obtain both Maxwell equations and their
Hamiltonian description in non-inertial frames. Then by means of a
non-covariant decomposition we define the non-inertial radiation
gauge and we find the form of the non-covariant Coulomb potential.
We identify the coordinate-dependent relativistic inertial
potentials and we show that they have the correct Newtonian limit.

In the second paper we will study properties of Maxwell equations in
non-inertial frames like the wrap-up effect and the Faraday rotation
in astrophysics. Also the 3+1 description without
coordinate-singularities of the rotating disk and the Sagnac effect
will be given, with added  comments on pulsar magnetosphere and on a
relativistic extension of the Earth-fixed coordinate system.

\end{abstract}

\maketitle

\vfill\eject

\section{Introduction}

As a consequence of many years of research devoted to try to
establish a consistent formulation of relativistic mechanics, we
have now a description of every isolated system (particles, strings,
fields, fluids), admitting a Lagrangian formulation, in arbitrary
global inertial or non-inertial frames in Minkowski space-time by
means of {\it parametrized Minkowski theories} \cite{1,2,3,4} (see
Ref.\cite{5} for a review). They allow one to get a Hamiltonian
description of the relativistic isolated systems, in which the
transition from a non-inertial (or inertial) frame to another one is
a gauge transformation generated by suitable first-class Dirac
constraints. Therefore, all the admissible conventions for clock
synchronization, identifying the instantaneous 3-spaces containing
the system and allowing a formulation of the Cauchy problem for the
equations of the fields present in the system, turn out to be gauge
equivalent.\medskip

The only known way to have a global description of non-inertial
frames is to choose an arbitrary time-like observer and a 3+1
splitting of Minkowski space-time, namely a foliation with
space-like hyper-surfaces (namely an arbitrary clock synchronization
convention) with a set of 4-coordinates (observer-dependent
Lorentz-scalar radar 4-coordinates $\sigma^A = (\tau ; \sigma^r)$,
$A = \{\tau , r\}$) adapted to the foliation and having the observer
as origin of the 3-coordinates $\sigma^r$ on each instantaneous
3-space $\Sigma_{\tau}$. The time parameter $\tau$, labeling the
leaves of the foliation, is an arbitrary monotonically increasing
function of the proper time of the observer. Each such foliation
defines a {\it global non-inertial frame centered on the given
observer} if it satisfies the M$\o$ller admissibility conditions
\cite{6}, \cite{3,5}, and if the instantaneous (in general
non-Euclidean) 3-spaces, described by the functions $z^{\mu}(\tau
,\sigma^r)$ giving their embedding in a reference inertial frame in
Minkowski space-time, tend to space-like hyper-planes at spatial
infinity \cite{3}. The 4-metric $g_{AB}(\tau ,\sigma^r) =
z^{\mu}_A(\tau ,\sigma^r)\, \eta_{\mu\nu}\, z^{\nu}_B(\tau
,\sigma^r)$, $z^{\mu}_A(\tau ,\sigma^r) = {{\partial\, z^{\mu}(\tau
,\sigma^r)}\over {\partial\, \sigma^A}}$, in the non-inertial frame
is a function of the embedding obtained from the flat metric
$\eta_{\mu\nu}$ in  inertial Cartesian 4-coordinates $x^{\mu}$ by
means of a general coordinate transformation $x^{\mu}\, \mapsto\,
\sigma^A = (\tau ; \sigma^r)$ with inverse transformation
$\sigma^A\, \mapsto\, x^{\mu} = z^{\mu}(\tau ,\sigma^r)$.
\medskip

If we couple the Lagrangian of the isolated system to an external
gravitational field, we replace the external gravitational 4-metric
with the embedding-dependent 4-metric of a non-inertial frame and we
re-express the components of the isolated system in adapted radar
4-coordinates knowing the instantaneous 3-spaces \footnote{For a
scalar field $\tilde \phi (x)$ we get $\phi (\tau ,\sigma^r) =
\tilde \phi (z(\tau ,\sigma^r))$. For the electro-magnetic potential
${\tilde A}_{\mu}(x)$ and field strength ${\tilde F}_{\mu\nu}(x)$ we
get the Lorentz-scalar fields $A_A(\tau ,\sigma^r) = {\tilde
A}_{\mu}(z(\tau ,\sigma^r))\, z^{\mu}_A(\tau ,\sigma^r)$,
$F_{AB}(\tau ,\sigma^r) = (\partial_A\, A_B -
\partial_B\, A_A)(\tau ,\sigma^r) = {\tilde F}_{\mu\nu}(z(\tau
,\sigma^r)\, z^{\mu}_A(\tau ,\sigma^r)\, z^{\nu}_B(\tau ,\sigma^r)$.
Differently from $\tilde \phi (x)$ and ${\tilde A}_{\mu}(x)$, the
fields $\phi (\tau ,\sigma^r)$ and $A_A(\tau ,\sigma^r)$ {\it know}
the whole instantaneous 3-space $\Sigma_{\tau}$. Scalar particles
are described with Lorentz-scalar 3-coordinates ${\vec \eta}_i(\tau
)$ in $\Sigma_{\tau}$ defined by $x^{\mu}_i(\tau ) = z^{\mu}(\tau
,{\vec \eta}_i(\tau ))$, $i=1,..,N$, i.e. by the intersection of
their world-lines $x^{\mu}_i(\tau )$ (parametrized not with their
proper time, but with the observer's one) with $\Sigma_{\tau}$. As a
consequence, each particle must have a well defined sign of the
energy. Both the world-lines $x^{\mu}_i(\tau )$ and the associated
4-momenta $p_i^{\mu}(\tau )$, satisfying $p_i^2(\tau ) = \sgn\,
m_i^2$ even in presence of interactions, are derived quantities
\cite{3,4}. }, we get the Lagrangian of the parametrized Minkowski
theory for the given isolated system. It is a function of the matter
and fields of the isolated system (now described as Lorentz-scalar
quantities in a non-inertial frame) and of the embedding
$z^{\mu}(\tau ,\sigma^r)$ of the instantaneous 3-spaces of the
non-inertial frame in Minkowski space-time. The main property of the
action functional associated with these Lagrangians is the
invariance \cite{1,3,5} under frame-preserving diffeomorphisms
\footnote{Schmutzer and Plebanski \cite{7} were the only ones
emphasizing the relevance of this subgroup of diffeomorphisms in
their attempt to obtain the theory of non-inertial frames in
Minkowski space-time as a limit from Einstein's general relativity.}
: this implies that the embeddings are {\it gauge variables}, so
that all M$\o$ller-admissible clock synchronization conventions
(i.e. any definition of instantaneous 3-spaces in space-times with
Lorentz signature) are gauge equivalent.\bigskip

Inertial frames are the special class of frames, connected by the
transformations of the Poincare' group (the relativity principle),
selected by the law of inertia. For every configuration of an
isolated system there is a special inertial frame intrinsically
selected by the system itself, the {\it rest frame}, whose
instantaneous 3-spaces (the Wigner 3-spaces with Wigner covariance)
are orthogonal to the conserved 4-momentum of the configuration.
This gives rise to the {\it rest-frame instant form of the
dynamics}. In Ref. \cite{8} there is a full account of the
rest-frame instant form for arbitrary isolated systems, with special
emphasis on the system of "N charged positive-energy scalar
particles with mutual Coulomb interaction plus the transverse
electro-magnetic field of the radiation gauge" \cite{9}. The
particles have Grassmann-valued electric charges (each replaced by a
two-level system, charge $+e$ - charge $-e$, described by a Clifford
algebra, after quantization) so that it is possible\medskip

a) to make a ultraviolet regularization of the Coulomb self-energies
and to eliminate the loops;\hfill

b) to make a infrared regularization killing the emission of soft
photons;\hfill

c) to allow us to have the Lienard-Wiechert transverse potential and
electric field expressible as functions only of the 3-positions and
3-momenta of the particles, independently from the chosen Green
function (retarded, advanced, symmetric, ..).

\medskip

This allows us to have a description of the one-photon exchange
diagram by means of a potential in the framework of a well defined
Cauchy problem for Maxwell equations.\bigskip

In the rest-frame instant form there are two realizations of the
Poincare' algebra:\hfill\medskip

1) An {\it external} one, in which the isolated system is simulated
by means of a {\it decoupled point particle carrying a pole-dipole
structure}: the invariant mass $M$ and the rest spin ${\vec {\bar
S}}$ of the isolated system. This decoupled point particle is
described by the canonical frozen Jacobi data of the non-covariant
external relativistic 3-center of mass: a non-covariant variable
$\vec z = Mc\, {\vec x}_{NW}(0)$ (${\vec x}_{NW}(0)$ is the Cauchy
datum of the Newton-Wigner 3-position ${\vec x}_{NW}(\tau )$) and an
a-dimensional 3-velocity $\vec h = \vec P/ Mc$, $\{ z^i, h^j \} =
\delta^{ij}$. This universal (i.e. independent from the isolated
system) breaking of manifest Lorentz covariance is irrelevant since
the 3-center of mass is decoupled from the internal dynamics. Since
the Poincare' generators are global quantities, the relativistic
center of mass (a known function of such generators) is a {\it
global quantity} not locally determinable (see Ref.\cite{8} for the
non-local aspects of the Newton-Wigner position). The non-covariant
canonical external 4-center of mass (or center of spin) ${\tilde
x}^{\mu}(\tau ) = ({\tilde x}^o(\tau ); {\vec {\tilde x}}(\tau ))$,
the covariant non-canonical external Fokker-Pryce 4-center of
inertia $Y^{\mu}(\tau ) = ({\tilde x}^o(\tau ); \vec Y(\tau ))$ and
the non-covariant non-canonical external M$\o$ller 4-center of
energy $R^{\mu}(\tau ) = ({\tilde x}^o(\tau ); \vec R(\tau ))$ are
known functions of $\tau$, $\vec z$, $\vec h$, $M$, ${\vec {\bar
S}}$ given in Ref.\cite{8}. All these collective variables have the
same constant 4-velocity: ${\dot Y}^{\mu}(\tau ) = {\dot {\tilde
x}}^{\mu}(\tau ) = {\dot R}^{\mu}(\tau ) = P^{\mu}/Mc = h^{\mu}$.

\bigskip

The embedding identifying the Wigner 3-spaces is ($\tau = c T$ is
the Lorentz-scalar rest time)

\beq
 z^{\mu}_W(\tau ,\sigma^u) = Y^{\mu}(\tau ) +
\epsilon^{\mu}_r(\vec h)\, \sigma^r,
 \label{1.1}
 \eeq

\noindent where $Y^{\mu}(\tau )$ is the covariant non-canonical
Fokker-Pryce external 4-center of inertia  and the 3 space-like
4-vectors $\epsilon^{\mu}_r(\vec h)$ are determined by the standard
Wigner boost $L^{\mu}{}_{\nu}(P, {\buildrel \circ \over P})$ for
time-like orbits sending the rest form ${\buildrel \circ \over
P}^{\mu} = Mc\, (1; \vec 0)$ of the total momentum into $P^{\mu} =
Mc\, u^{\mu}(P) = Mc\, \epsilon_{\tau}^{\mu}(\vec h) = Mc\, (\sqrt{1
+ {\vec h}^2}; \vec h) = Mc\, h^{\mu}$ (we collect here the various
notations used in previous papers), i.e. $\epsilon^{\mu}_A(\vec h) =
L^{\mu}{}_{\nu = A}(P, {\buildrel \circ \over P})$. We have
$\epsilon^o_{\tau}(\vec h) = \sqrt{1 + {\vec h}^2}$,
$\epsilon^i_{\tau}(\vec h) = h^i$, $\epsilon^o_r(\vec h) = - \sgn\,
h_r$, $\epsilon^i_r(\vec h) = \delta^i_r - \sgn\, {{h^i\, h_r}\over
{1 + \sqrt{1 + {\vec h}^2} }}$ (see the next Section for the
conventions on the 4-metric).
\bigskip

2) A {\it unfaithful internal} one inside the Wigner 3-spaces, whose
generators are determined by the energy-momentum tensor, obtained
from the Lagrangian of the parametrized Minkowski theory associated
with the given isolated system. The only non-vanishing generators
are $M$ and ${\vec {\bar S}}$. The vanishing of the internal
3-momentum is the {\it rest-frame condition}, while the vanishing of
the internal (interaction-dependent) Lorentz boosts {\it eliminates
the internal 3-center of mass} (this avoids a double counting of the
center of mass). As a consequence, the dynamics inside the
instantaneous Wigner 3-spaces is described {\it only by
Wigner-covariant relative variable and momenta} (${\vec \rho}_a(\tau
)$, ${\vec \pi}_a(\tau )$, $a=1,..,N-1$, for particles). The
invariant mass $M$ is the Hamiltonian for the internal Hamilton
equations. It is possible to make an orbit reconstruction \cite{4}
for the particles in the form ${\vec \eta}_i(\tau ) = {\vec
f}_i({\vec \rho}_a(\tau ), {\vec \pi}_a(\tau ))$ and to determine
the world-lines  \footnote{They turn out to be {\it covariant
non-canonical predictive coordinates}: $\{x^{\mu}_i(\tau ),
x^{\nu}_j(\tau )  \} \not= 0$ for all $i$ and $j$, $\mu$ and $\nu$.
Let us remark that this does not imply a breaking of microcausality,
which is preserved at the level of the 3-coordinates ${\vec
\eta}_i(\tau )$.},

\beq
 x^{\mu}_i(\tau ) = z^{\mu}_W(\tau, {\vec
\eta}_i(\tau )) = Y^{\mu}(\tau ) + \epsilon^{\mu}_r(\vec h)\,
f^r_i({\vec \rho}_a(\tau ), {\vec \pi}_a(\tau )).
 \label{1.2}
 \eeq

\bigskip

In this paper we study in detail the properties of global admissible
non-inertial frames in Minkowski space-time, generalizing the
notions defined in the inertial rest-frame instant form of dynamics.
We show that also in non-inertial frames every isolated system can
be described as an external decoupled non-covariant canonical center
of mass (described by frozen Jacobi data) carrying a pole-dipole
structure: the invariant mass and an effective spin. Moreover,
following the same methods developed for the inertial rest frame, we
identify the constraints eliminating the internal 3-center of mass
inside the instantaneous 3-spaces.\medskip

In the admissible non-inertial frames  the instantaneous 3-spaces
are orthogonal to a given fixed 4-vector $l^{\mu}_{(\infty )}$ at
spatial infinity \footnote{A preliminary description of particles
and of their quantization in a class of such frames was given in
Ref.\cite{10}. There we introduced an auxiliary decoupled scalar
particle whose 4-momentum coincides with $l^{\mu}_{(\infty )}$. Here
we will avoid to use this method.}.\medskip

Then we will restrict the description to the special family of
non-inertial frames, in which the instantaneous 3-spaces tend to
Wigner 3-spaces, orthogonal to the conserved 4-momentum of the
isolated system, at spatial infinity (i.e. $l^{\mu}_{(\infty )} =
h^{\mu} = P^{\mu}/Mc$): they are the {\it non-inertial rest frames},
a non-inertial extension of the inertial ones. This will allow us to
define the non-inertial rest-frame instant form of dynamics. The
non-inertial rest frame are the only ones allowed by the equivalence
principle in the treatment of canonical metric and tetrad gravity in
asymptotically flat and globally hyperbolic space-times without
super-translations as shown in Refs. \cite{5,11}.\bigskip

Even if in a non-covariant way, which is however consistent with the
coordinate-dependence of the inertial effects, we will give a
unified special relativistic description of many properties of
isolated systems in accelerated frames, which are scattered in the
literature and treated without a global interpretative framework.
\bigskip

Then, as in Ref.\cite{8}, we consider the description of the
isolated system of positive-energy scalar particles with
Grassmann-valued electric charges plus the electro-magnetic field as
a parametrized Minkowski theory. As a consequence we obtain both
Maxwell equations and their Hamiltonian description in non-inertial
frames.\medskip

By means of a non-covariant decomposition we define the {\it
non-inertial non-covariant radiation gauge}: this allows to
visualize the non-inertial dynamics of transverse electro-magnetic
fields, the electro-magnetic Dirac observables. We find the
modification of the Coulomb potential in a non-inertial frame: its
non-covariance is due to the same type of coordinate-dependence
present in the {\it relativistic inertial potentials}, which are
explicitly identified for the first time (they are the components of
the 4-metric $g_{AB}(\tau, \sigma^r)$) and shown to have the correct
Newtonian limit. The final Dirac Hamiltonian will contain not only
the invariant mass $Mc$ but also the modifications induced by the
potentials associated with the inertial effects present in the given
non-inertial frame.

\bigskip

In a second paper we will study applications of this 3+1 framework
to the description of the rotating disk, the Sagnac effect, the
Faraday rotation, the wrap-up effect. A preliminary version of the
material contained in these two paper is present in arXiv:
0812.3057.

\bigskip

In Section II we review the admissible 3+1 splittings of Minkowski
space-time and the properties of the associated global non-inertial
frames (Subsection A), we compare them with the accelerated
coordinate systems associated with the 1+3 point of view (Subsection
B) and we define the non-covariant notations for the
electro-magnetic field in non-inertial frames (Subsection
C).\medskip

In Section III we study the description of the isolated system
"charged scalar positive-energy particles plus the electro-magnetic
field" in the framework of parametrized Minkowski theories. In
particular we show that in non-inertial frames and also in inertial
frames with non-Cartesian coordinates there is no true conservation
law for the energy-momentum tensor: like in general relativity one
could introduce a coordinate-dependent energy-momentum pseudo-tensor
describing the contribution of the foliation associated with the
admissible 3+1 splitting of Minkowski space-time. However, reverting
to inertial frames, it is possible to find the conserved (Poincare'
4-vector) 4-momentum of the isolated system.
\medskip

In Section IV we give the Hamiltonian description and the Hamilton
equations of the isolated system "charged scalar positive-energy
particles plus the electro-magnetic field" in admissible
non-inertial frames (Subsection A). Then we introduce the
non-covariant radiation gauge for the electro-magnetic field and we
find both the inertial forces and the non-inertial expression of the
Coulomb potential (Subsection B). Finally we evaluate the
non-relativistic limit recovering the Newtonian apparent inertial
forces (Subsection C).\medskip

In Section V we review the determination of the internal Poincare'
generators and of the constraints eliminating the internal 3-center
of mass in the inertial rest frames (Subsection A). Then we show how
these results are modified  in the special family of the
non-inertial rest frames (Subsections B and C) and in arbitrary
admissible non-inertial frames (Subsection D) .\medskip

In the Conclusions we give an overview of the results obtained in
this paper and we list the applications to be discussed in the
second paper.

\bigskip

In Appendix A there is the expression of the Landau-Lifschitz
non-inertial electro-magnetic fields in the 3+1 point of
view.\medskip

In Appendix B there is a comparison of the covariant and
non-covariant decompositions of the electro-magnetic field in
non-inertial frames and the definition of the non-covariant
radiation gauge.

\vfill\eject

\section{Admissible 3+1 Splittings of Minkowski Space-Time and Notations}

We use the signature convention $\eta_{\mu\nu} = \sgn\, (+---)$,
$\sgn = \pm 1$, for the flat Minkowski metric ($\sgn = +1$ is the
particle physics convention, while $\sgn = - 1$ is the one of
general relativity), since it has been used in Refs.\cite{11} for
canonical gravity. Since in Ref. \cite{8} the convention $\sgn = +
1$ was used, in this Section we also introduce the notations needed
for the treatment of the electro-magnetic field in non-inertial
frames.\bigskip

\subsection{Admissible 3+1 Splittings of Minkowski Space-Time}

Let us consider an admissible 3+1 splitting of Minkowski space-time,
whose instantaneous 3-spaces $\Sigma_{\tau}$ are identified by the
embedding $z^{\mu}(\tau ,\sigma^r)$. The radar 4-coordinates
$\sigma^A = (\tau ;\sigma^r)$ are adapted to an arbitrary time-like
observer with world-line $x^{\mu}(\tau )$ in the reference inertial
frame, chosen as the origin of the curvilinear 3-coordinates
$\sigma^r$ on each $\Sigma_{\tau}$. The Lorentz-scalar time $\tau$,
with dimensions $[\tau ] = [c\, t] = [l]$, is a monotonically
increasing function of the proper time of the observer. Therefore,
we can put the embeddings in the following form

\bea
 z^{\mu}(\tau ,\sigma^u ) &=& x^{\mu}(\tau ) + F^{\mu}(\tau ,
 \sigma^u ) = x^{\mu}_o + \epsilon^{\mu}_A\, \Big[f^A(\tau ) + F^A(\tau
 ,\sigma^u )\Big],\qquad F^{\mu}(\tau ,\vec o) = 0,\nonumber \\
 &&{}\nonumber \\
 x^{\mu}(\tau ) &=& x^{\mu}_o + \epsilon^{\mu}_A\, f^A(\tau ).
 \label{2.1}
 \eea

\noindent At spatial infinity $z^{\mu}(\tau ,\sigma^r)$ must tend in
a direction-independent way to a space-like hyper-plane with unit
time-like normal $l^{\mu}_{(\infty )} = \epsilon^{\mu}_{\tau}$: this
implies $F^{\mu}(\tau ,\sigma^s)\, \rightarrow\,
\epsilon^{\mu}_{(\infty )\, r}\, \sigma^r$ with the 3 space-like
4-vectors $\epsilon^{\mu}_{(\infty )\, r} = \epsilon^{\mu}_r$
orthogonal to $l^{\mu}_{(\infty )}$. The asymptotic orthonormal
tetrads $\epsilon^{\mu}_A$ are associated to  asymptotic inertial
observers and   satisfy $\epsilon^{\mu}_A\, \eta_{\mu\nu}\,
\epsilon^{\nu}_B = \eta_{AB}$. Let us remark that the natural
notation for the asymptotic tetrads would be $\epsilon^{\mu}_{(A)}$.
However, for the sake of simplicity we shall use the notation
$\epsilon^{\mu}_A$ for $\delta_A^{(B)}\,
\epsilon^{\mu}_{(B)}$.\medskip

The time-like observer $x^{\mu}(\tau )$, origin of the 3-coordinates
on the instantaneous 3-spaces $\Sigma_{\tau}$, has the following
unit 4-velocity and 4-acceleration (we use the notation ${\dot
x}^{\mu}(\tau ) = {{d\, x^{\mu}(\tau )}\over {d\tau}}$; it must be
$\sgn\, {\dot x}^2(\tau ) > 0$)

\begin{eqnarray*}
 u^{\mu}(\tau ) &=& {{{\dot x}^{\mu}(\tau )}\over {\sqrt{\sgn\, {\dot x}^2(\tau )}}}
 = \epsilon^{\mu}_A\, u^A(\tau ),\qquad u^2(\tau ) = \sgn,\nonumber \\
 &&{}\nonumber \\
 &&u^A(\tau ) = {{{\dot f}^A(\tau)}\over {\sqrt{\Big({\dot f}^{\tau}(\tau )\Big)^2 -
 \sum_u\, \Big({\dot f}^u(\tau )\Big)^2}}},\qquad
 \Big({\dot f}^{\tau}(\tau )\Big)^2\, >\, \sum_u\, \Big({\dot f}^u(\tau
 )\Big)^2,\end{eqnarray*}

 \bea
 a^{\mu}(\tau ) &=& {{d u^{\mu}(\tau )}\over {d\tau}} = \epsilon^{\mu}_A\, a^A(\tau
 ),\qquad a_{\mu}(\tau )\, u^{\mu}(\tau ) = 0,\nonumber \\
 &&{}\nonumber \\
 &&a^A(\tau) = {{{\ddot f}^A(\tau )\, \Big(\Big({\dot f}^{\tau}(\tau )\Big)^2 -
 \sum_u\, \Big({\dot f}^u(\tau )\Big)^2\Big) - {\dot f}^A(\tau )\,
 \Big({\dot f}^{\tau}(\tau )\, {\ddot f}^{\tau}(\tau ) - \sum_u\,
 {\dot f}^u(\tau )\, {\ddot f}^u(\tau )\Big)}\over {\Big(\Big({\dot f}^{\tau}(\tau )\Big)^2 -
 \sum_u\, \Big({\dot f}^u(\tau )\Big)^2\Big)^{3/2}}}.\nonumber \\
 &&{}
 \label{2.2}
 \eea

As a consequence we can write $u^{\mu}(\tau ) =
L^{\mu}{}_{\nu}(u(\tau ), {\buildrel \circ \over u})\, {\buildrel
\circ \over u}^{\nu}$, ${\buildrel \circ \over u}^{\mu} = \sgn\, (1;
\vec 0)$, by using the standard Wigner boost for time-like
4-vectors.
\bigskip

Eqs.(\ref{2.1}) imply

\bea
 z^{\mu}_{\tau}(\tau ,\sigma^u ) &=& \partial_{\tau}\, z^{\mu}(\tau
 ,\sigma^u) = {\dot x}^{\mu}(\tau ) + \partial_{\tau}\,
 F^{\mu}(\tau ,\sigma^u ) = \epsilon^{\mu}_A\, \Big({\dot f}^A(\tau ) +
 \partial_{\tau}\, F^A(\tau ,\sigma^u )\Big) =\nonumber \\
 &=& (1 + n(\tau ,\sigma^u ))\, l^{\mu}(\tau ,\sigma^u ) + h^{rs}(\tau
 ,\sigma^u )\, n_r(\tau ,\sigma^u )\, z^{\mu}_s(\tau ,\sigma^u ),
 \nonumber \\
  &&{}\nonumber \\
 z^{\mu}_r(\tau ,\sigma^u ) &=& \partial_r\, z^{\mu}(\tau ,\sigma^u) =
 \partial_r\, F^{\mu}(\tau , \sigma^u ) = \epsilon^{\mu}_A\,
 \partial_r\, F^A(\tau ,\sigma^u).
 \label{2.3}
 \eea

\bigskip

While the 3 independent space-like 4-vectors $z^{\mu}_r(\tau
,\sigma^u)$ are tangent to $\Sigma_{\tau}$,  the time-like 4-vector
$z^{\mu}_{\tau}(\tau ,\sigma^u)$ has been decomposed on them and on
the unit normal $l^{\mu}(\tau ,\sigma^u)$, $l^2(\tau ,\sigma^u) =
\sgn$, to $\Sigma_{\tau}$ ($l_{\mu}(\tau ,\sigma^u)\, z^{\mu}_r(\tau
,\sigma^u) = 0$). This decomposition defines the {\it lapse and
shift functions} $N(\tau ,\sigma^u) = 1 + n(\tau ,\sigma^u)
> 0$ and $N^r(\tau ,\sigma^u) = n^r(\tau ,\sigma^u)$ (we use the
notation of Ref.\cite{11}). At spatial infinity we have:
$l^{\mu}(\tau ,\sigma^u)\, \rightarrow \, l^{\mu}_{(\infty )} =
\epsilon^{\mu}_{\tau}$, $N(\tau ,\sigma^u)\, \rightarrow\, 1$
($n(\tau ,\sigma^u)\, \rightarrow\, 0$), $n^r(\tau ,\sigma^u)\,
\rightarrow 0$.\bigskip

The 4-metric induced by the 3+1 splitting is $g_{AB}(\tau ,\sigma^u)
= z^{\mu}_A(\tau ,\sigma^u)\, \eta_{\mu\nu}\, z^{\nu}_B(\tau
,\sigma^u)$ and we have

\begin{eqnarray*}
 g_{\tau\tau}(\tau ,\sigma^u) &=& \Big[z^{\mu}_{\tau}\, \eta_{\mu\nu}\,
 z^{\nu}_{\tau}\Big](\tau ,\sigma^u) = \nonumber \\
 &=& \sgn\, \Big[ \Big({\dot f}^{\tau}(\tau )+ \partial_{\tau}\,
 F^{\tau}(\tau ,\sigma^v )\Big)^2 - \sum_u\, \Big({\dot f}^u(\tau )+ \partial_{\tau}\,
 F^u(\tau ,\sigma^v )\Big)^2\Big] =\nonumber \\
 &=&\sgn\, \Big[\Big(1 + n(\tau ,\sigma^v )\Big)^2 - h^{rs}(\tau
 ,\sigma^v)\, n_r(\tau ,\sigma^v )\, n_s(\tau ,\sigma^v )\Big],
 \end{eqnarray*}

\bea
 g_{\tau r}(\tau ,\sigma^v) &=& \Big[z^{\mu}_{\tau}\, \eta_{\mu\nu}\,
 z^{\nu}_r\Big](\tau ,\sigma^v) =\nonumber \\
 &=& - \sgn\, \Big[\sum_u\, \Big({\dot f}^u(\tau ) + \partial_{\tau}\, F^u(\tau
 ,\sigma^v )\Big)\, \partial_r\, F^u(\tau ,\sigma^v ) -\nonumber \\
 &-& \Big({\dot f}^{\tau}(\tau ) + \partial_{\tau}\, F^{\tau}(\tau ,\sigma^v )\Big)\,
 \partial_r\, F^{\tau}(\tau ,\sigma^v )\Big] =\nonumber \\
 &=&  - \sgn\, n_r(\tau ,\sigma^v) = g_{rs}(\tau ,\sigma^v)\, n^s(\tau ,\sigma^v) =
 - \sgn\, h_{rs}(\tau ,\sigma^v)\, n^s(\tau ,\sigma^v),\nonumber \\
 &&{}\nonumber \\
 g_{rs}(\tau ,\sigma^v) &=& \Big[z^{\mu}_r\, \eta_{\mu\nu}\, z^{\nu}_s\Big](\tau
 ,\sigma^v) =\nonumber \\
 &=& - \sgn\, \Big[\sum_u\, \partial_r\, F^u(\tau ,\sigma^v )\, \partial_s\,
 F^u(\tau ,\sigma^v ) - \partial_r\, F^{\tau}(\tau ,\sigma^v )\, \partial_s\,
 F^{\tau}(\tau ,\sigma^v )\Big] =\nonumber \\
 &=&  - \sgn\, h_{rs}(\tau ,\sigma^v).
 \label{2.4}
 \eea

\noindent While the 3-metric $g_{rs}$ in $\Sigma_{\tau}$ and its
inverse $\gamma^{rs}$ ($\gamma^{ru}\, g_{us} = \delta^r_s$) have
signature $\sgn\, (---)$, the 3-metric $h_{rs}$ and its inverse
$h^{rs} = - \sgn\, \gamma^{rs}$ ($h^{ru}\, h_{us} = \delta^r_s$)
have signature $(+++)$.\medskip

For the inverse 4-metric $g^{AB}$ ($g^{AC}\, g_{CB} = \delta^A_B$)
we have

\bea
 g^{\tau\tau} &=& {{\sgn}\over {(1 + n)^2}},\qquad g^{\tau\tau}\, g^{rs}
 - g^{\tau r}\, g^{\tau s} = - {{h^{rs}}\over {(1 + n)^2}}, \nonumber \\
 g^{\tau r} &=& - \sgn\, {{n^r}\over {(1 + n)^2}},\qquad
 g^{rs} = - \sgn\, \Big(h^{rs} - {{n^r\, n^s}\over
 {(1 + n)^2}}\Big).
 \label{2.5}
 \eea

For the determinants we have

\bea
 &&\gamma = - \sgn\, det\, g_{rs} = det\, h_{rs} > 0,\qquad
 g = det\, g_{AB}  < 0,\quad \Rightarrow\,\, \sqrt{- g} = (1 +
 n)\, \sqrt{\gamma}.\nonumber \\
 &&{}
 \label{2.6}
 \eea

Finally the unit normal to the simultaneity surfaces $\Sigma_{\tau}$
has the expression

\begin{eqnarray*}
 l^{\mu}(\tau ,\sigma^u) &=& \Big[\eta^{\mu}{}_{\alpha\beta\gamma}\, z_1^{\alpha}\,
 z_2^{\beta}\, z_3^{\gamma}\Big](\tau ,\sigma^u) = \Big[{{1}\over {\sqrt{\gamma}}}\,
 \epsilon^{\mu}{}_{\alpha\beta\gamma}\, z_1^{\alpha}\, z_2^{\beta}\,
 z_3^{\gamma}\Big](\tau ,\sigma^u) =\nonumber \\
 &=& \epsilon^{\mu}_A\, l^A(\tau ,\sigma^v ) = \epsilon^{\mu}_A\, \eta^{AE}\,
 \Big({{\epsilon_{EBCD}}\over {\sqrt{\gamma}}}\, \partial_1\, F^B\,
 \partial_2\, F^C\, \partial_3\, F^D\Big)(\tau , \sigma^v ) =\nonumber \\
 &=& L^{\mu}{}_{\nu}(l(\tau ,\sigma^v ), {\buildrel \circ \over l})\,
 {\buildrel \circ \over l}^{\nu},\qquad {\buildrel \circ \over l}^{\mu}
 = \sgn\, (1; \vec 0),
  \end{eqnarray*}

 \bea
 l^2(\tau ,\sigma^u ) = \sgn, &&\Rightarrow  \Big(l^{\tau}(\tau
 ,\sigma^u )\Big)^2\, >\, \sum_u\, \Big(l^u(\tau ,\sigma^v
 )\Big)^2,\nonumber \\
 &&{}\nonumber \\
 \Rightarrow&& \eta_{\mu\nu} = \sgn\, \Big(l_{\mu}\, l_{\nu} -
 z_{r\mu}\, h^{rs}\, z_{s\nu}\Big)(\tau ,\sigma^v).
  \label{2.7}
  \eea
\medskip

The 3+1 splitting for which $l^{\mu}$ is constant, i.e. $\tau$- and
$\sigma^r$-independent, have the instantaneous 3-spaces
corresponding to parallel space-like hyper-planes: when the frame is
non-inertial these hyper-planes are not equally spaced due to linear
acceleration and/or have rotating 3-coordinates, so that they are
not Euclidean 3-spaces.

\medskip

The Wigner boost sending ${\buildrel \circ \over l}^{\mu}$ into
$l^{\mu}(\tau ,\sigma^u )$ ($\beta_l^i = - \sgn\, \beta_{l\, i}$)
has the following expression

\bea
 &&L^{\mu}{}_{\nu}(l(\tau ,\sigma^u ), {\buildrel \circ \over l}) =
  \begin{array}{|ll|} \gamma_l & \gamma_l\, \beta_l^i \\
 \gamma_l\, \beta_l^j & \delta^{ij} + (\gamma_l - 1)\, {{\beta_l^i\, \beta_l^j}\over
 {\sum_k\, (\beta_l^k)^2}} \end{array}(\tau ,\sigma^u ),\nonumber \\
 &&{}\nonumber \\
 &&l^{\mu}(\tau ,\sigma^u ) = L^{\mu}{}_o(l(\tau ,\sigma^u ),
 {\buildrel \circ \over l}) = \gamma_l(\tau ,\sigma^u )\, \Big(1;
 \beta_l^i(\tau ,\sigma^u )\Big) =
   \epsilon^{\mu}_A\, l^A(\tau ,\sigma^u ) {\buildrel {def}\over
 =} \epsilon^{\mu}_o(l(\tau ,\sigma^u )),\nonumber \\
 &&\epsilon^{\mu}_j(l(\tau ,\sigma^u )) {\buildrel {def}\over =}
 L^{\mu}{}_j(l(\tau ,\sigma^u ), {\buildrel \circ \over
 l}),\nonumber \\
 &&{}\nonumber \\
 &&\gamma_l = {1\over {\sqrt{1 - \sum_u\, (\beta_l^u)^2}}} = l^o
 = {1\over {\sqrt{\gamma}}}\, \epsilon^o_A\, \eta^{AE}\,
 \epsilon_{EBCD}\, \partial_1\, F^B\, \partial_2\, F^C\,
 \partial_3\, F^D,\nonumber \\
 &&\beta^i_l = \gamma_l^{-1}\, l^i = {{\epsilon^i_A\, \eta^{AE}\,
  \epsilon_{EBCD}\, \partial_1\, F^B\, \partial_2\, F^C\,
   \partial_3\, F^D}\over {\epsilon^o_A\, \eta^{AE}\,  \epsilon_{EBCD}\,
   \partial_1\, F^B\, \partial_2\, F^C\,  \partial_3\, F^D}}.
 \label{2.8}
 \eea

\noindent The orthonormal tetrads $\epsilon^{\mu}_A(l(\tau ,\sigma^u
)) = L^{\mu}{}_A(l(\tau ,\sigma^u ), {\buildrel \circ \over l})$,
$\eta_{\mu\nu}\, \epsilon^{\mu}_A(l(\tau ,\sigma^u ))\,
\epsilon^{\mu}_B(l(\tau ,\sigma^u )) = \eta_{AB}$, are the columns
of the Wigner boost.
\medskip

The Wigner boosts $L^{\mu}{}_{\nu}(u(\tau ), {\buildrel \circ\over
u})$ has a similar parametrization in terms of parameters
$\beta^i_u(\tau )$.

\medskip

The M$\o$ller admissibility conditions \cite{6}, \cite{3}, implying
that the 3+1 splitting gives rise to a nice foliation of Minkowski
space-time with space-like leaves identifying the instantaneous
3-spaces $\Sigma_{\tau}$, are

\bea
 && \sgn\, g_{\tau\tau}(\tau ,\sigma^u) = \Big[(1 + n)^2 - h^{rs}\, n_r\,
n_s\Big](\tau ,\sigma^u)\,\, > 0,\qquad  \sgn\, g_{rr}(\tau
,\sigma^u ) = - h_{rr}(\tau
,\sigma^u) < 0,\nonumber \\
 &&{}\nonumber \\
 && \begin{array}{|ll|}
 g_{rr}(\tau ,\sigma^u ) & g_{rs}(\tau ,\sigma^u ) \\ g_{sr}(\tau
 ,\sigma^u ) & g_{ss}(\tau ,\sigma^u ) \end{array} = \begin{array}{|ll|}
 h_{rr}(\tau ,\sigma^u ) & h_{rs}(\tau ,\sigma^u ) \\ h_{sr}(\tau
 ,\sigma^u ) & h_{ss}(\tau ,\sigma^u ) \end{array}\, > 0, \nonumber \\
 &&{}\nonumber \\
 &&\sgn\, det\, [g_{rs}(\tau ,\sigma^u )] = - \gamma (\tau ,\sigma^u)\, <
 0,\qquad \Rightarrow det\, [g_{AB}(\tau ,\sigma^u )]\, <
 0.\nonumber \\
 &&{}
 \label{2.9}
 \eea
 \bigskip

\noindent They are restrictions on the functions $F^{\mu}(\tau
,\sigma^r)$ of Eqs.(\ref{2.1}). When they are satisfied,
Eqs.(\ref{2.1}) define a {\it global (in general non-rigid)
non-inertial frame}. While linear accelerations are not restricted
by Eqs.(\ref{2.9}), {\it rigid rotations are forbidden} \cite{3}.
The condition $\sgn\, g_{\tau\tau}(\tau ,\sigma^u) > 0$ implies that
in each point $\sigma^u$ the tangential velocity $\omega (\tau
,\sigma^u)\, r(\tau ,\sigma^u)$ is less than $c$: instead with
$\omega = \omega (\tau )$, like it happens in rigidly  rotating
coordinate systems, we get $\sgn\, g_{\tau\tau}(\tau ,R^u) = 0$ at
the distance $R^u$ from the rotation axis where $\omega\, R = c$, so
that the time-like vector $z^{\mu}_{\tau}(\tau ,\sigma^u)$ would
become a null vector (the so-called {\it horizon problem} of the
rotating disk).\medskip

Since $1 + n(\tau ,\sigma^u ) > 0$ gives the proper time distance
from $\Sigma_{\tau}$ to $\Sigma_{\tau + d\tau}$ along the world-line
of the Eulerian observer through $(\tau ,\sigma^u )$ with tangent
vector $l^{\mu}(\tau ,\sigma^u )$, the condition $1 + n(\tau ,
\sigma^u ) > 0$ implies that $\Sigma_{\tau}$ and $\Sigma_{\tau +
d\tau}$ {\it intersect nowhere}. By continuity this implies that the
M$\o$ller-admissible 3+1 splittings are nice foliations with
space-like leaves tending to space-like hyper-planes at spatial
infinity in a direction-independent way.
\bigskip

Since the 3-metric $h_{rs}(\tau ,\sigma^u)$ is a real symmetric
matrix, it can be diagonalized with a rotation matrix
$V(\theta^i(\tau ,\sigma^u))$, $V^T = V^{-1}$ ($\theta^i(\tau
,\sigma^u)$ are Euler angles). Therefore, by using the notations of
Ref.\cite{12} for canonical gravity in the York canonical basis, we
can parametrize the 3-metric in the following form \footnote{ As
shown in Ref.\cite{12} the basic variables of tetrad gravity are not
the embedding $z^{\mu}(\tau ,\sigma^u)$ but tetrads
$E^{\mu}_{(\alpha )}(\tau ,\sigma^u)$, defined after an admissible
3+1 splitting of the space-time identifying the instantaneous
3-spaces $\Sigma_{\tau}$. The quantities $z^{\mu}_A(\tau ,\sigma^u)$
are now the {\it transition coefficients} from world components of
tensors to $\Sigma_{\tau}$-adapted components in radar coordinates
$\sigma^A = (\tau ,\sigma^u)$: $E^{\mu}_{(\alpha )} = z^{\mu}_A\,
E^A_{(\alpha )}$. The 4-metric tensor is defined by the associated
cotetrads: $g_{AB} = E^{(\alpha )}_A\, \eta_{(\alpha )(\beta )}\,
E_B^{(\beta )}$. The gauge variables of tetrad gravity in the York
canonical basis are six parameters of the Lorentz group acting on
the flat $(\alpha )$ indices of the tetrads $E^{\mu}_{(\alpha )}$,
the lapse ($1 + n$) and shift ($n_r$) functions, the Euler angles
$\theta^i$ and the momentum variable conjugate to $\phi^6 =
\gamma^{1/2}$, i.e. the trace ${}^3K$ of the extrinsic curvature of
the instantaneous 3-space $\Sigma_{\tau}$. The volume variable $\phi
= \gamma^{1/12}$ is determined by the super-hamiltonian constraint.
The momenta $\pi_i^{(\theta )}$, conjugate to $\theta^i$, are
determined by the super-momentum constraints. The symmetric 3-metric
$h_{rs} = - \sgn\, g_{rs}$ can be put in the form $h_{rs} = \sum_a\,
\lambda_a\, V_{ra}(\theta^i)\, V_{sa}(\theta^i)$, where  the
eigenvalues (assumed non degenerate) have the expression $\lambda_a
= \phi^4\, e^{2\, \sum_{\bar a}\, \gamma_{\bar aa}\, R_{\bar a}}$.
The two functions $R_{\bar a}$ describe the two physical degrees of
freedom of the gravitational field. A gauge fixing for $\theta^i$
and ${}^3K$ implies the determination of the lapse and shift
functions. \hfill\break
 Instead in non-inertial frames in Minkowski space-time, where gravity
is absent, all the functions ($n$, $n_r$, $\gamma = \phi^{12}$,
$\theta^i$, $R_{\bar a}$ ) parametrizing the components of the
4-metric $g_{AB}$ of Eq.(\ref{2.4}) are {\it gauge variables}
globally described by the embedding $z^{\mu}(\tau ,\sigma^u)$ of
Eq.(\ref{2.1}). \hfill\break
 In parametrized Minkowski theories (see the next Section), where the
embedding is the basic variable, in absence of matter the
super-hamiltonian and super-momentum constraints are replaced by the
vanishing of the momentum $\rho_{\mu}(\tau ,\sigma^u) \approx 0$,
see Eq.(\ref{3.10}), conjugated to $z^{\mu}(\tau ,\sigma^u)$. If we
fix $z^{\mu}(\tau ,\sigma^u)$ like in Eq.(\ref{4.1}), so that the
3-metric is completely fixed ($\theta^i$, $\gamma$ and $R_{\bar a}$
are given), then Eqs.(\ref{4.2}) of Section IV determine the lapse
and shift functions. The extrinsic curvature is determined either
from the variation of the unit normal $l^{\mu}$ to $\Sigma_{\tau}$
or from ${}^3K_{rs} = {1\over {2\, (1 + n)}}\, (n_{r|s} + n_{s|r} -
\partial_{\tau}\, h_{rs})$.}:

\bea
  h_{rs}(\tau ,\sigma^u ) &=& - \sgn\, g_{rs}(\tau ,\sigma^u )
 = \Big(\gamma^{1/3}\, \sum_a\, Q_a^2\, V_{ra}(\theta^i)\, V_{sa}(\theta^i)
 \Big)(\tau ,\sigma^u ) = \nonumber \\
 &=& \sum_a\, e_{(a)r}(\tau ,\sigma^u )\, e_{(a)s}(\tau ,
 \sigma^u ), \nonumber \\
 &&{}\nonumber \\
 &&e_{(a)r} = \gamma^{1/6}\, Q_a\, V_{ra}(\theta^i),\qquad e^r_{(a)}
 = \gamma^{-1/6}\, Q^{-1}_a\, V_{ra}(\theta^i),\nonumber \\
 &&\gamma = det\, h_{rs},\qquad Q_a = e^{\sum_{\bar a}\, \gamma_{\bar aa}\, R_{\bar
 a}},
 \label{2.10}
 \eea

\noindent where $e_{(a)r}(\tau ,\sigma^u)$ and $e^r_{(a)}(\tau
,\sigma^u)$, ($\sum_a\, e^r_{(a)}\, e_{(a)s} = \delta^r_s$,
$\sum_r\, e^r_{(a)}\, e_{(b)r} = \delta_{ab}$) are cotriads and
triads on $\Sigma_{\tau}$, respectively. At spatial infinity we have
$e^r_{(a)}(\tau ,\sigma^u)\, \rightarrow\, \delta^r_a$,
$e_{(a)r}(\tau ,\sigma^u)\, \rightarrow\, \delta_{ra}$. To express
$e_{(a)r}$ in terms of $\partial_r\, F^A$, we must find the
eigenvalues and the eigenvectors of the matrix $h_{rs}$ in the form
given in Eqs.(\ref{2.4}).\medskip

The three eigenvalues of the 3-metric are $\lambda_a =
\gamma^{1/3}\, Q_a^2> 0$. The positivity of the eigenvalues is
implied by the M$\o$ller conditions (\ref{2.9}): $\lambda_1\,
\lambda_2\, \lambda_3 = \gamma
> 0$, $\lambda_1 + \lambda_2 + \lambda_3 = h_{11} + h_{22} + h_{33}
> 0$, $\lambda_1\, \lambda_2 + \lambda_2\, \lambda_3 + \lambda_3\,
\lambda_1 =  \begin{array}{|ll|} h_{11} & h_{12} \\ h_{21} & h_{22}
\end{array} +  \begin{array}{|ll|} h_{11} & h_{13} \\ h_{31} & h_{33}
\end{array} + \begin{array}{|ll|} h_{22} & h_{23} \\ h_{32} & h_{33}
\end{array} > 0$.\medskip

This implies that the three 4-vectors $z^{\mu}_r(\tau ,\sigma^u )$
are space-like for every $\vec \sigma$, so that the unit normal
$l^{\mu}(\tau ,\sigma^u )$ is time-like everywhere on the
instantaneous 3-spaces.
\medskip

The M$\o$ller condition $\sgn\, g_{\tau\tau}(\tau ,\sigma^u) > 0$ of
Eqs.(\ref{2.9}) implies that $z^{\mu}_{\tau}(\tau ,\sigma^u )$ is
everywhere time-like on the instantaneous 3-spaces $\Sigma_{\tau}$.

\medskip

Let us remark that while the generic 3-spaces $\Sigma_{\tau}$ have a
3-metric with 3 distinct eigenvalues, there is a family of 3+1
splittings with two coinciding eigenvalues of $h_{rs}(\tau ,
\sigma^u)$ and another family with all the 3 eigenvalues coinciding:
they correspond to the existence of symmetries corresponding to the
Killing symmetries of Einstein general relativity.

\bigskip

The lapse and shift functions have the following expressions

\bea
 1 + n(\tau ,\sigma^u ) &=& \sgn\, z^{\mu}_{\tau}(\tau ,
 \sigma^u )\, l_{\mu}(\tau ,\sigma^u ) = \Big({{\sgn}\over {\sqrt{\gamma}}}\,
 \epsilon_{\mu\alpha\beta\gamma}\, z^{\mu}_{\tau}\, z^{\alpha}_1\,
 z^{\beta}_2\, z^{\gamma}_3\Big)(\tau ,\sigma^u) =\nonumber \\
 &=& \Big({\dot f}^{\tau}(\tau ) +
 \partial_{\tau}\, F^{\tau}(\tau ,\sigma^u )\Big)\,
 l^{\tau}(\tau ,\sigma^u ) -\nonumber \\
 &-& \sum_u\, \Big({\dot f}^u(\tau ) + \partial_{\tau}\,
 F^u(\tau ,\sigma^u )\Big)\, l^u(\tau ,\sigma^u )\, > 0,\nonumber \\
 &&{}\nonumber \\
  n_r(\tau ,\sigma^u) &=& h_{rs}(\tau ,\sigma^u)\, n^s(\tau ,\sigma^u) =
  \sum_u\, \Big({\dot f}^u(\tau ) + \partial_{\tau}\, F^u(\tau ,\sigma^v )\Big)\,
 \partial_r\, F^u(\tau ,\sigma^v ) -\nonumber \\
 &-& \Big({\dot f}^{\tau}(\tau ) + \partial_{\tau}\, F^{\tau}(\tau ,\sigma^v )\Big)\,
 \partial_r\, F^{\tau}(\tau ,\sigma^v ).
 \label{2.11}
 \eea
\medskip

Let us also remark that all the information carried by
$\epsilon^{\mu}_A\, f^A(\tau )$, i.e. the velocity and acceleration
of the time-like observer $x^{\mu}(\tau )$, is hidden in the lapse
and shift functions.

\medskip
 The extrinsic curvature of the instantaneous 3-space $\Sigma{\tau}$
 can be evaluated by means of the formula ${}^3K_{rs} = {1\over {2\, (1 + n)}}\,
(n_{r|s} + n_{s|r} - \partial_{\tau}\, h_{rs})$, by using the
Christoffel symbols associated to $h_{rs}$ for the 3-covariant
derivatives $n_{r|s}$.

\bigskip

In conclusion the relevant conditions on the functions $f^A(\tau )$,
$F^A(\tau ,\sigma^u)$ of an admissible 3+1 splitting of Minkowski
space-time are $\sgn\, {\dot x}^2(\tau ) > 0$, $1 + n(\tau
,\sigma^u) > 0$, $\sgn\, g_{\tau\tau}(\tau ,\sigma^u) > 0$ and
$\lambda_a(\tau ,\sigma^u) > 0$.

\bigskip\bigskip

Finally Eq.(\ref{2.10}) suggests that it must be $z^{\mu}_r(\tau
,\sigma^u) = \Lambda^{\mu}{}_a(\tau ,\sigma^u)\, e_{(a)r}(\tau
,\sigma^u)$, where $\Lambda(\tau ,\sigma^u)$ is some Lorentz matrix,
so that $- \sgn\, g_{rs} = \sgn\, \eta_{\mu\nu}\,
\Lambda^{\mu}{}_{a}\, \Lambda^{\nu}{}_{b}\, e_{(a)r}\, e_{(b)s} = -
\sgn\, \eta_{ab}\, e_{(a)r}\, e_{(b)s} = h_{rs}$.

To find $\Lambda(\tau ,\sigma^u)$ let us remember that in tetrad
gravity in the York canonical basis (see Ref.\cite{12}) the
expression of the tetrads adapted to $\Sigma_{\tau}$ (Schwinger time
gauge) in terms of the unit normal $l^A$ and of the triads
$e^r_{(a)}$ are ${\buildrel \circ \over E}^A_{(o)} = l^A$,
${\buildrel \circ \over E}^A_{(a)} = (0; e^r_{(a)})$. In terms of
them we have ${\buildrel \circ \over V}^A = (1 + n)\, {\buildrel
\circ \over E}^A_{(o)} + e^s_{(a)}\, n_s\, {\buildrel \circ \over
E}^A_{(a)} = (1; 0)^A$. The world components of this vector are
${\buildrel \circ \over V}^{\mu} = z^{\mu}_A\, {\buildrel \circ
\over V}^A = z^{\mu}_{\tau}$, while those of ${\buildrel \circ \over
E}^A_{(a)}$ are ${\buildrel \circ \over E}^{\mu}_{(a)} = z^{\mu}_A\,
{\buildrel \circ \over E}^A_{(a)} = z^{\mu}_r\, e^r_{(a)}$, so that
we get $z^{\mu}_r = e_{(a)r}\, {\buildrel \circ \over
E}^{\mu}_{(a)}$. For the unit normal we have $l^{\mu} = z^{\mu}_A\,
l^A$.

In Minkowski space-time our parametrization of the embedding uses
the asymptotic tetrads $\epsilon^{\mu}_A$ and we have $z^{\mu}_A =
\epsilon^{\mu}_B\, \partial_A\, F^B$ and $l^{\mu} =
\epsilon^{\mu}_A\, l^A = \epsilon^{\mu}_o(l)$. Therefore a set of
tetrads adapted to $\Sigma_{\tau}$ in the point $(\tau ,\sigma^u)$
is given by the orthonormal tetrads $\epsilon^{\mu}_A(l(\tau
,\sigma^u))$ defined in Eqs.(\ref{2.8}): they replace the adapted
tetrads $l^{\mu}$, ${\buildrel \circ \over E}^{\mu}_{(a)}$ of tetrad
gravity. Therefore, consistently with Eq.(\ref{2.10}), we must have

\beq
 z^{\mu}_r(\tau ,\sigma^u) = \epsilon^{\mu}_A\, \partial_r\,
 F^A(\tau ,\sigma^u) = \epsilon^{\mu}_a(l(\tau ,\sigma^u))\,
 e_{(a)r}(\tau ,\sigma^u).
 \label{2.12}
 \eeq
 \medskip

\noindent This implies $z^{\mu}_{\tau} = \Big[(1 + n)\, l^A +
\epsilon^s_{(a)}\, n_s\, \epsilon^{\mu}_a(l)\Big](\tau ,\sigma^u) =
L^{\mu}{}_{\nu}(l(\tau ,\sigma^u), {\buildrel \circ \over l})\,
G^{\nu}(\tau ,\sigma^u)$ with $G^{\mu} = (1 + n; e^s_{(r)}\, n_s)$.
Eqs.(\ref{2.12}) are a set of non-linear partial differential
equations for $\partial_r\, F^A(\tau ,\vec \sigma )$.\bigskip

It is difficult to construct explicit examples of admissible 3+1
splittings. Let us consider the following two examples in which the
instantaneous 3-spaces are space-like hyper-planes.\medskip

A)  {\it Rigid non-inertial reference frames with translational
acceleration}. An example are the following embeddings

\medskip

\bea
 z^{\mu}(\tau ,\sigma^u ) &=& x^{\mu}_o +
\epsilon^{\mu}_{\tau}\, f(\tau ) + \epsilon^{\mu}_r\,
\sigma^r,\nonumber \\
 &&{}\nonumber \\
 &&g_{\tau\tau}(\tau ,\sigma^u ) = \sgn\,
 \Big({{d f(\tau )}\over {d\tau}}\Big)^2,\quad g_{\tau r}(\tau ,\sigma^u )
 =0,\quad g_{rs}(\tau ,\sigma^u ) = -\sgn\, \delta_{rs}.
 \label{2.13}
 \eea

\medskip

This is a foliation with parallel hyper-planes with normal $l^{\mu}
= \epsilon^{\mu}_{\tau} = const.$ and with the time-like observer
$x^{\mu}(\tau ) = x^{\mu}_o + \epsilon^{\mu}_{\tau}\, f(\tau )$ as
origin of the 3-coordinates. The hyper-planes have translational
acceleration ${\ddot x}^{\mu}(\tau ) = \epsilon^{\mu}_{\tau}\, \ddot
f(\tau )$, so that they are not uniformly distributed like in the
inertial case $f(\tau ) = \tau$.

\bigskip

B) As shown in Refs.\cite{3}, the simplest example of 3+1 splitting,
whose instantaneous 3-spaces are space-like hyper-planes carrying
admissible differentially rotating 3-coordinates \footnote{As shown
in Refs.\cite{3}, if we use the embedding $z^{\mu}(\tau ,\sigma^u)=
x^{\mu}(\tau ) + \epsilon^{\mu}_r\, R^r{}_s(\tau )\, \sigma^s$ such
that $\Omega^r = \Omega^r(\tau )$, then the resulting
$g_{\tau\tau}(\tau ,\sigma^u)$ violates M$\o$ller conditions,
because it vanishes at $\sigma = \sigma_R = {1\over {\Omega (\tau
)}}\, \Big[- {\dot x}_{\mu}(\tau )\, \epsilon^{\mu}_r\, R^r{}_s(\tau
)\, (\hat \sigma \times \hat \Omega (\tau ))^r + \sqrt{{\dot
x}^2(\tau ) + [{\dot x}_{\mu}(\tau )\, \epsilon^{\mu}_r\,
R^r{}_s(\tau )\, (\hat \sigma \times \hat \Omega (\tau
))^r]^2}\Big]$. We use the notations $\sigma^u = \sigma\, {\hat
\sigma}^u$, $\Omega^r = \Omega\, {\hat \Omega}^r$, ${\hat \sigma}^2
= {\hat \Omega}^2 = 1$. At this distance from the rotation axis the
tangential rotational velocity becomes equal to the velocity of
light. This is the {\it horizon problem} of the rotating disk. This
pathology is common to most of the rotating coordinate systems
quoted after Eq.(\ref{2.16}) and in Appendices A (see also th first
Section of the next paper on the rotating disk). Let us remark that
an analogous pathology happens on the event horizon of the
Schwarzschild black hole, where the time-like Killing vector of the
static space-time becomes light-like: in this case we do not have a
coordinate singularity but an intrinsic geometric property of the
solution of Einstein's equations. For the rotating Kerr black hole
the same phenomenon happens already at the boundary of the
ergosphere \cite{13}, as a consequence of the Killing vectors own by
this solution. Let us remark that in the existing theory of rotating
relativistic stars \cite{14}, where differential rotations are
replacing the rigid ones in model building, it is assumed that in
certain rotation regimes an ergosphere may form \cite{15}: however
in this case it is not known whether Killing vectors and a dynamical
ergosphere exist, so that the horizon problem,arising if one uses
4-coordinates adapted to the Killing vectors, could be associated to
a coordinate singularity like for the rotating disk. In the study of
the magnetosphere of pulsars the horizon of the rotating disk is
named the {\it light cylinder} (see Appendix A). }, is given by the
embedding ($\sigma = |\vec \sigma |$; $\epsilon^{\mu}_r$ are the
asymptotic space-like axes and the unit normal is $l^{\mu} =
\epsilon^{\mu}_{\tau} = const.$; $\alpha_i(\tau ,\vec \sigma ) =
F(\sigma )\, {\tilde \alpha}_i(\tau )$, $i=1,2,3$, are Euler angles;
$R^r{}_s(\alpha_i(\tau ,\sigma ))$ is a rotation matrix satisfying
the asymptotic conditions $R^r{}_s(\tau , \sigma)\,
{\rightarrow}_{\sigma \rightarrow \infty} \delta^r_s$, $\partial_A\,
R^r{}_s(\tau ,\sigma )\, {\rightarrow}_{\sigma \rightarrow
 \infty}\, 0$)

\begin{eqnarray*}
 z^{\mu}(\tau ,\sigma^u ) &=& x^{\mu}(\tau ) + \epsilon^{\mu}_r\,
R^r{}_s(\tau , \sigma )\, \sigma^s,\qquad x^{\mu}(\tau) = x^{\mu}_o
+ f^A(\tau)\, \epsilon^{\mu}_A,\nonumber \\
 &&{}\nonumber \\
 R^r{}_s(\tau ,\sigma ) &=& R^r{}_s(\alpha_i(\tau,\sigma )) =
 R^r{}_s(F(\sigma )\, {\tilde \alpha}_i(\tau)),\nonumber \\
 &&{}\nonumber \\
 &&0 < F(\sigma ) < {1\over {A\, \sigma}},\qquad {{d\, F(\sigma
 )}\over {d\sigma}} \not= 0\,\, (Moller\,\, conditions),
 \end{eqnarray*}

\begin{eqnarray*}
 z^{\mu}_{\tau}(\tau ,\sigma^u) &=& {\dot x}^{\mu}(\tau ) -
 \epsilon^{\mu}_r\,  R^r{}_s(\tau
 ,\sigma )\, \delta^{sw}\, \epsilon_{wuv}\, \sigma^u\, {{\Omega^v(\tau
 ,\sigma )}\over c},\nonumber \\
  z^{\mu}_r(\tau ,\sigma^u) &=& \epsilon^{\mu}_k\, R^k{}_v(\tau
 ,\sigma )\, \Big(\delta^v_r + \Omega^v_{(r) u}(\tau ,\sigma )\,
 \sigma^u\Big),
 \end{eqnarray*}

 \bea
 \sgn\, g_{\tau\tau}(\tau ,\sigma^u) &=& \sgn\, {\dot x}^2(\tau ) -
 2\, \sgn\, {\dot x}_{\mu}(\tau )\ \epsilon^{\mu}_r\,  R^r{}_s(\tau
 ,\sigma )\, \delta^{sw}\, \epsilon_{wuv}\, \sigma^u\, {{\Omega^v(\tau
 ,\sigma )}\over c} -\nonumber \\
 &-&{1\over {c^2}}\, \sum_k\, \epsilon_{krs}\, \sigma^r\, \Omega^s(\tau ,\sigma )\,
\epsilon_{kuv}\, \sigma^u\, \Omega^v(\tau ,\sigma ),\nonumber \\
 n_r(\tau ,\vec \sigma ) &=& - \sgn\, g_{\tau r}(\tau ,\sigma ) = -
 \sgn\, {\dot x}_{\mu}(\tau )\ \epsilon^{\mu}_k\,  R^k{}_v(\tau
 ,\sigma )\, \Big(\delta^v_r + \Omega^v_{(r) u}(\tau ,\sigma )\,
 \sigma^u\Big) -\nonumber \\
 &-& \epsilon_{smn}\, \sigma^m\, {{\Omega^n(\tau ,\sigma )}\over c}\,
 \Big(\delta^s_r + \Omega^s_{(r) u}(\tau ,\sigma )\, \sigma^u\Big),
 \nonumber \\
 h_{rs}(\tau ,\sigma^u) &=& - \sgn\, g_{rs}(\tau ,\sigma^u) =
 \delta_{rs} + \Big(\Omega^r_{(s) u}(\tau ,\sigma ) + \Omega^s_{(r)
 u}(\tau ,\sigma )\Big)\, \sigma^u +\nonumber \\
 &+& \sum_w\, \Omega^w_{(r) u}(\tau ,\sigma )\, \Omega^w_{(s) v}(\tau
 ,\sigma )\, \sigma^u\, \sigma^v,
 \label{2.14}
 \eea

\noindent where $\Big(R^{-1}(\tau ,\sigma )\,
\partial_{\tau}\, R(\tau ,\sigma )\Big)^u{}_v = \delta^{um}\,
\epsilon_{mvr}\, {{\Omega^r(\tau ,\sigma)}\over c}$,
${\partial_{\tau}\, R(\tau ,\sigma)}^u{}_v = R^u{}_n(\tau ,\sigma)\,
\delta^{nm}\, \epsilon_{mvr}\, {{\Omega^r(\tau ,\sigma)}\over c}$
with $\Omega^r(\tau ,\sigma ) = F(\sigma )\, \tilde \Omega (\tau
,\sigma )\, {\hat n}^r(\tau ,\sigma )$ \footnote{${\hat n}^r(\tau
,\sigma )$ defines the instantaneous rotation axis and $0 < \tilde
\Omega (\tau ,\sigma ) < 2\, max\, \Big({\dot {\tilde \alpha}}(\tau
), {\dot {\tilde \beta}}(\tau ), {\dot {\tilde \gamma}}(\tau
)\Big)$.} being the angular velocity and with $\Omega_{(r)}(\tau
,\sigma ) = R^{-1}(\tau ,\vec \sigma )\, \partial_r\, R(\tau ,\sigma
)$. The angular velocity vanishes at spatial infinity and has an
upper bound proportional to the minimum of the linear velocity
$v_l(\tau ) = {\dot x}_{\mu}\, l^{\mu}$ orthogonal to the space-like
hyper-planes. When the rotation axis is fixed and $\tilde \Omega
(\tau ,\sigma ) = \omega = const.$, a simple choice for the function
$F(\sigma )$ is $F(\sigma ) = {1\over {1 + {{\omega^2\,
\sigma^2}\over {c^2}}}}$ \footnote{Nearly rigid rotating systems,
like a rotating disk of radius $\sigma_o$, can be described by using
a function $F(\sigma )$ approximating the step function $\theta
(\sigma - \sigma_o)$.}.\medskip

Let us remark that the unit normal is $l^{\mu}(\tau ,\sigma^u) =
\epsilon^{\mu}_{\tau} = const.$ and the lapse function is $1 +
n(\tau ,\sigma^u) = \sgn\, \Big(z^{\mu}_{\tau}\, l_{\mu}\Big)(\tau
,\sigma^u) = \sgn\, \epsilon^{\mu}_{\tau}\, {\dot x}_{\mu}(\tau )$.

\medskip

The embedding (\ref{2.14}) has been used in the first paper of
Ref.\cite{10}, on quantum mechanics in non-inertial frames, in the
form $z^{\mu}(\tau ,\sigma^u) = x^{\mu}(\tau ) + F^{\mu}(\tau
,\sigma^u) = \theta (\tau )\, \epsilon^{\mu}_{\tau} + {\cal
A}^r(\tau ,\sigma^u)\, \epsilon^{\mu}_r$ with $x^{\mu}_o = 0$,
$\theta (\tau ) = f^{\tau}(\tau )$, ${\cal A}^r(\tau ,\sigma^u) =
f^r(\tau) + R^r{}_s(\tau ,\sigma)\, \sigma^s$, describing the
freedom in the choice of the mathematical time $\tau$ and with the
world-line of the time-like observer having the expression
$x^{\mu}(\tau ) = \epsilon_{\tau}^{\mu}\, \theta (\tau ) +
\epsilon^{\mu}_r\, {\cal A}^r(\tau ,0)$, namely with $f^r(\tau ) =
{\cal A}^r(\tau ,o)$ and ${\dot f}^r(\tau ) = {{w^r(\tau )}\over
{c}}$ ($\vec w(\tau )$ is the ordinary 3-velocity). If we choose
$\theta (\tau ) = \tau$, we get from Eq.(\ref{2.2}) $u^{\mu}(\tau )
= \epsilon^{\mu}_A\, u^A(\tau ) = {{\epsilon^{\mu}_{\tau} +
\epsilon^{\mu}_r\, {{w^r(\tau)}\over c}}\over {\sqrt{1 - {{{\vec
w}^2(\tau )}\over {c^2}} }}}$, $a^{\mu}(\tau ) = \epsilon^{\mu}_A\,
u^A(\tau ) = {1\over {c^2}}\, \sum_u\, {\dot w}^u(\tau )\, {\ddot
w}^u(\tau )\, \Big(1 - {{{\vec w}^2(\tau )}\over
{c^2}}\Big)^{-3/2}\, \Big(\epsilon^{\mu}_{\tau} + \epsilon^{\mu}_r\,
{{w^r(\tau)}\over c}\Big)$. The lapse function is $1 + n(\tau ) =
{\dot f}^{\tau}(\tau )$.
\bigskip

To evaluate the non-relativistic limit for $c \rightarrow \infty$,
where $\tau = c\, t$ with $t$ the absolute Newtonian time and
$\partial_{\tau} = {1\over c}\, \partial_t$, we choose the gauge
function $F(\sigma ) = {1\over {1 + {{\omega^2\, \sigma^2}\over
{c^2}}}}\, \rightarrow_{c \rightarrow \infty}\, 1 - {{ \omega^2\,
\sigma^2}\over {c^2}} + O(c^{-4})$. This implies

\bea
 R^a{}_r(\tau ,\sigma ) &\rightarrow_{c \rightarrow \infty}&
 R^a{}_r(\tau ) - {{\omega^2\, \sigma^2}\over {c^2}}\,
 \sum_i\, {\tilde \alpha}_i(\tau )\, {{\partial\, R^a{}_r(\tau
 ,\sigma )}\over {\partial\, \alpha_i}}{|}_{F(\sigma ) = 1} +
 O(c^{-4}) =\nonumber \\
 &{\buildrel {def}\over =}& R^a{}_r(\tau ) -
 {{\omega^2\, \sigma^2}\over {c^2}}\,
 R^{(1)a}{}_r(\tau ) + O(c^{-4}),
 \label{2.15}
 \eea

\noindent and we can introduce a new 3-velocity $\vec v(\tau )$ by
means of  $w^r(\tau ) = c\, {\dot f}^r(\tau) = R^r{}_s(\tau )\,
v^s(\tau )$.  We have $\Omega^r(\tau ,\sigma) =  \tilde \Omega
(\tau) {\hat n}^r(\tau ) + O(c^{-1})$ for the angular velocity and
$\Omega_{(r)}(\tau ,\sigma) = 0 + O(c^{-2})$.
\medskip

Therefore the corrections to rigidly-rotating non-inertial frames
coming from M$\o$ller conditions are of order $O(c^{-2})$ and become
important at the distance from the rotation axis where the horizon
problem for rigid rotations appears.\medskip

Then, from Eqs. (\ref{2.14}), (\ref{2.4}), (\ref{2.7}) and
(\ref{2.11}) we get

\bea
 z^{\mu}(\tau ,\sigma^u) &\rightarrow& x^{\mu}(\tau ) +
 \epsilon^{\mu}_r\, R^r{}_s(\tau )\, \sigma^s - {{\omega^2\,
 \sigma^2}\over {c^2}}\, \epsilon^{\mu}_r\, R^{(1)r}{}_s(\tau )\,
 \sigma^s + O(c^{-4}),\nonumber \\
 &&{}\nonumber \\
 z^{\mu}_{\tau}(\tau ,\sigma^u) &\rightarrow& {\dot x}^{\mu}(\tau )
 + \epsilon^{\mu}_r\, \partial_{\tau}\, R^r{}_s(\tau )\, \sigma^s
 + O(c^{-3}) =\nonumber \\
 &=& \epsilon^{\mu}_{\tau} + \epsilon^{\mu}_r\, {\dot f}^r(\tau ) +
 {1\over c}\, \epsilon^{\mu}_r\, R^r{}_s(\tau)\, \epsilon_{suv}\,
 \Omega^u(\tau )\, \sigma^v + O(c^{-3}),\nonumber \\
 z^{\mu}_r(\tau ,\sigma^u) &\rightarrow& \epsilon^{\mu}_s\,
 \Big[R^s{}_r(\tau ) - {{\omega^2}\over {c^2}}\, R^{(1)s}{}_u(\tau )\,
 (\delta^u_r\, \sigma^2 + 2\, \sigma^u\, \sigma^v\, \delta_{vr})\Big]
 + O(c^{-4}),\nonumber \\
 &&{}\nonumber \\
 h_{rs}(\tau ,\sigma^u) &\rightarrow& \delta_{rs} - 2\, {{\omega^2}\over
 {c^2}}\, \sum_u\, R^u{}_r(\tau )\, R^{(1)u}{}_v(\tau )\, (\delta^v_s + 2\, \sigma^v\,
 \sigma^n\, \delta_{ns}) + O(c^{-4}), \nonumber \\
 &&{}\nonumber \\
 n(\tau ) &=& 0,\qquad
 n_r(\tau ,\sigma^u) \rightarrow {1\over c}\, \Big(\delta_{rs}\, v^s(\tau ) +
 \epsilon_{ruv}\, \Omega^u(\tau )\, \sigma^v\Big) +
 O(c^{-3}).\nonumber \\
 &&{}
 \label{2.16}
 \eea

\bigskip

There is the enormous amount of bibliography, reviewed in
Ref.\cite{16}, about the problems of the {\it rotating disk} and of
the {rotating coordinate systems}. Independently from the fact
whether the disk is a material extended object or a geometrical
congruence of time-like world-lines (integral lines of some
time-like unit vector field), the idea followed by many researchers
\cite{6,17,18} (in Refs.\cite{18} are quoted the attempts to develop
electro-dynamics in rotating frames) is to start from the Cartesian
4-coordinates of a given inertial system, to pass to cylindrical
3-coordinates and then to make a either Galilean (assuming a
non-relativistic behaviour of rotations at the relativistic level)
or Lorentz transformation to comoving rotating 4-coordinates (see
the locality hypothesis in the next Subsection), with a subsequent
evaluation of the 4-metric in the new coordinates. In other cases
\cite{19} a suitable global 4-coordinate transformation is
postulated, which avoids the horizon problem. Various authors (see
for instance Refs.\cite{20}) do not define a coordinate
transformation but only a rotating 4-metric. Just starting from
M$\o$ller rotating 4-metric \cite{6}, Nelson (see the second paper
in Ref.\cite{13}) was able to deduce a 4-coordinate transformation
implying it.

\medskip

See the first Section of the second paper for the description of the
rotating disk and of the Sagnac effect in the 3+1 framework.

\subsection{Congruences of Time-Like Observers Associated with an
Admissible 3+1 Splitting, the 1+3 Point of View and the Locality
Hypothesis}

Each admissible 3+1 splitting of Minkowski space-time, having the
time-like observer $x^{\mu}(\tau )$ as origin of the 3-coordinates
on the instantaneous 3-spaces $\Sigma_{\tau}$, automatically
determines two time-like vector fields and therefore {\it two
congruences of (in general) non-inertial time-like observers}:

i) The time-like vector field $l^{\mu}(\tau ,\sigma^u )\,
\partial_{\mu}$ of the normals to the simultaneity surfaces $\Sigma_{\tau}$
(by construction surface-forming, i.e. irrotational), whose flux
lines are the world-lines $x^{\mu}_{l,\tau_o,\sigma^u_o}(\tau )$,
$u^{\mu}_{}(\tau) = {{{\dot x}^{\mu}_{l,\tau_o,\sigma^u_o}}\over
{\sqrt{\sgn\, {\dot x}^2_{l.\tau_o,\sigma^u_o}(\tau)}}}$,
$u^{\mu}_{l,\tau_o,\sigma^u_o}(\tau_o) = l^{\mu}(\tau_o,
\sigma^u_o)$, of the so-called (in general non-inertial) {\it
Eulerian observers}. The simultaneity surfaces $\Sigma_{\tau}$ are
(in general non-flat) Riemannian 3-spaces in which every physical
system is visualized and in each point the {\it tangent space} to
$\Sigma_{\tau}$ is the {\it local observer rest frame} of the
Eulerian observer through that point. The 3+1 viewpoint of these
observers is called {\it hyper-surface 3+1 splitting}.

ii) The time-like evolution vector field ${{z^{\mu}_{\tau}(\tau
,\vec \sigma )}\over { \sqrt{\sgn\, g_{\tau\tau}(\tau ,\vec \sigma )
} }}\, \partial_{\mu}$, which in general is not surface-forming
(i.e. it has non-zero vorticity like in the case of the rotating
disk). The observers associated to its flux lines
$x^{\mu}_{z,\sigma^u_o}(\tau ) = z^{\mu}(\tau ,\sigma^u_o)$,
$u^{\mu}_{z,\sigma^u_o}(\tau ) = {{z^{\mu}_{\tau}(\tau ,\vec \sigma
)}\over { \sqrt{\sgn\, g_{\tau\tau}(\tau ,\vec \sigma ) } }}$, have
the {\it local observer rest frames}, the tangent 3-spaces
orthogonal to the evolution vector field, {\it not tangent} to
$\Sigma_{\tau}$: there is no notion of 3-space for these observers
(1+3 point of view or {\it threading splitting}) and no
visualization of the physical system in large. However these
observers can use the notion of simultaneity associated to the
embedding $z^{\mu}(\tau ,\vec \sigma )$, which determines their
4-velocity. Like for the observer $x^{\mu}(\tau )$, their 4-velocity
is not parallel to $l^{\mu}(\tau ,\sigma^u)$. The 3+1 viewpoint of
these observers is called {\it slicing 3+1 splitting}.

\bigskip

Every 1+3 point of view considers only a time-like observer (either
$x^{\mu}(\tau )$ or $x^{\mu}_{l,\tau_o,\sigma^u_o}(\tau )$ or
$x^{\mu}_{z,\sigma^u_o}(\tau )$) and tries to give a description of
the physics in a region around the observer's world-line assumed
known. Since there is no global notion of simultaneity, namely of
instantaneous 3-space, one identifies the space-like hyper-planes
orthogonal to the observer unit 4-velocity $u^{\mu}_{obs}(\tau )$ at
every instant $\tau$ (the observer local rest frames) as local
instantaneous 3-spaces $\Sigma_{obs\, \tau}$ (strictly speaking it
is a tangent space and not a 3-space). Then one makes a choice of a
tetrad $V^{\mu}_{obs\, A}((\tau)) = \Big(u^{\mu}_{obs}(\tau );
V^{\mu}_{obs\, (r)}(\tau )\Big)$, $\eta_{\mu\nu}\, V^{\mu}_{obs\,
(A)}(\tau )\, V^{\nu}_{obs\, (B)}(\tau ) = \eta_{(A)(B)}$. The space
axes $V^{\mu}_{obs\, (r)}(\tau )$ can be chosen arbitrarily, even if
often they are chosen as the tangents to three space-like geodesics
on $\Sigma_{obs\, \tau}$ at the observer position. After parallel
transport of the tetrad to the points of $\Sigma_{obs\, \tau}$ not
on the observer world-line one tries to build an {\it accelerated
4-coordinate system} having the observer as origin of the
3-coordinates \cite{21}. In the case of the tangents to space-like
geodesics one  builds a local system of Fermi coordinates around the
observer world-line \cite{22} (see also Ref.\cite{23} for an updated
discussion of Fermi-Walker and Fermi normal coordinates).

The drawback of this construction is that the $\tau$-dependent
family of hyper-planes $\Sigma_{obs\, \tau}$ will have hyper-planes
at different $\tau$'s {\it intersecting} at some distance from the
observer world-line, usually estimated by using the so-called {\it
acceleration radii} of the observer. This implies that every system
of accelerated 4-coordinates of this type will develop {\it
coordinate singularities} when the hyper-planes intersect. As a
consequence it is not possible to formulate a well-posed Cauchy
problem for Maxwell equations in these accelerated coordinate
systems: they can only be used for evaluating local
semi-relativistic inertial effects.
\medskip

At each instant $\tau$ the tetrads $V^{\mu}_{obs\, (A)}(\tau )$
coincide with some Lorentz matrix $V^{\mu}_{obs\, (A)}(\tau ) =
\Lambda^{\mu}{}_{\nu = A}(\tau )$, which connects the reference
inertial frame to the {\it instantaneous comoving inertial frame}
associated with the accelerated observer at $\tau$. A possibility is
to use the tetrads $\epsilon^{\mu}_A(u_{obs}(\tau ))$ associated
with the Wigner boost $L^{\mu}{}_{\nu}(u_{obs}(\tau ), {\buildrel
\circ \over u}_{obs})$. This fact is at the heart of the {\it
locality hypothesis} \cite{24} according to which an accelerated
observer is physically equivalent (for measurements) to a continuous
family of hypothetical momentarily comoving inertial
observers.\medskip

If we parametrize the Lorentz transformation $\Lambda (\tau )$ as
the product of a pure boost with a pure rotation $\Lambda (\tau ) =
B(\vec \beta (\tau ))\, {\cal R}(\alpha (\tau ), \beta (\tau ),
\gamma (\tau ))$ and we call $R^r_s(\tau ) = {\cal R}^r{}_s(\tau )$,
we can write (from Eq.(\ref{2.8}) we have $B^{jk}(\vec \beta (\tau
)) = \delta^{jk} + (\gamma(\tau ) - 1)\, {{\beta^j(\tau )\,
\beta^k(\tau )}\over {\sum_n\, (\beta^n(\tau ))^2}}$)

\bea
 V^\mu_{obs\, (A)}(\tau ) = \Lambda^\mu_{\nu=A}
(\tau)= \left(
\begin{array}{cc}
\frac{1}{\sqrt{1 - \vec{\beta}^2(\tau)}}& \frac{R^i_k(\tau )\,
\beta^k(\tau)}{\sqrt{1 - \vec{\beta}^2(\tau)}}\\
\frac{\beta^j(\tau)}{\sqrt{1 - \vec{\beta}^2(\tau)}}&\,
R^i_k(\tau)\, B^{jk}({\vec \beta} (\tau))
\end{array}
\right).
 \label{2.17}
\eea

Let us define the angular velocity $\omega_r(\tau )$ by means of
${{d\, R^r_s(\tau )}\over {d\tau}}\, {\buildrel {def}\over =}\,
\epsilon_{ruv}\, \omega_u(\tau)\, R^v_s(\tau )$. Even if the
observer is connected with the embedding $z^{\mu}(\tau ,\vec
\sigma)$, this angular velocity is not related to the angular
velocity defined after Eq.(\ref{2.14}).\medskip

Finally, if we write

\bea
 && {{d V^{\mu}_{obs\, (A)}(\tau )}\over {d\tau}} =
 {{\cal A}_{obs\, (A)}}^{(B)}(\tau )\, V^{\mu}_{obs\, (B)}(\tau ),
 \nonumber \\
 &&{}\nonumber \\
 &\Rightarrow & {\cal A}_{obs\, (A)(B)}(\tau ) = - {\cal
 A}_{obs\, (B)(A)}(\tau ) = {{d V^{\mu}_{obs\, (A)}(\tau
 )}\over {d\tau}}\,\eta_{\mu\nu}\,  V^\nu_{obs\, (B)}(\tau ),
 \label{2.18}
  \eea

\noindent and we introduce the definitions $a_{obs\, r}(\tau ) =
{\cal A}_{obs\, (\tau)(r)}(\tau )$, $\Omega_{obs\, r}(\tau ) =
{1\over 2}\, \epsilon_{ruv}\, {\cal A}_{obs\, (u)(v)}(\tau )$, then
the acceleration radii have the following definition \cite{24}:
$I_1(\tau) = \sum_r\, \Big(\Omega^2_{obs\, r}(\tau ) - a^2_{obs\,
r}(\tau )\Big)$, $I_2(\tau ) = \sum_r\, a_{obs\, r}(\tau )\,
\Omega_{obs\, r}(\tau )$. By means of Eq.(\ref{2.17}) they can be
expressed in terms of the parameters of the Lorentz transformation
and their $\tau$-derivatives.

\bigskip

Let us remark that, since each instantaneous 3-space $\Sigma_{\tau}$
centered on an accelerated observer is in general a non-flat
Riemannian 3-manifold with 3-metric $g_{rs}(\tau, \sigma^u)$ and
Riemann 3-curvature tensor ${}^3R_{rsmn}(\tau, \sigma^u)$, we can
look for special 3-coordinate systems on $\Sigma_{\tau}$ such that
the curvature effects are second order near the observer, mimicking
what is done in general relativity \cite{21,22,23} to define local
inertial frames and to visualize inertial effects. On
$\sigma_{\tau}$ around the observer, origin of the 3-coordinates and
whose world-line (in the flat Minkowski space-time) is a 4-geodesics
if the observer is inertial, we can introduce:\medskip
 a) Riemann 3-coordinates such that $\partial_v\, g_{rs}(\tau,
 \sigma^u){|}_{\sigma^u =0} = 0$;\medskip
 b) Riemann normal 3-coordintes (the three coordinate lines are
 3-geodesics of $\Sigma_{\tau}$) such that $\partial_m\, \partial_n\, g_{rs}(\tau,
 \sigma^u){|}_{\sigma^u =0} = 0$ and $\partial_m\,
 {}^3\Gamma^r_{uv}(\tau, \sigma^u){|}_{\sigma^u =0}$ are
 proportional to suitable combinations of the 3-curvature tensor.

In this way it is possible to simplify the expression of the special
relativistic effects in non-inertial frames near the observer as
first order corrections depending on the observer acceleration.
\bigskip

Finally let us remark that given an admissible 3+1 splitting of
Minkowski space-time, the infinitesimal spatial length $dl$ in the
instantaneous 3-spaces $\Sigma_{\tau}$ is defined by putting $d\tau
= 0$ in the line element $ds^2 = g_{AB}(\tau ,\sigma^u)\,
d\sigma^A\, d\sigma^B$, namely we have $dl^2 = g_{rs}(\tau
,\sigma^u)\, d\sigma^r\, d\sigma^s$. This global, but
coordinate-dependent, definition has to be contrasted with the
local, but coordinate-independent, definition used in the 1+3 point
of view as it is done for instance in Landau-Lifschitz \cite{17}.
This definition is only locally valid in the local rest frame of an
observer: since there is no notion of instantaneous 3-space it
cannot be used in a global way. For a detailed comparison of these
two notions of spatial length see Section II of the first paper of
Ref.\cite{3}.

\bigskip

\subsection{Notations for the Electro-Magnetic Field in Non-Inertial Frames}

Let us add some notations for the electro-magnetic field in the
non-inertial frames, where the instantaneous 3-space is either
curved or flat but with rotating coordinates [in both cases it is
not Euclidean and has the 3-metric $h_{rs}$ of signature $(+++)$].
\medskip

The basic field is the electro-magnetic potential $A_A = (A_{\tau};
A_r)$. We have $A^A = (A^{\tau}; A^A) = g^{AB}\, A_B = g^{A\tau}\,
A_{\tau} + g^{As}\, A_s$. Instead in  inertial frames we have
$A^{\tau} = \sgn\, A_{\tau}$, $A^r = - \sgn\, A_r$.

\medskip

In non-inertial frames it is convenient to introduce the following
"Euclidean" notation: ${\tilde A}^r = h^{rs}\, A_s \not= A^r$ (in
inertial frames: ${\tilde A}^r = A_r = - \sgn\, A^r$)

\bigskip

We shall adopt the following conventions for the electric and
magnetic fields in terms of $F_{AB} = \partial_A\, A_B -
\partial_B\, A_A$ \footnote{In the inertial case, where $h_{rs} = \delta_{rs}$ implies
$V^r {\buildrel {def}\over =}\, {\tilde V}^r = V_r$ for the
components of 3-vector $\vec V$ not being the vector part of a
4-vector (like $\vec E$ and $\vec B$), we can use the vector
notation $\vec E = \{ E_r\} = \{{\tilde E}^r \}$, $\vec B = \{ B_r\}
= \{ {\tilde B}^r \}$, ${\vec E}^2 = \sum_r\, E_r^2 = \sum_r\,
({\tilde E}^r)^2$, ${\vec B}^2 = \sum_r\, B_r^2 = \sum_r\, ({\tilde
B}^r)^2$, $({\dot {\vec \eta}}_i \times \vec B)_r = \sum_{uv}\,
\epsilon_{ruv}\, {\dot \eta}^u_i\, B_v = \sum_{uv}\,
\epsilon_{ruv}\, {\dot \eta}_i^u\, {\tilde B}^v$, $(\vec E \times
\vec B)_r = \sum_{uv}\, \epsilon_{ruv}\, E_u\, B_v = \sum_{uv}\,
\epsilon_{ruv}\, {\tilde E}^u\, {\tilde B}^v$. Since ${\tilde V}^r =
h^{rs}\, V_s \not= V^r$, {\it we are not going to use the vector
notation in non-inertial frames}.} :\medskip

a) In  inertial frames we have \footnote{$\epsilon_{uvr}$ is the
Euclidean Levi-Civita tensor with $\epsilon_{123} = 1$;
$\epsilon^{uvr}$ is never introduced.}\medskip

\bea
 E_r &=& - F_{\tau r} = F^{\tau r}\, = {\tilde E}^r,\nonumber \\
 &&{}\nonumber \\
 B_r &=& {1\over 2}\, \epsilon_{ruv}\, F_{uv} = {1\over 2}\,
\epsilon_{ruv}\, F^{uv}\, = {\tilde B}^r, \qquad F_{uv} = F^{uv} =
\epsilon_{uvr}\, B_r = \epsilon_{uvr}\, {\tilde B}^r.
 \label{2.19}
 \eea
\medskip

b) In  non-inertial frames  we put the definitions

\beq
 E_r\, {\buildrel {def}\over =}\, - F_{\tau r},\qquad
 B_r\, {\buildrel {def}\over =}\, {1\over 2}\, \epsilon_{ruv}\, F_{uv},
 \qquad F_{rs} = \epsilon_{rsu}\, B_u.
 \label{2.20}
 \eeq

\noindent Since we have

\bea
 F^{AB} &=& g^{AC}\, g^{BD}\, F_{CD} = (g^{A\tau}\, g^{Br} - g^{Ar}\,
g^{B\tau})\, F_{\tau r} + g^{Ar}\, g^{Bs}\, F_{rs} =\nonumber \\
 &=&(g^{Ar}\, g^{B\tau} - g^{A\tau}\, g^{Br})\, E_r + \epsilon_{rsu}\,
g^{Ar}\, g^{Bs}\, B_u,\nonumber \\
 &&{}\nonumber \\
 F^{\tau u} &=& (g^{\tau r}\, g^{\tau u} - g^{\tau\tau}\, g^{ur})\, E_r
+ \epsilon_{rsn}\, g^{\tau r}\, g^{us}\, B_n =\nonumber \\
 &=& h^{ur}\, E_r + {1\over {(1 + n)^2}}\, \epsilon_{rsn}\, n^r\,
 h^{us}\, B_n,\nonumber \\
 F^{uv} &=& (g^{ur}\, g^{\tau v} - g^{\tau u}\, g^{vr})\, E_r +
\epsilon_{rsn}\, g^{ur}\, g^{vs}\, B_n =\nonumber \\
 &=& {{(h^{ur}\, n^v - h^{vr}\, n^u)\, E_r}\over {(1 + n)^2}} +
 \epsilon_{rsn}\, \Big(h^{ur}\, h^{vs} - {{n^r\, (n^v\, h^{us} -
 n^u\,  h^{us})}\over {(1 + n)^2}}\Big)\, B_n,
 \label{2.21}
 \eea

\noindent by analogy with inertial frames we can put

\bea
 F^{\tau r} &{\buildrel {def}\over =}& {\check E}^r,\qquad
 {\check E}^r = {\tilde E}^r + {{\epsilon_{uvn}\, n^u\,
 h^{rv}\, h_{nm}\, {\tilde B}^m}\over {(1 + n)^2}} \not= {\tilde E}^r = h^{rs}\, E_s,
 \nonumber \\
 &&{}\nonumber \\
 F^{uv} &{\buildrel {def}\over =}& \epsilon_{uvr}\, {\check B}^r,\qquad
 {\check B}^r = {2\over {(1 + n)^2}}\,
 \epsilon_{ruv}\, {\tilde E}^u\, n^v +\nonumber \\
 &+& \epsilon_{ruv}\, \epsilon_{ksn}\, \Big(h^{uk}\, h^{vs} - {{n^k\,
 (n^v\, h^{us} - n^u\, h^{vs})}\over {(1 + n)^2}}\Big)\, h_{nm}\, {\tilde
 B}^m \not= {\tilde B}^r = h^{rs}\, B_s.
 \label{2.22}
 \eea

\vfill\eject

\section{Parametrized Minkowski Theories and the
Inertial Rest-Frame Instant Form for Charged Particles plus the
Electro-Magnetic Field.}

In this Section we will give a review of the description of the
isolated system "N charged positive-energy scalar particles with
Grassmann-valued electric charges plus the electro-magnetic field"
\cite{9}  in the framework of parametrized Minkowski theories
\cite{1,5} (see also the Appendix of the first paper in
Refs.\cite{11}).
\bigskip

Let be given an admissible 3+1 splitting of Minkowski space-time
centered on a time-like observer $x^{\mu}(\tau )$. Let $\sigma^A =
(\tau ; \sigma^u)$ be the adapted observer-dependent radar
4-coordinates and $z^{\mu}(\tau ,\sigma^u)$ the embedding of the
instantaneous 3-spaces $\Sigma_{\tau}$ into Minkowski space-time as
seen from an arbitrary reference inertial observer. Let $
g_{AB}(\tau ,\sigma^u) = z^{\mu}_A(\tau ,\sigma^u)\, \eta_{\mu\nu}\,
z^{\nu}_B(\tau ,\sigma^u)$ be the associated 4-metric.\medskip

The electro-magnetic field is described by the Lorentz-scalar
potential $A_A(\tau ,\sigma^u)$ knowing the equal-time surface. The
field strength is $F_{AB}(\tau ,\sigma^u) = \Big(\partial_A\, A_B -
\partial_B\, A_A\Big)(\tau ,\sigma^u)$.\medskip

The scalar positive-energy particles are described by the
Lorentz-scalar 3-coordinates $\eta^r_i(\tau )$ defined by
$x^{\mu}_i(\tau ) = z^{\mu}(\tau ,\eta^u_i(\tau ))$, where
$x^{\mu}_i(\tau )$ are their world-lines. $Q_i$ are the
Grassmann-valued electric charges satisfying $Q^2_i = 0$, $Q_i\, Q_j
= Q_j\, Q_i \not= 0$ for $i \not= j$. Each $Q_i$ is an even bilinear
function of a complex Grassmann variable $\theta_i(\tau )$: $Q_i =
e\, \theta^*_i(\tau )\, \theta_i(\tau )$.

\bigskip

As shown in Ref.\cite{9} the description of N scalar positive-energy
particles with Grassmann-valued electric charges plus the
electro-magnetic field is done in parametrized Minkowski theories
with the action

\begin{eqnarray*}
  S &=&\int d\tau\, d^{3}\sigma \,{\cal L}(\tau ,\sigma^u) = \int d\tau\, L(\tau
),  \nonumber \\
&&\nonumber\\
 {\cal L}(\tau ,\sigma^u) &=&{\frac{i}{2}}\sum_{i=1}^{N}\,
\delta ^{3}(\sigma^u - \eta^u_i(\tau ))\, \Big[\theta _{i}^{\ast
}(\tau ){\dot{\theta}}_{i}(\tau ) -
{\dot{\theta}}_{i}^{\ast }(\tau )\theta _{i}(\tau )\Big]-  \nonumber \\
&&\nonumber\\
 &-&\sum_{i=1}^{N}\, \delta ^{3}(\sigma^u - \eta^u_i(\tau
))\, \Big[m_i\, c\, \sqrt{\sgn\, [g_{\tau \tau }(\tau ,\sigma^u) +
2\, g_{\tau r}(\tau ,\sigma^u)\, {\dot{\eta}}_i^r(\tau ) + g_{rs
}(\tau ,\sigma^u)\, {\dot{\eta}}_i^r(\tau )\, {\dot{\eta}}_i^s(\tau
)]} -  \end{eqnarray*}

\bea
 &-&{{Q_i(\tau )}\over c}\, \Big(A_{\tau }(\tau ,\sigma^u) +
A_{r}(\tau ,\sigma^u)\, {\dot{\eta}}_i^r(\tau )\Big)\Big]-  \nonumber \\
&&\nonumber\\
&-&{\frac{1}{4c}}\,\sqrt{- g(\tau ,\sigma^u)}\, g^{AC }(\tau
,\sigma^u)\, g^{BD}(\tau ,\sigma^u)\, F_{AB}(\tau ,\sigma^u)\,
F_{CD}(\tau , \sigma^u).
 \label{3.1}
 \eea

\medskip

The canonical momenta are (for dimensional convenience we introduce
a $c$ factor in the definition of the electro-magnetic momenta)

\begin{eqnarray*}
\rho _{\mu }(\tau ,\sigma^u) &=& - \sgn\,{\frac{{\partial {\cal
L}(\tau ,\sigma^u)}}{{\partial z_{\tau }^{\mu }(\tau
,\sigma^u)}}}=\nonumber \\
&&\nonumber\\
&=&\sum_{i=1}^{N}\delta ^{3}(\sigma^u - \eta^u_i(\tau ))\, m_i\,
c{\frac{{z_{\tau \mu }(\tau ,\sigma^u) + z_{{r} \mu }(\tau ,
\sigma^u)\, {\dot{\eta}}_{i}^{r}(\tau )}}{\sqrt{\sgn\, [g_{\tau \tau
}(\tau , \sigma^u) + 2\, g_{\tau {r}}(\tau,\sigma^u)\,
{\dot{\eta}}_i^r(\tau ) + g_{{r}{s}}(\tau ,\sigma^u)\, {\dot{\eta}}
_i^r(\tau )\, {\dot{\eta}}_i^s(\tau )]}}} +  \nonumber \\
&&\nonumber\\
 &+&\sgn\,{\frac{\sqrt{- g(\tau ,\sigma^u)}}{4c}}\, \Big[(g^{\tau \tau }\, z_{\tau
\mu } + g^{\tau {r}}\, z_{r\mu })\, g^{AC}\, g^{BD}\, F_{AB}\,
F_{CD}-  \nonumber \\
&&\nonumber\\
&-&2\, \Big(z_{\tau \mu }\, (g^{A\tau}\, g^{\tau C}\, g^{ BD} +
g^{AC}\, g^{B\tau }\, g^{\tau D})+  \nonumber \\
&&\nonumber\\
&+&z_{{r}\mu }\, (g^{Ar}\, g^{\tau C} + g^{A\tau }\, g^{rC})\,
g^{BD}\Big)\, F_{AB }\, F_{CD}
)\Big](\tau ,\sigma^u) =  \nonumber \\
&&\nonumber\\
&=&[(\rho _{\nu }\, l^{\nu })\, l_{\mu } + (\rho _{\nu }\,
z_{r}^{\nu })\, \gamma ^{rs}\, z_{s\mu }](\tau ,\sigma^u),
\end{eqnarray*}

\begin{eqnarray*}
 \pi ^{\tau }(\tau ,\sigma^u) &=&c\, {\frac{{\partial L}}{{\partial
\partial_{\tau }A_{\tau }(\tau ,\sigma^u)}}} = 0,  \nonumber \\
&&\nonumber\\
 \pi ^r(\tau ,\sigma^u) &=&c\, {\frac{{\partial L}}{{\partial
\partial _{\tau }A_{r}(\tau ,\sigma^u)}}} =  {{\gamma (\tau
, \sigma^u)}\over {\sqrt{- g(\tau ,\sigma^u)} }}\, \h^{rs}(\tau
,\sigma^u)\, (F_{\tau s} - n^u\, F_{us})(\tau ,\sigma^u) =\nonumber \\
 &=& - {{\sqrt{\gamma}(\tau ,\sigma^u)}\over {1 + n(\tau ,\sigma^u)}}\,
 h^{rs}(\tau ,\sigma^u)\, \Big(E_s -
\epsilon_{suv}\, n^u\,B_v\Big)(\tau ,\sigma^u),\end{eqnarray*}

\begin{eqnarray*}
 \kappa _{i{r}}(\tau ) &=&+{\frac{{\partial L(\tau )}}{{\partial {\
\dot{\eta}}_{i}^{r}(\tau )}}}= {{Q_i}\over c}\, A_{r}(\tau
,\eta^u_i(\tau )) -\nonumber \\
&&\nonumber\\
&-&\sgn\, m_i\, c\, {{g_{\tau r}(\tau ,\eta^u_i(\tau )) + g_{
rs}(\tau ,\eta^u_i(\tau ))\, {\dot{\eta}}_i^s(\tau )}\over
{\sqrt{\sgn\, [g_{\tau \tau }(\tau ,\eta^u_i(\tau )) + 2\, g_{\tau
r}(\tau ,\eta^u_i(\tau ))\, {\dot{\eta}}_i^r(\tau ) + g_{rs}(\tau
,\eta^u_i(\tau ))\, { \dot{\eta}}_i^r(\tau )\, {\dot{\eta}}_i^s(\tau
)]} }},
 \end{eqnarray*}

\bea
 \pi _{\theta \,i}(\tau ) &=&{\frac{{\partial L(\tau )}}{{\partial
{\dot{ \theta}}_{i}(\tau )}}} = - {\frac{i}{2}}\, \theta _{i}^{\ast
}(\tau ),  \qquad \pi _{\theta ^{\ast }\,i}(\tau ) =
{\frac{{\partial L(\tau )}}{{\partial {\ \dot{\theta}}_{i}^{\ast
}(\tau )}}} = - {\frac{i}{2}}\, \theta _{i}(\tau ).
 \label{3.2}
\end{eqnarray}

\noindent The following Poisson brackets are assumed

\begin{eqnarray}
&&\{z^{\mu }(\tau ,\sigma^u),\rho _{\nu }(\tau ,{\sigma}^{^{\prime
}\, u}\} = - \sgn\, \eta _{\nu }^{\mu }\, \delta ^{3}(\sigma^u -
{\sigma}^{^{\prime} u}),  \nonumber \\
&&\nonumber\\
&&\{A_{A}(\tau ,\sigma^u),\pi ^{B}(\tau,{\sigma}^{^{\prime } u})\} =
c\, \eta _A^B\, \delta ^{3}(\sigma^u - {\sigma}^{^{\prime } u}),
\qquad \{\eta _{i}^r(\tau ),\kappa _{j s}(\tau )\} = + \delta
_{ij}\, \delta _s^r,  \nonumber \\
&&\nonumber\\
&&\{\theta _{i}(\tau ),\pi _{\theta \,j}(\tau )\}=-\delta _{ij},
\qquad \{\theta _{i}^{\ast }(\tau ),\pi _{\theta ^{\ast }\,j}(\tau
)\}=-\delta_{ij}.
 \label{3.3}
\end{eqnarray}

The Grassmann momenta give rise to the second class constraints

\bea
 &&\pi_{\theta \, i}+{\frac{i}{2}}\theta^{*}_i\approx 0,\qquad \pi_{\theta^{*}\, i}
 + {\frac{i}{2} } \theta_i\approx 0,\qquad
\lbrace \pi_{\theta \, i}+{\frac{i}{2}}\theta^{*}_i,
\pi_{\theta^{*}\, j}+{\frac{i}{2}}\theta_j\rbrace =-i\delta_{ij},
 \label{3.4}
 \eea

\noindent so that $\pi _{\theta \, i}$ and $\pi_{\theta^{*}\, i}$
can be eliminated with the help of Dirac brackets

\begin{equation}
\lbrace A,B\rbrace {}^{*}=\lbrace A,B\rbrace - i\,[\lbrace
A,\pi_{\theta \, i} + {\frac{i}{2}}\theta^{*}_i\rbrace \lbrace
\pi_{\theta^{*}\, i} + {\frac{i}{2}} \theta_i,B\rbrace + \lbrace
A,\pi_{\theta^{*}\, i} + {\frac{i}{2}} \theta_i \rbrace \lbrace
\pi_{\theta \, i} + {\frac{i}{2}}\theta^{*}_i,B\rbrace ].
 \label{3.5}
\end{equation}

\noindent As a consequence, the  Grassmann variables $\theta_i(\tau
)$, $\theta^*_i(\tau )$, have the fundamental Dirac brackets ( we
will still denote it as $\lbrace .,.\rbrace$ for the sake of
simplicity)

\beq
 \{\theta _{i}(\tau ), \theta _{j}(\tau )\} = \{\theta _{i}^{\ast
}(\tau ), \theta _{j}^{\ast }(\tau )\} = 0,  \qquad \{\theta
_{i}(\tau ), \theta_{j}^{\ast }(\tau )\} = - i\, \delta _{ij}.
 \label{3.6}
\eeq

\bigskip

If we introduce the energy-momentum tensor of the isolated system
(in inertial frames we have $T_{\perp\perp} = T^{\tau\tau}$ and
$T_{\perp r} = \delta_{rs}\, T^{\tau s}$)

\begin{eqnarray*}
 T^{AB}(\tau ,\sigma^u ) &=& - {2\over {\sqrt{g(\tau ,\sigma^u
)}}}\, {{\delta\, S}\over {\delta\, g_{AB}(\tau ,\sigma^u )}},
\nonumber \\
 &&{}\nonumber \\
 T^{\mu\nu} &=& z_A^{\mu}\, z_B^{\nu}\, T^{AB} =
  l^{\mu}\, l^{\nu}\, T_{\perp\perp} + (l^{\mu}\, z^{\nu}_r +
 l^{\nu}\, z^{\mu}_r)\, \gamma^{rs}\, T_{\perp s} + z^{\mu}_r\,
 z^{\mu}_s\, T^{rs},\nonumber \\
 &&{}\nonumber \\
 &&T_{\perp\perp} = l_{\mu}\, l_{\nu}\, T^{\mu\nu} = (1 + n)^2\,
 T^{\tau\tau},\nonumber \\
 &&T_{\perp r} = l_{\mu}\, z_{r\, \nu}\, T^{\mu\nu} = - (1 + n)\,
 h_{rs}\, (T^{\tau\tau}\, n^s + T^{\tau s}),\nonumber \\
 &&T_{rs} = z_{r\, \mu}\, z_{s\, \nu}\, T^{\mu\nu} = n_r\, n_s\,
 T^{\tau\tau} + (n_r\, h_{su} + n_s\, h_{ru})\, T^{\tau u} + h_{ru}\,
 h_{sv}\, T^{uv},
 \end{eqnarray*}

\bea
  T_{\perp \perp}(\tau ,\sigma^u ) &=& \Big({1\over {2\, c\, \sqrt{\gamma}}}\,
 \Big[{1\over {\sqrt{\gamma}}}\, h_{rs}\, \pi^r\, \pi^s + {{\sqrt{\gamma}}\over
 2}\, h^{rs}\, h^{uv}\, F_{ru}\, F_{sv}\Big]\Big)(\tau ,\sigma^u)+\nonumber\\
&&\nonumber\\
&+& \sum_{i=1}^{N}\, {{\delta^3(\sigma^u - \eta^u_i(\tau ))}\over
{\sqrt{\gamma (\tau ,\sigma^u)}}}\, \Big(\sqrt{m_i^2\, c^2 +
\h^{rs}\, \Big[\kappa _{i r}(\tau ) - {{Q_i}\over c}\, A_r\Big]\,
\Big[\kappa _{i s}(\tau) - {{Q_i}\over c}\,
A_s\Big)(\tau ,\sigma^u)\Big]},\nonumber \\
&&\nonumber\\
 T_{\perp s}(\tau ,\sigma^u ) &=& \Big({{F_{rs}\, \pi^s}\over {c\,
 \sqrt{\gamma}}}\Big)(\tau ,\vec \sigma) -
 \sum_{i=1}^{N}\, {{\delta^3(\sigma^u - \eta^u_i(\tau
))}\over {\sqrt{\gamma (\tau ,\sigma^u)}}}\, \Big[\kappa _{i\,s} -
{{Q_i}\over c}\, A_s(\tau ,\sigma^u)\Big],\nonumber \\
&&\nonumber\\
 T_{rs}(\tau ,\sigma^u ) &=& \Big(h_{ru}\, h_{sv}\, \Big[- {{\pi^u\, \pi^v}\over
 {\gamma}} + {{n^u\, n^v}\over {(1 + n)^2}}\, \Big({{n_m\, \pi^m}\over {(1 + n)\,
 \sqrt{\gamma}}}\Big)^2\Big] +\nonumber \\
 &+& {1\over 2}\, h_{rs}\, \Big[{{h_{lm}\, \pi^l\, \pi^m}\over {\gamma}} +
 {1\over 2}\, h^{lm}\, h^{uv}\, F_{lu}\, F_{mv}\Big] +
  \Big[h^{lm} - {{n^l\, n^m}\over {(1 + n)^2}}\Big]\, F_{rl}\,
 F_{sm}\Big)(\tau ,\sigma^u) +\nonumber \\
 &+& \sum_{i=1}^{N}\, {{\delta^3(\sigma^u - \eta^u_i(\tau
))}\over {\sqrt{\gamma (\tau ,\sigma^u)}}}\, \Big({{\Big[\kappa
_{i\,r} - {{Q_i}\over c}\, A_r\Big]\, \Big[\kappa _{i\,s} -
{{Q_i}\over c}\, A_s\Big] }\over {\sqrt{m_i^2\, c^2 + \h^{uv}\,
\Big[\kappa_{i u}(\tau ) - {{Q_i}\over c}\, A_u\Big]\,
\Big[\kappa_{i v}(\tau) - {{Q_i}\over c}\, A_v\Big]} }}\Big)(\tau
,\sigma^u),\nonumber \\
 &&{}
 \label{3.7}
 \eea

\noindent then from Eq.(\ref{3.2}) we get

\bea
 \rho_{\mu}(\tau ,\sigma^u) &=& \Big(\sqrt{- g}\, z_{A\, \mu}\,
 T^{\tau A}\Big)(\tau ,\sigma^u) =\nonumber \\
 &=& \Big((1 + n)^2\, \sqrt{\gamma}\, T^{\tau\tau}\, l_{\mu} +
 (1 + n)\, \sqrt{\gamma}\, \Big[T^{\tau r} + T^{\tau\tau}\, n^r\Big]\,
 z_{r\, \mu}\Big)(\tau ,\sigma^u) =\nonumber \\
 &=& \Big(\sqrt{\gamma}\, \Big[l_{\mu}\, T_{\perp\perp} - z_{r\, \mu}\,
 h^{rs}\, T_{\perp s}\Big]\Big)(\tau ,\sigma^u).
 \label{3.8}
 \eea

\bigskip

Let us remark that, since all the dependence on the embeddings is in
the 4-metric, the Euler-Lagrange equations for the embeddings
$z^{\mu}(\tau ,\sigma^u )$ associated with the Lagrangian
(\ref{3.1}) are  (the symbol '${\buildrel \circ \over =}$' means
evaluated on the solutions of the equations of motion)

\bea
 {{\delta\, S }\over {\delta\, z^{\mu}(\tau ,\sigma^u)}}
 &=&\Big( {{\partial {\cal L}}\over {\partial
z^{\mu}}}-\partial_A {{\partial {\cal L}}\over {\partial
z^{\mu}_A}}\Big) (\tau ,\sigma^u ) =
 2\, \eta_{\mu\nu}\, \partial_A\, \Big[\sqrt{-g}\, T^{AB}\,
 z_B^{\nu}\Big](\tau ,\sigma^u ) =\nonumber \\
 &=& \Big(\sqrt{-g}\, z^C_{\mu}\, g_{CD}\, T^{DA}{}_{;A}\Big)(\tau
 ,\sigma^u)\, {\buildrel \circ \over =}\, 0,
 \label{3.9}
\eea

\noindent where $T^{AB}{}_{;B}(\tau ,\sigma^u)$ is the covariant
derivative associated to the 4-metric $g_{AB}(\tau ,\sigma^u)$
induced by the admissible 3+1 splitting of Minkowski
space-time.\medskip

They may be rewritten in a form valid for every isolated system
$\Big(\partial_A\, T^{AB}\, z^{\mu}_B\Big)(\tau ,\sigma^u)\,
{\buildrel \circ \over =}\, - \Big({1\over {\sqrt{-g}}}\,
\partial_A\, [\sqrt{-g}\, z^{\mu}_B]\, T^{AB}\Big)(\tau ,\sigma^u)$.
When $\partial_A\, [\sqrt{-g}\, z^{\mu}_B](\tau ,\sigma^u) = 0$, as
it happens in inertial frames in inertial Cartesian coordinates, we
get the conservation of the energy-momentum tensor
$T^{AB}{|}_{inertial}$, i.e. $\partial_A\, T^{AB}{|}_{inertial}\,
{\buildrel \circ \over =}\, 0$. Then, after integrating over a
4-volume bounded by a 3-volume $V_1$ at $\tau_1$, a 3-volume $V_2$
at $\tau_2 > \tau_1$ and a time-like 3-surface $S_{12}$ joining them
and with section $S_{\tau}$, boundary of a 3-volume $V_{\tau}$, at
$\tau$, we get ${d\over {d\tau}}\, \int_{V_{\tau}} d^3\sigma\,
T^{A\tau}{|}_{inertial}(\tau ,\sigma^u) = - \int_{S_{\tau}}
d^2\Sigma_B\, T^{AB}{|}_{inertial}(\tau ,\sigma^u)$, namely the
time-variation of the 4-momentum contained in $V_{\tau}$ is balanced
by the flux of energy-momentum through the boundary $S_{\tau}$. For
infinite volume and suitable boundary conditions we get the
conservation of the 4-momentum $P^A = \int_{\Sigma_{\tau}}
d^3\sigma\, T^{A\tau}{|}_{inertial}(\tau ,\sigma^u)$. \medskip

Otherwise, in non-inertial frames and also in inertial frames with
non-Cartesian coordinates we do not have a real conservation law,
but the equation $T^{AB}{}_{;B}(\tau ,\sigma^u) \cir 0$, which, like
in general relativity, could be rewritten as a conservation law
$\partial_B\, \Big(T^{AB} + t^{AB}\Big)(\tau ,\sigma^u) \cir 0$
involving a coordinate-dependent energy-momentum pseudo-tensor
describing the "energy-momentum" of the foliation associated to the
3+1 splitting. Moreover a quantity as $\int_{\Sigma_{\tau}}
d^3\Sigma_B\, T^{AB}{|}_{non-inertial}(\tau ,\sigma^u)$ is not a
tensor under frame-preserving diffeomorphisms (even when
$T^{AB}_{non-inertial}$ transforms correctly as a tensor density),
so that it cannot give rise to a well defined coordinate-independent
quantity. However, differently from general relativity where the
equivalence principle says that global inertial frames do not exist,
in Minkowski space-time it is always possible to revert to inertial
frames and to find the standard 4-momentum constant of motion, which
is a 4-vector under the Poincare' transformations connecting
inertial frames.

\medskip

\bigskip

At the Hamiltonian level from Eqs.(\ref{3.2}) we obtain the
following five primary constraints

\bea
 \pi^{\tau}(\tau ,\sigma^u) &\approx& 0,\nonumber \\
 &&{}\nonumber \\
 {\cal H}_{\mu }(\tau ,\sigma^u) &=& \rho _{\mu }(\tau
,\sigma^u ) - l_{\mu }(\tau ,\sigma^u)\, \sqrt{\gamma (\tau
,\sigma^u)}\, T_{\perp \perp }(\tau ,\sigma^u) +\nonumber \\
 &+& z_{r \mu }(\tau ,\sigma^u)\, \h ^{rs}(\tau
,\sigma^u)\, \sqrt{\gamma (\tau ,\sigma^u)}\, T_{\perp s}(\tau
,\sigma^u )\approx 0,
 \label{3.10}
 \eea

\medskip
The Lorentz-scalar primary constraint $\pi^{\tau}(\tau ,\sigma^u)
\approx 0$ is a consequence of the invariance of the action under
electro-magnetic gauge transformations.
\medskip

\bigskip

The canonical Hamiltonian $H_c$ is

\bea H_{c} &=&+ \sum_{i=1}^{N}\, \kappa _{i r}(\tau )\,
{\dot{\eta}}_i^r(\tau ) + \int d^3\sigma\, \Big[ {1\over c}\, \pi
^A\, \partial _{\tau }\, A_A - \rho _{\mu }\, z_{\tau }^{\mu
} - {\cal L}\Big](\tau ,\sigma^u) =\nonumber \\
&=&{1\over c}\, \int d^3\sigma\, \Big[ \partial _r\, \Big(\pi
^r(\tau , \sigma^u)\, A_{\tau }(\tau ,\sigma^u)\Big) - A_{\tau
}(\tau ,\sigma^u)\, \Gamma (\tau ,\sigma^u)\Big] = - {1\over c}\,
\int d^3\sigma\, A_{\tau }(\tau ,\sigma^u)\, \Gamma (\tau
,\sigma^u),\nonumber \\
 &&{}
 \label{3.11}
\eea

\noindent after the elimination of a surface term and the
introduction of the quantity

\beq
 \Gamma (\tau ,\sigma^u) \equiv \partial _r\, \pi ^r(\tau
,\sigma^u) + \sum_{i=1}^{N}\, Q_i\, \delta ^3(\sigma^u -
\eta^u_i(\tau )).
 \label{3.12}
\eeq

As a consequence,  the Dirac Hamiltonian is

\beq H_{D} = \int d^3\sigma\, \Big[ \lambda ^{\mu }\, {\cal H} _{\mu
} + \mu \, \pi ^{\tau } - {1\over c}\, A_{\tau }\, \Gamma \Big](\tau
,\sigma^u).
 \label{3.13}
\eeq

Here $\lambda ^{\mu }(\tau ,\sigma^u)$ and $\mu(\tau ,\sigma^u)$ are
the Dirac multipliers associated with the primary
constraints.\medskip

The requirement that the five primary constraints be $\tau
$-independent, i.e. $\{\pi ^{\tau }(\tau ,\sigma^u), H_{D}\}\approx
0$, $\{ {\cal H}^{\mu}(\tau ,\sigma^u ), H_D \} \approx 0$, implies
only  the Gauss' law secondary constraint

\beq \Gamma (\tau ,\sigma^u)\approx 0.
 \label{3.14}
\eeq

\medskip

The 6 constraints are all first class, since they satisfy the
following Poisson brackets

\bea
 &&\{\Gamma (\tau ,\sigma^u), \pi^\tau(\tau,\sigma^{'\, u})\}
 = \{\Gamma (\tau ,\sigma^u), {\cal H}_{\mu}(\tau ,\sigma^{'\, u} )\}
= \{\pi^\tau(\tau,\sigma^u), {\cal H}_{\mu}(\tau ,\sigma^{'\, u} )\} = 0\nonumber\\
 &&\nonumber\\
 &&\lbrace {\cal H}_{\mu}(\tau ,\sigma^u ),{\cal H}_{\nu}(\tau
,\sigma^{'\, u} )\rbrace = {1\over c}\,
 \Big( [l_{\mu}\, z_{{r}\nu} - l_{\nu}\, z_{{r}\mu}]\, { {\pi^{ r}}
 \over {\sqrt{\gamma }} } -\nonumber \\
 &&\nonumber\\
 &&\qquad -z_{u\mu}\, \h^{{u}{r}}\, F_{{r}{s}}
\, \h^{{s}{v}}\, z_{ v\nu}\Big)(\tau ,\sigma^u) \, \Gamma (\tau
,\sigma^u )\, \delta^3( \sigma^u - \sigma^{'\, u})\approx 0.
 \label{3.15}
 \eea

\bigskip

The constraints $\pi^{\tau}(\tau ,\sigma^u ) \approx 0$ and $\Gamma
(\tau ,\sigma^u ) \approx 0$ are the canonical generators of the
electro-magnetic gauge transformations.\medskip

Instead the constraints ${\cal H}_{\mu}(\tau ,\sigma^u ) \approx 0$
generate the gauge transformations from an admissible 3+1 splitting
of Minkowski space-time to another one. These constraints  can be
replaced with their projections ${\cal H}_r(\tau ,\sigma^u )  =
{\cal H}_{\mu}(\tau ,\sigma^u )\, z_r^{ \mu}(\tau ,\sigma^u )
\approx 0$, ${\cal H}_{\perp}(\tau ,\sigma^u ) = {\cal H}_{\mu}(\tau
,\sigma^u )\, l^{\mu}(\tau ,\sigma^u ) \approx 0$, tangent and
normal to the instantaneous 3-space $\Sigma_{\tau}$ respectively.
Modulo the Gauss law constraint $\Gamma (\tau ,\sigma^u ) \approx
0$, the new constraints satisfy the universal Dirac algebra of the
super-hamiltonian and super-momentum constraints of canonical metric
gravity (see the first paper in Refs.\cite{11}). The gauge
transformations generated by the constraint ${\cal H}_{\perp}(\tau
,\sigma^u )$ change the instantaneous 3-spaces $\Sigma_{\tau}$ (i.e.
the clock synchronization convention), while those generated by the
constraints ${\cal H}_r(\tau ,\sigma^u )$ change the 3-coordinates
on $\Sigma_{\tau}$.

\medskip

The  Hamilton-Dirac equations are

\begin{eqnarray*}
 {{\partial\, z^{\mu}(\tau ,\sigma^u)}\over {\partial\, \tau}} &=&
 \Big((1 + n)\, l^{\mu} + n^r\, z^{\mu}_r\Big)(\tau ,\sigma^u)\,\, \cir
 \,\, - \sgn\, \lambda^{\mu}(\tau ,\sigma^u),
 \end{eqnarray*}

 \begin{eqnarray*}
 {{\partial A_{\tau}(\tau ,\sigma^u )}\over {\partial \tau}} &\cir&
 \{ A_{\tau}(\tau ,\sigma^u ), H_D\} = \mu(\tau
 ,\sigma^u ),\nonumber \\
 &&{}\nonumber \\
  {{\partial A_r(\tau ,\sigma^u )}\over {\partial \tau}} &\cir&
  \{ A_r(\tau ,\sigma^u ), H_D \} =
 - \int d^3\sigma^{'}\,\Big[\Big( \lambda_\mu\, l^\mu\,
\sqrt{\gamma}\Big)(\tau,\sigma^{'\, u})\,
\{A_r(\tau, \sigma^u), T_{\perp\perp}(\tau,\sigma^{'\, u})\} -\nonumber\\
&&\nonumber\\
&-&\Big(\lambda_\mu\, z^{\mu}_u\, h^{us}\,
\sqrt{\gamma}\Big)(\tau,\sigma^{'\, u})\,
\{A_r(\tau,\sigma^u), T_{\perp s}(\tau,\sigma^{'\, u})\} +\nonumber\\
&&\nonumber\\
  &+& {1\over c}\, A_{\tau}(\tau ,\sigma^{'\, u})\, \{ A_r(\tau
  ,\sigma^u  ), \Gamma (\tau ,\sigma^{'\, u})\} \Big],
  \end{eqnarray*}

\begin{eqnarray*}
 {{\partial \pi^r(\tau ,\sigma^u )}\over {\partial \tau}} &\cir&
  \{ \pi^r(\tau ,\sigma^u ), H_D \} =
 - \int d^3\sigma^{'}\,\Big[\Big( \lambda_\mu\, l^\mu\,
\sqrt{\gamma}\Big)(\tau,\sigma^{'\, u})\,
\{\pi^r(\tau,\sigma^u), T_{\perp\perp}(\tau,\sigma^{'\, u})\} -\nonumber\\
&&\nonumber\\
&-&\Big(\lambda_\mu\,z^{\mu}_u\, h^{us}\, \sqrt{\gamma}
\Big)(\tau,\sigma^{'\, u})\, \{\pi^r(\tau,\sigma^u), T_{\perp
s}(\tau,\sigma^{'\, u})\}\Big],
 \end{eqnarray*}

\bea
 \frac{d\eta^r_i(\tau)}{d\tau}&\cir&
  \{ \eta_i^r(\tau ), H_D \} =
 - \int d^3\sigma^{'}\,\Big[\Big( \lambda_\mu\, l^{mu}\,
\sqrt{\gamma}\Big)(\tau,\sigma^{'\, u})\,
\{\eta^r_i(\tau), T_{\perp\perp}(\tau,\sigma^{'\, u})\} -\nonumber\\
&&\nonumber\\
&-&\Big(\lambda_\mu\,z^{\mu}_u\, h^{us}\, \sqrt{\gamma}
\Big)(\tau,\sigma^{'\, u})\,
\{\eta_i^r(\tau), T_{\perp s}(\tau,\sigma^{'\, u})\},\nonumber\\
&&\nonumber\\
\frac{d\kappa_{ir}(\tau)}{d\tau}&\cir&
  \{ \kappa_{ir}(\tau), H_D \} =
 - \int d^3\sigma^{'}\,\Big[\Big( \lambda_\mu\, l^{mu}\,
\sqrt{\gamma} \Big)(\tau,\sigma^{'\, u})\,
\{\kappa_{ir}(\tau), T_{\perp\perp}(\tau,\sigma^{'\, u})\} -\nonumber\\
&&\nonumber\\
&-&\Big(\lambda_\mu\, z^{\mu}_u\, h^{us}\, \sqrt{\gamma}
\Big)(\tau,\sigma^{'\, u})\,
\{\kappa_{ir}(\tau), T_{\perp s}(\tau,\sigma^{'\, u})\} +\nonumber\\
&&\nonumber\\
  &+& {1\over c}\, A_{\tau}(\tau ,\sigma^{'\, u})\, \{
  \kappa_{ir}(\tau ), \Gamma (\tau ,\sigma^{'\, u})\} \Big].
 \label{3.16}
 \eea
\medskip

The Grassmann-valued electric charges are constants of the motion,
${{d\, Q_i(\tau )}\over {d \tau}} \, \cir\, 0$.

Since the embedding variables $z^{\mu}(\tau ,\sigma^u )$ are the
only configuration variables with Lorentz indices, the ten conserved
generators of the Poincar\'{e} transformations are:

\beq
 P^{\mu } = \int d^{3}\sigma \rho ^{\mu }(\tau ,\sigma^u),
\qquad J^{\mu \nu } =  \int d^{3}\sigma (z^{\mu }\rho ^{\nu }-z^{\nu
}\rho ^{\mu })(\tau ,\sigma^u).
 \label{3.17}
\eeq
 \medskip

The determination of the radiation gauge of the electro-magnetic
field in non-inertial frames will be done in the next Section.

\vfill\eject

\section{The Hamiltonian Description of Charged Particles and the
Electro-Magnetic Field in Non-Inertial Frames }

In this Section we study the system of charged positive-energy
scalar particles plus the electro-magnetic field in a given
admissible non-inertial frame. Then we define the radiation gauge in
non-inertial frames.

\subsection{The Hamilton Equations in an Admissible Non-Inertial Frame.}

Let us choose an admissible 3+1 splitting of the type (\ref{2.1}) by
adding the gauge fixing constraints

\bea
 \chi (\tau ,\sigma^u) &=& z^{\mu}(\tau ,\sigma^u) - z^{\mu}_F(\tau
 ,\sigma^u) \approx 0,\nonumber \\
 &&{}\nonumber \\
 &&z^{\mu}_F(\tau ,\sigma^u) = x^{\mu}(\tau ) + F^{\mu}(\tau
 ,\sigma^u),\qquad F^{\mu}(\tau ,0) = 0,
 \label{4.1}
 \eea

\noindent to the first class constraints ${\cal H}_{\mu}(\tau
,\sigma^u) \approx 0$ of Eqs.(\ref{3.10}).\medskip

From the Hamilton-Dirac equations (\ref{3.16}) we have that the
Dirac multipliers $\lambda^{\mu}(\tau ,\sigma^u)$ in the Dirac
Hamiltonian (\ref{3.13}) take the form

\bea
 \lambda^{\mu}(\tau ,\sigma^u) &\cir& - \sgn\, \Big({\dot x}^{\mu}(\tau ) +
 {{\partial\, F^{\mu}(\tau ,\sigma^u)}\over {\partial\, \tau}}\Big) =
 - \sgn\, z^{\mu}_{F\, \tau}(\tau ,\sigma^u) =\nonumber \\
 &=& - \sgn\, \Big[(1 + n_F)\, l^{\mu}_F + n^r_F\, \partial_r\,F^{\mu}
 \Big](\tau ,\sigma^u),\nonumber \\
 &&{}\nonumber \\
 &&- \lambda_{\mu}\, l^{\mu}_F = 1 + n_F,\qquad \lambda_{\mu}\,
 z^{\mu}_{F\, s}\, h^{sr}_F = n^r_F.
 \label{4.2}
 \eea
 \medskip

${\cal H}_{\mu}(\tau ,\sigma^u) \approx 0$ and $\chi (\tau
 ,\sigma^u) \approx 0$ are second class constraints \footnote{We assume $\{
 {\cal H}_{\mu}(\tau ,\sigma^u_1), \chi (\tau ,\sigma^u_2) \} \not=
 0$ as a restriction of $F^{\mu}(\tau ,\sigma^u)$}, which eliminate
 the variables $z^{\mu}(\tau ,\vec \sigma )$ and $\rho_{\mu}(\tau
 ,\sigma^u)$. If we go to Dirac brackets, so that these constraints
 become strongly zero, the Dirac Hamiltonian does not depend any
 more upon the constraints ${\cal H}_{\mu}(\tau ,\sigma^u) \approx
 0$.\medskip

To find the new Dirac Hamiltonian $H_{D\, F}$ at the level of Dirac
brackets (still denoted $\{ .,.\}$) let us put the Dirac multiplier
(\ref{4.2}) in the Hamilton-Dirac equations (\ref{3.16}) for all the
variables ${\cal F} = A_{\tau}, A_r, \pi^r, \eta^r_i, \kappa_{ir}$
independent from the embeddings and their momenta

\bea
 {{\partial\, {\cal F}(..)}\over {\partial\, \tau}} &\cir& \{ {\cal
 F}(..), H_D \} =\nonumber \\
 &=& \int d^3\sigma\, \{ {\cal F}(..), \Big(\lambda^{\mu}\, {\cal
 H}_{\mu} + \mu\, \pi^{\tau} - {1\over c}\, A_{\tau}\, \Gamma\Big)(\tau
 ,\sigma^u) \} =\nonumber \\
 &&{}\nonumber \\
 &\cir& \int d^3\sigma\, \{ {\cal F}(..), \Big((1 + n_F)\, \sqrt{\gamma_F}\,
 T_{\perp\perp} + n^r_F\, \sqrt{\gamma_F}\, T_{\perp r} + \mu\, \pi^{\tau} -
 {1\over c}\, A_{\tau}\, \Gamma\Big)(\tau ,\sigma^u) \} =\nonumber \\
 &{\buildrel {def}\over =}& \{ {\cal F}(..), H_{D\, F} \}.
 \label{4.3}
 \eea

As a consequence the new Dirac Hamiltonian is

\bea
 H_{D\, F} &=& \int d^3\sigma\, \Big((1 + n_F)\, \sqrt{\gamma_F}\, T_{\perp\perp}
 + n^r_F\, \sqrt{\gamma_F}\, T_{\perp r} + \mu\, \pi^{\tau} - {1\over c}\, A_{\tau}\,
 \Gamma\Big)(\tau ,\sigma^u) =\nonumber \\
 &&{}\nonumber \\
 &=&\int d^3\sigma\, \Big((1 + n_F(\tau ,\sigma^u))\,
 \Big[\sqrt{\gamma_F(\tau ,\sigma^u)}\, T^{\prime}_{\perp\perp}(\tau ,\sigma^u) +\nonumber \\
 &+& \sum_i\, \delta^3(\sigma^u - \eta^u_i(\tau ))\, \Big(\sqrt{m_i^2\, c^2 +
\h^{rs}_F\, \Big(\kappa_{ir}(\tau ) - {{Q_i}\over c}\, A_r\Big)\,
\Big(\kappa_{is}(\tau ) - {{Q_i}\over c}\, A_s\Big)}\Big)(\tau ,\sigma^u )\,\Big]\nonumber\\
&&\nonumber\\
&+&n_F^r(\tau ,\sigma^u)\, \left[\, {1\over c}\, F_{rs}(\tau
,\sigma^u)\, \pi^s(\tau ,\sigma^u) - \sum_i\, \delta^3(\sigma^u -
\eta^u_i(\tau ))\, \Big(\kappa_{ir}(\tau ) - {{Q_i}\over c}\,
A_r(\tau ,\sigma^u)\Big)\,
\right]\nonumber\\
&&\nonumber\\
&+&\mu(\tau ,\sigma^u)\, \pi^\tau(\tau ,\sigma^u) - {1\over c}\,
A_\tau(\tau ,\sigma^u)\, \Gamma(\tau ,\sigma^u) \Big),
 \label{4.4}
 \eea

\noindent where the energy-momentum tensor is evaluated at
$z^{\mu}(\tau ,\sigma^u) = z^{\mu}_F(\tau ,\sigma^u)$

 \bea
  \Big(\sqrt{\gamma_F}\, T^{\prime}_{\perp\perp}\Big)(\tau ,\sigma^u) &=&
\frac{1}{2c}\, \Big(\frac{1}{\sqrt{\gamma_F(\tau ,\sigma^u)}}\,
\h_{F\,rs}(\tau ,\sigma^u) \pi^r(\tau ,\sigma^u)\,
\pi^s(\tau ,\sigma^u) +\nonumber \\
&+& \frac{\sqrt{\gamma_F(\tau ,\sigma^u)}}{2}\, \h^{rs}_F(\tau
,\sigma^u)\, h^{uv}_F(\tau ,\sigma^u)\, F_{ru}(\tau ,\sigma^u)\,
F_{sv}(\tau ,\sigma^u)\Big).
 \label{4.5}
 \eea

\bigskip

The Hamilton-Dirac equations for the particle positions take the
form

 \bea
  \dot{\eta}^r_i(\tau ) &\cir&
   \left(\Big(1 + n_F\Big)\,\, \frac{\h^{rs}_F\, \Big(\kappa_{is}(\tau )
- {{Q_i}\over c}\, A_s\Big)}{\sqrt{m_i^2\, c^2 + \h^{uv}_F\,
\Big(\kappa_{iu}(\tau ) - {{Q_i}\over c}\, A_u\Big)\,
\Big(\kappa_{iv}(\tau ) - {{Q_i}\over c}\, A_v\Big)}} \right) (\tau
,\eta^u_i(\tau ))
- \nonumber \\
&-&n^r_F(\tau ,\eta^u_i(\tau )),
 \label{4.6}
  \eea

\noindent which can be inverted to get

\bea
 \kappa_{ir}(\tau ) &=& \Big(\frac{\h_{F\,rs}(\tau ,\eta^u_i(\tau ))\,
 m_ic\, \Big(\dot{\eta}^s_i(\tau ) + n_F^s\Big)}
 {\sqrt{\Big(1 + n_F\Big)^2 -
\h_{F\,uv}\, \Big(\dot{\eta}^u_i(\tau ) + n_F^u\Big)\,
\Big(\dot{\eta}^v_i(\tau ) + n_F^v\Big)}}\Big)(\tau ,\eta^u_i(\tau
)) +\nonumber \\
 &+& {{Q_i}\over c}\, A_r(\tau ,\eta^u_i(\tau )).
 \label{4.7}
 \eea

For the particle momenta we get the Hamilton-Dirac equations

 \bea
 &&\frac{d}{d\tau}\kappa_{ir}(\tau )\cir\,\,  {{Q_i}\over c}\, \dot{\eta}^u_i(\tau
)\, \frac{\partial\, A_u(\tau ,\eta^u_i(\tau ))}{\partial\eta_i^r} +
{{Q_i}\over c}\, \frac{\partial\, A_\tau(\tau ,\eta^u_i(\tau
))}{\partial \eta^r_i} + {\cal
F}_{ir}(\tau ),\nonumber\\
 &&{}\nonumber \\
 &&{}\nonumber\\
 &&{\cal F}_{ir}(\tau ) = \Big({ { m_i\, c\, \Big[1 + n_F\Big]^{-1}}\over
 {\sqrt{\Big(1 + n_F\Big)^2 - \h_{F\,uv}\,
\Big(\dot{\eta}^u_i(\tau ) + n_F^u\Big)\, \Big(\dot{\eta}^v_i(\tau )
+ n_F^v\Big)} }}\Big)(\tau ,\eta^u_i(\tau ))\nonumber \\
 &&{}\nonumber \\
 &&\Big(  \frac{\partial \h_{F\,st}(\tau ,\eta^u_i(\tau
))}{\partial \eta^r_i}\,\,  \Big(\dot{\eta}^s_i(\tau ) + n_F^s(\tau
,\eta^u_i(\tau ))\Big)\, \Big(\dot{\eta}^t_i(\tau ) + n_F^t(\tau
,\eta^u_i(\tau ))\Big) - \frac{\partial n_F(\tau
,\eta^u_i(\tau ))}{\partial \eta^r_i} +\nonumber\\
 &&\nonumber\\
 &+& \frac{\partial n_F^s(\tau ,\eta^u_i(\tau ))}{\partial
\eta^r_i}\, \h_{F\,st}(\tau ,\eta^u_i(\tau ))\,\,
\Big(\dot{\eta}^t_i(\tau ) + n_F^t(\tau ,\eta^u_i(\tau ))\Big)
\,\,\Big),
 \label{4.8}
 \eea

\noindent where ${\cal F}_{ir}(\tau )$ denotes a set of {\it
relativistic inertial forces}.

As a consequence, the second order form of the particle equations of
motion implied by Eqs. (\ref{4.7}) and (\ref{4.8}) is

\begin{eqnarray*}
 &&\frac{d}{d\tau}\left(
\frac{ \h_{F\,rs}\,\, m_i\, c\, \Big(\dot{\eta}^s_i(\tau ) +
n_F^s\Big)}{\sqrt{\Big(1 + n_F\Big)^2 - \h_{F\,uv}\,
\Big(\dot{\eta}^u_i(\tau ) + n_F^u\Big)\, \Big(\dot{\eta}^v_i(\tau )
+ n_F^v\Big)}} \right)(\tau ,\eta^u_i(\tau ))\,\on\nonumber \\
 &&\cir\,\,  {{Q_i}\over c}\, \left[\dot{\eta}^u_i(\tau )\,\left( \frac{\partial
A_u(\tau ,\eta^u_i(\tau ))}{\partial\eta_i^r} - \frac{\partial
A_r(\tau ,\eta^u_i(\tau ))}{\partial\eta_i^u}\right) + \left(
\frac{\partial A_\tau(\tau ,\eta^U_i(\tau ))}{\partial \eta^r_i} -
\frac{\partial A_r(\tau ,\eta^u_i(\tau
))}{\partial\tau}\right) \right] +\nonumber\\
&&\nonumber\\
&+&{\cal F}_{ir}(\tau ),
 \end{eqnarray*}

\bea
 &&or\nonumber \\
 &&{}\nonumber \\
 &&m_i\, c\, \frac{d}{d\tau}\left(
\frac{ \dot{\eta}^s_i(\tau ) + n_F^s}{\sqrt{\Big(1 + n_F\Big)^2 -
\h_{F\,uv}\, \Big(\dot{\eta}^u_i(\tau ) + n_F^u\Big)\,
\Big(\dot{\eta}^v_i(\tau ) + n_F^v\Big)}} \right)(\tau
,\eta^u_i(\tau
))\,\cir\nonumber \\
 &&\cir\,\,  {{Q_i}\over c}\, \h_F^{sr}(\tau ,\eta^u_i(\tau ))\, \Big[
\dot{\eta}^u_i(\tau )\,\left( \frac{\partial A_u(\tau ,
\eta^u_i(\tau ))}{\partial\eta_i^r} - \frac{\partial A_r(\tau ,
\eta^u_i(\tau ))}{\partial\eta_i^u}\right) +\nonumber \\
 &+& \left( \frac{\partial A_\tau(\tau ,\eta^u_i(\tau
))}{\partial \eta^r_i} - \frac{\partial A_r(\tau ,\eta^u_i(\tau
))}{\partial\tau}\right)\Big] + {\tilde {\cal F}}^s_i(\tau ),\nonumber \\
 &&{}\nonumber \\
 &&{}\nonumber \\
 &&{\tilde {\cal F}}^s_i(\tau ) =\nonumber \\
 &=& \Big({{m_i\, c\,\, \Big(1 + n_F\Big)^{-1}\, \h_F^{sr}}\over
 {\sqrt{\Big(1 + n_F\Big)^2 -
\h_{F\,uv}\, \Big(\dot{\eta}^u_i(\tau ) + n_F^u\Big)\,
\Big(\dot{\eta}^v_i(\tau ) + n_F^v\Big)}}}\Big)(\tau ,\eta^u_i(\tau
))\nonumber \\
 &&\Big[
 \Big(\frac{\partial \h_{F\,st}(\tau ,\eta^u_i(\tau ))}{\partial
\eta^r_i}\,\,  \Big(\dot{\eta}^s_i(\tau ) + n_F^s(\tau ,
\eta^u_i(\tau ))\Big)\, \Big(\dot{\eta}^t_i(\tau ) + n_F^t(\tau
,\eta^u_i(\tau ))\Big) -\nonumber \\
 &&{}\nonumber \\
 &-& \frac{\partial n_F(\tau
,\eta^u_i(\tau ))}{\partial \eta^r_i}+
  \frac{\partial n_F^s(\tau ,\eta^u_i(\tau ))}{\partial
\eta^r_i}\, \h_{F\,st}(\tau ,\eta^u_i(\tau ))\,\,
\Big(\dot{\eta}^t_i(\tau ) + n_F^t(\tau ,\eta^u_i(\tau ))\Big)
\,\, \Big) -\nonumber \\
 &-& \Big({{\partial\, h_{F\, ru}}\over {\partial\, \tau}} + {\dot
 \eta}^v_i(\tau )\, {{\partial\, h_{F\, ru}}\over {\partial\,
 \eta^v_i}}\Big)(\tau ,\eta^u_i(\tau ))\, \Big({\dot \eta}^u_i(\tau )
 + n^u_F(\tau ,\eta^u_i(\tau ))\Big) \Big].
 \label{4.9}
 \eea
\medskip

Here ${\tilde {\cal F}}_{ir}(\tau )$ is the form of inertial forces
whose non-relativistic limit to rigid non-inertial frames is
evaluated in Subsection C.

\bigskip

If, as in Eqs.(\ref{2.20}),  we {\it define} the non-inertial
electric and magnetic fields in the form \footnote{In the {\em
inertial case} Eqs.(\ref{2.19}) and (\ref{3.2}) imply  $\pi^s\on -
\delta^{sr}\, E^r = - {\tilde E}^s$, so that the components of the
energy-momentum tensor are $T_{\tau\tau}\, \on\, \frac{1}{2c}\,
\Big( {\vec E}^2 + {\vec B}^2 \Big)$, $T_{\tau r}\, \on \, {1\over
c}\, \Big( \vec{E}\times\vec{B} \Big)_r$.}

\bea
 E_r &\byd& \left(
\frac{\partial A_\tau}{\partial \eta^r_i}-\frac{\partial
A_r}{\partial\tau}\right)
= - F_{\tau r},\nonumber\\
 &&\nonumber\\
 B_r &\byd& \frac{1}{2}\, \varepsilon_{ruv}\,F_{uv} = \epsilon_{ruv}\,
 \partial_u\, A_{\perp\, v}\,\Rightarrow\, F_{uv} = \varepsilon_{uvr}\,B_r,
 \label{4.10}
 \eea

\noindent  the homogeneous Maxwell equations, allowing the
introduction of the electro-magnetic potentials, have the standard
inertial form $\epsilon_{ruv}\, \partial_u\,B_v = 0$,
$\epsilon_{ruv}\, \partial_u\,E_v + \frac{1}{c}\, \frac{\partial\,
B_r}{\partial \tau} = 0$.

Then also Eqs.(\ref{4.9}) take the standard inertial form plus
inertial forces

 \bea
 &&\frac{d}{d\tau}\left( \frac{ \h_{F\,rs}\,\,
 m_ic\, \Big(\dot{\eta}^s_i(\tau ) + n_F^s\Big)}{\sqrt{\Big(1 + n_F\Big)^2 -
\h_{F\,uv}\, \Big(\dot{\eta}^u_i(\tau ) + n_F^u\Big)\,
\Big(\dot{\eta}^v_i(\tau ) + n_F^v\Big)}}\right)(\tau ,\eta^u_i(\tau
))  \cir \nonumber \\
 &&{}\nonumber \\
 &\cir& {{Q_i}\over c}\, \left[E_r + \epsilon_{ruv}\, {\dot \eta}^u_i(\tau ) B_v\,
 \right](\tau , \eta^u_i(\tau )) + {\cal F}_{ir}(\tau ).
 \label{4.11}
  \eea

\bigskip

The Hamilton-Dirac equations for the electro-magnetic field are

\bea
 \frac{\partial}{\partial \tau}A_\tau(\tau,\sigma^u) &\cir&
c\, \mu(\tau,\sigma^u),\nonumber \\
 \frac{\partial}{\partial
\tau}\, A_r(\tau ,\sigma^u )&\cir& \Big(\frac{\partial}{\partial
\sigma^r}\, A_\tau + \frac{1 + n_F}{\sqrt{\gamma_F}}\, \h_{F\,rs}\,
\pi^s + n_F^s\, F_{sr}\Big)(\tau ,\sigma^u ),\nonumber\\
&&\nonumber\\
\frac{\partial}{\partial \tau}\, \pi^r(\tau ,\sigma^u ) &\cir&
\sum_i\, Q_i\, \dot{\eta}_i^r(\tau )\, \delta^3(\sigma^u
- \eta^u_i(\tau )) +\nonumber\\
&+&\Big(\frac{\partial}{\partial \sigma^s}\, \Big[ (1 +
n_F)\,\sqrt{\gamma_F}\, \h_F^{rs}\, \h_F^{uv} \, F_{uv} - (n_F^s\,
\pi^r - n_F^r\, \pi^s)\Big]\Big)(\tau ,\sigma^u ).
 \label{4.12}
  \eea

Eqs.(\ref{4.12}) imply

 \bea
\pi^s(\tau, \sigma^u)&=& -\left[- \frac{\sqrt{\gamma_{F}}}{1 +
n_F}\, \h_{F}^{sr}\left( F_{\tau r} - n^v_F\,
F_{vr}\right)\right](\tau, \sigma^u)=\nonumber\\
 &&\nonumber\\
 &=& -\sqrt{-g_{F}(\tau, \sigma^u)}\, g_{F}^{\tau
A}(\tau, \sigma^u)\, g_{F}^{s B}(\tau, \sigma^u)\,F_{AB}(\tau,
\sigma^u).
 \label{4.13}
 \eea

If we introduce the charge density $\bar \rho$, the charge current
density ${\bar j}^r$ and the total charge $Q_{tot} = \sum_i\, Q_i$
on $\Sigma_{\tau}$

 \bea
  \overline{\rho}(\tau, \sigma^u)&=&
\frac{1}{\sqrt{\gamma_{F}(\tau, \sigma^u)}} \sum_{i=1}^N
Q_i\,\delta^3(\sigma^u - \eta^u_i(\tau)),\nonumber\\
 &&\nonumber\\
  \overline{J}^r(\tau, \sigma^u)&=&
\frac{1}{\sqrt{\gamma_{F}(\tau, \sigma^u)}} \sum_{i=1}^N
Q_i\,\dot{\eta}^r_i(\tau)\, \delta^3(\sigma^u - \eta^u_i(\tau)),
\nonumber \\
  &&{}\nonumber \\
  \Rightarrow&& Q_{tot} = \int d^3\sigma\, \sqrt{\gamma_{F}(\tau, \sigma^u)}\,
\overline{\rho}(\tau, \sigma^u),
 \label{4.14}
  \eea

\noindent then the last of Eqs.(\ref{4.12}) can be rewritten in form

 \bea
\frac{\partial}{\partial\sigma^r}\,\pi^r(\tau, \sigma^u)&\approx&
 - \sqrt{\gamma_{F}(\tau, \sigma^u)}\,
\overline{\rho}(\tau, \sigma^u),\nonumber\\
 &&\nonumber\\
  \frac{\partial\,\pi^r(\tau, \sigma^u)} {\partial\tau} &\,\cir\,&
\frac{\partial}{\partial\sigma^s}\,
\left[\sqrt{-g_{F}}\,\h_{F}^{sv}\,\h_{F}^{ru}\, F_{vu} - (n_F^s\,
\pi^r - n_F^r\,\pi^s)\right](\tau, \sigma^u) +\nonumber\\
 &&\nonumber\\
&+&\sqrt{\gamma_{F}(\tau, \sigma^u)}\, \overline{J}^r(\tau,
\sigma^u).
 \label{4.15}
  \eea
\medskip

If we introduce the 4-current density  $s^A(\tau ,\sigma^u)$

 \bea
s^\tau(\tau, \sigma^u)&=& \frac{1}{\sqrt{-g_{F}(\tau, \sigma^u)}}
\sum_{i=1}^N\, Q_i\,\delta^3(\sigma^u - \eta^u_i(\tau)),\nonumber\\
 &&\nonumber\\
  s^r(\tau, \sigma^u)&=&
\frac{1}{\sqrt{-g_{F}(\tau, \sigma^u)}}\, \sum_{i=1}^N\,
Q_i\,\dot{\eta}^r_i(\tau)\, \delta^3(\sigma^u - \eta^u_i(\tau)),
 \label{4.16}
  \eea

\noindent and  we use (\ref{4.13}), then Eqs.(\ref{4.15}) can be
rewritten as manifestly covariant equations for the field strengths
as in Ref.\cite{25}

 \beq
\frac{1}{\sqrt{-g_{F}(\tau, \sigma^u)}}\, \frac{\partial}{\partial
\sigma^A}\left[\sqrt{-g_{F}(\tau, \sigma^u)}\, g_{F}^{AB}(\tau,
\sigma^u)\,  g_{F}^{CD}(\tau, \sigma^u)\, F_{BD} (\tau, \sigma^u)
\right]\, \cir\, - s^C(\tau, \sigma^u).
 \label{4.17}
  \eeq

Eqs.(\ref{4.17}) imply the following continuity equation due to the
skew-symmetry of $F_{AB}$

 \bea
 &&\frac{1}{\sqrt{-g_{F}(\tau, \sigma^u)}}\,
\frac{\partial}{\partial\sigma^C}\, \left[\sqrt{-g_{F}(\tau,
\sigma^u)}\, s^C(\tau, \sigma^u)\right]\, \cir\, 0,\nonumber \\
 &&{}\nonumber \\
 &&or\nonumber \\
 &&{}\nonumber \\
 &&\frac{1}{\sqrt{\gamma_{F}(\tau, \sigma^u)}}\,
\frac{\partial}{\partial\tau}\, \left[\sqrt{\gamma_{F}(\tau,
\sigma^u)}\, \overline{\rho}(\tau, \sigma^u)\right] +
\frac{1}{\sqrt{\gamma_{F}(\tau, \sigma^u)}}\,
\frac{\partial}{\partial\sigma^r}\, \left[\sqrt{\gamma_{F}(\tau,
\sigma^u)}\,\overline{J}^r(\tau, \sigma^u)\right]\, \cir\, 0, \nonumber \\
 &&{}
 \label{4.18}
  \eea
\medskip

\noindent so that consistently we recover $\frac{d}{d\tau}\,
Q_{tot}\, \cir\, 0$.

\bigskip

See Appendix A for the expression of the Landau-Lifschitz
non-inertial electro-magnetic fields \cite{17}.

\subsection{The Radiation Gauge for the Electro-Magnetic Field in
Non-Inertial Frames.}

In Appendix B there is a general discussion about the non-covariant
decomposition of the vector potential $\vec A(\tau ,\sigma^u)$ and
its conjugate momentum $\vec \pi (\tau ,\sigma^u)$ (the electric
field) into longitudinal and transverse parts in absence of matter.
Only with this decomposition we can define a Shanmugadhasan
canonical transformation adapted to the two first class constraints
generating electro-magnetic gauge transformations and identify the
physical degrees of freedom (Dirac observables) of the
electro-magnetic field without sources. This method  identifies the
{\it radiation gauge} as the natural one from the point of view of
constraint theory. Here we extend the construction to the case in
which there are charged particles: this will allow us to find the
expression of the mutual Coulomb interaction among the charges in
non-inertial frames .

\bigskip

As in Eq.(\ref{c3}) let us introduce the non-covariant flat
Laplacian $\Delta = \sum_r\, \partial^2_r$ in the instantaneous
non-Euclidean 3-space $\Sigma_{\tau}$. We use the non-covariant
notation ${\hat \partial}^r = \delta^{rs}\, \partial_s$  relying on
the positive signature of the 3-metric $h_{F\, rs}(\tau ,\sigma^u) =
- \sgn\, g_{F\, rs}(\tau ,\sigma^u)$. Since we have:

\beq
 \Delta\, \left(-\frac{1}{4\pi}\,\frac{1}{\sqrt{\sum_u\,
(\sigma^u - \sigma^{'\, u})^2} }\right) = \delta^3(\sigma^u,
\sigma^{'\, u}),\qquad
 or\,\, {1\over {\Delta}}\, \delta^3(\sigma^u, \sigma^{'\, u})
 = -\frac{1}{4\pi}\,\frac{1}{\sqrt{\sum_u\,
(\sigma^u - \sigma^{'\, u})^2} },
 \label{4.19}
 \eeq

\noindent with $\delta^3(\sigma^u, \sigma^{'\, u})$ the delta
function for $\Sigma_{\tau}$ \footnote{The delta functions are
defined in Appendix B after Eq.(\ref{c3}).}, we can introduce the
projectors

\bea
 {\bf P}^{rs}(\sigma^u, \sigma^{'\, u}) &=& \delta^{rs}\,
\delta^3(\sigma^u ,\sigma^{'\, u}) -
 {\hat \partial}^r\, {\hat \partial}^s\, \left(-\frac{1}{4\pi}\,
 \frac{1}{\sqrt{\sum_u\, (\sigma^u - \sigma^{'\, u})^2} }\right) =
  P^{rs}_{\perp}(\sigma^u )\, \delta^3(\sigma^u
 ,\sigma^{'\, u}),\nonumber \\
 &&{}\nonumber \\
 P^{rs}_{\perp}(\sigma^u ) &=& \delta^{rs} - {{{\hat \partial}^r\,
 {\hat \partial}^s}\over {\Delta}}.
 \label{4.20}
 \eea

As a consequence the transverse part of the electro-magnetic
quantities (${\hat \partial}^r\, A_{\perp r} = \partial_r\, A_{\perp
r} = 0$, $\partial_r\, \pi^r_{\perp} = 0$) are

 \bea
 A_{\perp r}(\tau ,\sigma^u ) &=& \delta_{ru}\, \int
 d^3\sigma^{'}\, {\bf P}^{rs}(\sigma^u ,\sigma^{'\, u})\,
 A_s(\tau ,\sigma^{'\, u}) = \delta_{ru}\, P^{us}_{\perp}(
 \sigma^u )\, A_s(\tau ,\sigma^u ),\nonumber \\
 \pi^r_\perp(\tau ,\sigma^u) &=& \sum_s\,\int d^3\sigma'\,{\bf
P}^{rs}(\sigma^u, \sigma^{'\, u})\, \pi^s(\tau ,\sigma^{'\, u})\, =
\sum_s\, P^{rs}_{\perp}(\sigma^u )\, \pi^s(\tau ,\sigma^u ).
 \label{4.21}
 \eea

\bigskip

Therefore the Gauss law constraint (\ref{3.12}) implies the
following decomposition of $\pi^r(\tau ,\sigma^u )$

\beq
 \pi^r(\tau ,\sigma^u) = \pi^r_\perp(\tau ,\sigma^u) +
 {\hat \partial}^r\, \int d^3\sigma'\,\left(-\frac{1}{4\pi}\,\frac{1}{
 \sqrt{\sum_u\, (\sigma^u - \sigma^{'\, u})^2}}\right)\,
 \left(\Gamma(\tau ,\sigma^{'\, u}) - \sum_i\, Q_i\,
\delta^3(\sigma^{'\, u} , \eta^u_i(\tau ))\right).
 \label{4.22}
  \eeq

If, following Dirac \cite{26}, we introduce the variable canonically
conjugate to $\Gamma (\tau ,\sigma^u )$ (it describes a Coulomb
cloud of longitudinal photons)

\bea
 \eta_{em}(\tau ,\sigma^u) &=& - \int
d^3\sigma'\, \left(-\frac{1}{4\pi}\,\frac{1}{ \sqrt{\sum_u\,
(\sigma^u - \sigma^{'\, u})^2} }\right)\,\left(\sum_r\, {\hat
\partial}^{'\, r}\, A_r(\tau ,\sigma^{'\, u})\right),\nonumber \\
&&\nonumber\\
&&\{\eta_{em}(\tau ,\sigma^u), \Gamma(\tau ,\sigma^{'\, u})\} =
\delta^3(\sigma^u, \sigma^{'\, u}),
 \label{4.23}
  \eea

\noindent we have the following non-covariant decomposition of the
vector potential

 \beq
 A_r(\tau ,\sigma^u) = A_{\perp\,r}(\tau ,\sigma^u) - \partial_r\,
 \eta_{em}(\tau ,\sigma^u).
  \label{4.24}
 \eeq

\bigskip

If we introduce the following new Coulomb-dressed momenta for the
particles

 \bea
\check{\kappa}_{ir}(\tau ) &=& \kappa_{ir}(\tau ) + {{Q_i}\over c}\,
\frac{\partial}{\partial\eta^r_i}\, \eta_{em}(\tau
,\eta^u_i(\tau )),\nonumber \\
 &&{}\nonumber \\
 \Rightarrow && \kappa_{ir}(\tau ) - {{Q_i}\over c}\, A_r(\tau ,\eta^u_i(\tau )) =
\check{\kappa}_{ir}(\tau ) - {{Q_i}\over c}\, A_{\perp\,r}(\tau
,\eta^u_i(\tau ))
 \label{4.25}
 \eea

\noindent we arrive at the following non-covariant Shanmugadhasan
canonical transformation in non-inertial frames

\bea
 &&\begin{array}{|c|c|}
\hline
&\\
A_r(\tau ,\sigma^u)&\eta^r_i(\tau )\\
&\\
\pi^r(\tau ,\sigma^u)&\kappa_{ir}(\tau )\\
&\\
\hline
\end{array}
\mapsto
\begin{array}{|cc|c|}
\hline
&&\\
A_{\perp\,r}(\tau ,\sigma^u)&\eta_{em}(\tau ,\sigma^u)&\eta^r_i(\tau )\\
&&\\
\pi^r_\perp(\tau ,\sigma^u)&\Gamma(\tau ,\sigma^u)
\approx 0&\check{\kappa}_{ir}(\tau )\\
&&\\
\hline
\end{array},\nonumber \\
 &&{}\nonumber \\
 &&\{A_{\perp r}(\tau ,\sigma^u), \pi_\perp^s(\tau ,\sigma^{'\, u})\}
= c\, {\bf P}^{rs}(\sigma^u, \sigma^{'\, u}) = c\,
P^{rs}_{\perp}(\sigma^u )\, \delta^3(\sigma^u, \sigma^{'\, u}),
\nonumber\\
 &&\nonumber\\
 &&\{\eta^r_i(\tau ), \check{\kappa}_{is}(\tau )\} =  \delta^r_s\,\delta_{ij}.
 \label{4.26}
  \eea

The electromagnetic part of the hamiltonian (4.4) can be expressed
in terms of the new canonical variables, since we have:

\begin{eqnarray}
&&\int d^3\sigma\,\sqrt{\gamma(\tau,\sigma^u)}\left[(1 + n_F)\,
T^{\prime}_{\perp\perp}+\frac{n^r_F}{c}\,
F_{rs}\,\pi^s\right](\tau ,\sigma^u ) = \nonumber\\
&&\nonumber\\
&=& {1\over c}\, {\cal W}(\eta^u_1(\tau ),...,\eta^u_N(\tau )) +
\int d^3\sigma\, \sqrt{\gamma(\tau,\sigma^u)} \Big[(1 + n_F)\,
\check{T}_{\perp\perp} + n^r_F\, \check{T}_{\perp r}\Big](\tau
,\sigma^u)+\nonumber\\
&&\nonumber\\
&+&\frac{1}{c}\int d^3\sigma\,
a_\tau(\tau,\sigma^u)\,\Gamma(\tau,\sigma^u)+{\cal O}(\Gamma^2),
 \label{4.27}
\end{eqnarray}

\noindent where the energy-momentum tensor has the form

 \bea
 \sqrt{\gamma(\tau,\sigma^u)}\,\check{T}_{\perp\perp}(\tau ,\sigma^u)&=&
+\frac{\h_{F\,rs}(\tau ,\sigma^u)}{2\, c\, \sqrt{\gamma_F(\tau
,\sigma^u)}}\,\pi_\perp^r(\tau ,\sigma^u)\,\pi_\perp^s(\tau
,\sigma^u) +\nonumber\\
&&\nonumber\\
&+&\frac{\sqrt{\gamma_F(\tau ,\sigma^u)}}{4\, c}\, \h^{rs}_F(\tau
,\sigma^u)\, h^{uv}_F(\tau ,\sigma^u)\,
F_{ru}(\tau ,\sigma^u)\, F_{sv}(\tau ,\sigma^u),\nonumber \\
 &&{}\nonumber \\
 &&{}\nonumber \\
 \sqrt{\gamma(\tau,\sigma^u)}\,\check{T}_{\perp r}(\tau ,\sigma^u) &=& {1\over c}\,
F_{rs}(\tau ,\sigma^u)\,\pi_\perp^s(\tau ,\sigma^u).
 \label{4.28}
 \eea

In Eq.(\ref{4.27})  we have introduced the potentials ($F_{rs} =
\partial_r\, A_{\perp s} - \partial_s\, A_{\perp r}$
)

\bea
 &&{\cal W}(\eta^u_1(\tau ),...,\eta^u_N(\tau ))=\nonumber\\
&&\nonumber\\
&=&+ \int d^3\sigma\, \frac{\h_{F\,rs}(\tau ,\sigma^u)\, \Big(1 +
n_F(\tau ,\sigma^u)\Big)}{2\sqrt{\gamma_F(\tau ,\sigma^u)}}\,
\left(2\, \pi_\perp^r(\tau ,\sigma^u) + \frac{1}{4\pi}\, \sum_i\,
\frac{\partial}{\partial\sigma^r}\, \frac{Q_i}{ \sqrt{\sum_u\,
(\sigma^u - \eta_i^u(\tau ))^2} }\right)
\nonumber \\
 && \left( \frac{1}{4\pi}\, \sum_j\,
\frac{\partial}{\partial\sigma^s}\, \frac{Q_j}{ \sqrt{\sum_u\,
(\sigma^u - \eta_j^u(\tau ))^2}}\right) +\nonumber\\
&&\nonumber\\
&+&n_F^r(\tau ,\sigma^u)\, F_{rs}(\tau ,\sigma^u)\, \left(
\frac{1}{4\pi}\, \sum_j\, \frac{\partial}{\partial\sigma^s}\,
\frac{Q_j}{ \sqrt{\sum_u\, (\sigma^u - \eta_j^u(\tau ))^2} }\right),
 \label{4.29}
 \eea

\noindent and the function

\bea
 a_\tau(\tau ,\sigma^u)&=&  \int d^3\sigma'\,
 \frac{1}{4\pi\, \sqrt{\sum_u\, (\sigma^u - \sigma^{'\, u})^2}
 } \,\, \frac{\partial}{\partial\sigma^{\prime\,r}}\, \Big[
 n^s_F(\tau ,\sigma^{'\, u})\, F_{sr}(\tau ,\sigma^{'\, u})
 +\nonumber \\
 &+& \frac{(1 + n_F(\tau ,\sigma^{'\, u}))\, \h_{F\,rs}(\tau
,\sigma^{'\, })}{\sqrt{\gamma_F(\tau ,\sigma^{'\, u})}}\,
\left(\pi_\perp^s(\tau ,\sigma^{'\, u}) + \frac{1}{4\pi}\, \sum_j\,
\frac{\partial}{\partial\sigma^{\prime\,s}}\,
\frac{Q_j}{\sqrt{\sum_u\, (\sigma^u - \eta_j^u(\tau ))^2}}\right)
\Big].\nonumber\\
&&{}
 \label{4.30}
  \eea

\medskip

Then, the Dirac Hamiltonian (\ref{4.4}) has the following form in
the new variables

\bea H_{D\, F}&=&\sum_i\,\Big(1 + n_F(\tau
 ,\eta^u_i(\tau ))\Big)\times\nonumber\\
&&\nonumber\\
&&\times\sqrt{m_i^2\, c^2 + \h^{rs}_F(\tau , \eta^u_i(\tau ))\,
(\check{\kappa}_{ir}(\tau ) - {{Q_i}\over c}\, A_{\perp\,r}(\tau
,\eta^u_i(\tau )))\, (\check{\kappa}_{is}(\tau )
- {{Q_i}\over c}\, A_{\perp\,s}(\tau ,\eta^u_i(\tau )))} -\nonumber\\
&&\nonumber\\
&-&\sum_i\,n_F^r(\tau ,\eta^u_i(\tau ))\,(\check{\kappa}_{ir}(\tau )
- {{Q_i}\over c}\, A_{\perp\,r}(\tau ,\eta^u_i(\tau ))) +\nonumber\\
&&\nonumber\\
&+&{1\over c}\, {\cal W}(\eta^u_1(\tau ),...,\eta^u_N(\tau )) + \int
d^3\sigma\, \sqrt{\gamma(\tau,\sigma^u)} \Big[(1 + n_F)\,
\check{T}_{\perp\perp} + n^r_F\, \check{T}_{\perp r}\Big](\tau
,\sigma^u)\nonumber\\
&&\nonumber\\
&+&\int d^3\sigma\,\mu(\tau ,\sigma^u)\, \pi^\tau(\tau ,\sigma^u) -
{1\over c}\, \Big( A_\tau(\tau,\sigma^u) - a_\tau(\tau
,\sigma^u)\Big)\, \Gamma(\tau ,\sigma^u)\Big) + {\cal O}(\Gamma^2),
 \label{4.31}
\eea

\noindent In Eq.(\ref{4.31}) we can discard the term quadratic in
the constraint $\Gamma (\tau ,\sigma^u ) \approx 0$, because it is
strongly zero according to constraint theory: it does never
contribute to the dynamics on the constraint sub-manifold (the only
relevant region of phase space for constrained systems).\medskip

\bigskip

To get the non-covariant radiation gauge we add the gauge fixing

\beq
 \eta_{em}(\tau,\sigma^u)\approx 0,
 \label{4.32}
 \eeq

\noindent implying $A_r \approx A_{\perp r}$ due to Eq.(\ref{4.26}).
The $\tau$-constancy, $\frac{\partial\eta_{em}(\tau,
\sigma^u)}{\partial\tau} \approx 0$, of this gauge fixing, together
with the Gauss law constraint $\Gamma(\tau,\sigma^u) \approx 0$,
implies the secondary gauge fixing

\beq
 A_\tau(\tau,\sigma^u) - a_\tau(\tau,\sigma^u) \approx 0,
 \label{4.33}
 \eeq

\noindent so that we get

 \bea
&&A_\tau(\tau ,\sigma^u) \approx  \int d^3\sigma'\, \frac{1}{4\pi\,
\sqrt{\sum_u\, (\sigma^u - \sigma^{'\, u})^2} }\,\,
\frac{\partial}{\partial\sigma^{\prime\,r}}\, \Big[
n^s_F(\tau ,\sigma^{'\, u})\, F_{sr}(\tau ,\sigma^{'\, u}) +\nonumber\\
 &+&\frac{\Big(1 + n_F(\tau ,\sigma^{'\, u})\Big)\, \h_{F\,rs}(\tau
,\sigma^{'\, u})}{\sqrt{\gamma_F(\tau ,\sigma^{'\, u})}}\,
\left(\pi_\perp^s(\tau ,\sigma^{'\, u}) + \frac{1}{4\pi}\, \sum_j\,
\frac{\partial}{\partial\sigma^{\prime\,s}}\, \frac{Q_j}{
\sqrt{\sum_u\, (\sigma^{'\, u} - \eta_j^u(\tau ))^2} }\right)
\Big).\nonumber \\
 &&{}
 \label{4.34}
\eea

\bigskip

Therefore, in the radiation gauge the magnetic field of
Eqs.(\ref{2.19}) is transverse: $B_r = \epsilon_{ruv}\, \partial_u\,
A_{\perp\, v}$. But the electric field $E_r = - F_{\tau r} = -
\partial_{\tau}\, A_{\perp\, r} + \partial_r\, A_{\tau}$ is not
transverse: it has $E_{\perp\, r} = - \partial_{\tau}\, A_{\perp\,
r}$ as a transverse component. Instead the transverse quantity is
$\pi^r_{\perp}$, which coincides with $\delta^{rs}\, E_{\perp\, s}$
only in inertial frames, and whose expression in terms of the
electric and magnetic fields, determined by Eqs.(\ref{4.22}) and
(\ref{3.2}), is $ \pi^r_{\perp}(\tau ,\sigma^u) =
\Big[{{\sqrt{\gamma}}\over {1 + n}}\, h^{rs}\, (E_s -
\epsilon_{suv}\, n^u\, B_v)\Big](\tau ,\sigma^u) + {\hat
\partial}^r\, \Big(\sum_i\, {{Q_i}\over {4\pi\, \sqrt{\sum_u\,
(\sigma^u - \eta_i^u(\tau))^2} }}\Big)$.

\bigskip

The final form of the Dirac Hamiltonian in the radiation gauge
(after the elimination of the variables $\eta_{em}$, $\Gamma$,
$A_{\tau}$, $\pi^{\tau}$ by going to Dirac brackets) is

\bea
 H_{D\, F}&=&\sum_i\,\Big(1 + n_F(\tau
 ,\eta^u_i(\tau ))\Big)\times\nonumber\\
&&\nonumber\\
&&\times\sqrt{m_i^2\, c^2 + \h^{rs}_F(\tau , \eta^u_i(\tau ))\,
(\check{\kappa}_{ir}(\tau ) - {{Q_i}\over c}\, A_{\perp\,r}(\tau
,\eta^u_i(\tau )))\, (\check{\kappa}_{is}(\tau )
- {{Q_i}\over c}\, A_{\perp\,s}(\tau ,\eta^u_i(\tau )))} -\nonumber\\
&&\nonumber\\
&-&\sum_i\,n_F^r(\tau ,\eta^u_i(\tau ))\,(\check{\kappa}_{ir}(\tau )
- {{Q_i}\over c}\, A_{\perp\,r}(\tau ,\eta^u_i(\tau ))) +\nonumber\\
&&\nonumber\\
&+&{1\over c}\, {\cal W}(\eta^u_1(\tau ),...,\eta^u_N(\tau )) + \int
d^3\sigma\, \sqrt{\gamma(\tau,\sigma^u)} \Big[(1 + n_F)\,
\check{T}_{\perp \perp} + n^r_F\, \check{T}_{\perp r}\Big](\tau
,\sigma^u)
 \label{4.35}
\eea

\noindent where $\check{T}_{AB}$ is given in Eq.(\ref{4.28}). In
$H_{DF}$ the components of $g_{AB}(\tau ,\sigma^u)$ are {\it the
inertial potentials giving rise to the relativistic inertial
forces}.

\hfill

The Hamilton-Dirac equations for the particles are (${\cal
F}_{ir}(\tau )$ is defined in Eq.(\ref{4.8}))

\bea
 \dot{\eta}^r_i(\tau ) &\cir&\frac{\Big(1 + n_F(\tau ,\eta^u_i(\tau ))\Big)\,
\h^{rs}_F(\tau ,\eta^u_i(\tau ))\, \Big(\check{\kappa}_{is}(\tau ) -
{{Q_i}\over c}\, A_{\perp\,s}(\tau ,\eta^u_i(\tau
))\Big)}{\sqrt{m_i^2\, c^2 + \, h^{uv}_F(\tau ,\eta^u_i(\tau ))\,
\Big(\check{\kappa}_{iu}(\tau ) - {{Q_i}\over c}\, A_{\perp\,u}(\tau
,\eta^u_i(\tau ))\Big)\, \Big(\check{\kappa}_{iv}(\tau ) -
{{Q_i}\over c}\, A_{\perp\,v}(\tau ,\eta^u_i(\tau ))\Big)}} -\nonumber \\
&-& n^r_F(\tau ,\eta^u_i(\tau )),\nonumber \\
 &&{}\nonumber \\
 \frac{d}{d\tau}\check{\kappa}_{ir}(\tau ) &\cir&  {{Q_i}\over c}\, \dot{\eta}^u_i(\tau )\,
 \frac{\partial\, A_{\perp\,u}(\tau ,\eta^u_i(\tau ))}{\partial\, \eta_i^r} -
 {1\over c}\, \frac{\partial}{\partial \eta^r_i}{\cal
W}(\eta^u_1(\tau ),...,\eta^u_N(\tau )) + {\cal F}_{ir}(\tau ).
 \label{4.36}
  \eea

\medskip

In the second half of Eqs.(\ref{4.36}) the sum of the inertial
2-body Coulomb potentials is replaced by the non-inertial N-body
potential ${\cal W}(\eta^u_1(\tau ),...,\eta^u_N(\tau ))$ of
Eq.(\ref{4.29}), which can be shown to have the following property
due to Eq.(\ref{4.30})

 \beq
  \frac{\partial{\cal W}}{\partial\eta^r_i} = - Q_i\,
\left(\frac{\partial a_\tau}{\partial\sigma^r}\right)_{\sigma^u =
\eta^u_i} \approx - Q_i\,\left(\frac{\partial
A_\tau}{\partial\sigma^r}\right)_{\sigma^u = \eta^u_i}.
 \label{4.37}
 \eeq

In the radiation gauge the electric field of Eq.(\ref{2.19}) is $E_r
\approx - \partial_{\tau}\, A_{\perp r} + \partial_r\, A_{\tau}$.
Consistently with Eq.(\ref{4.11}) we have

\bea
 Q_i\,E_r(\tau ,\eta^u_i(\tau )) &=& - Q_i \frac{\partial
A_{\perp r}(\tau ,\eta^u_i(\tau ))}{\partial\tau} +
Q_i\left(\frac{\partial A_\tau(\tau
,\sigma^u)}{\partial\sigma^r}\right)_{\sigma^u = \eta^u_i}
\approx\nonumber \\
 &\approx& - Q_i \frac{\partial
A_{\perp r}(\tau ,\eta^u_i(\tau ))}{\partial\tau} -
\frac{\partial{\cal W}(\eta^u_1(\tau ),...,\eta^u_N(\tau
))}{\partial\eta^r_i} =\nonumber \\
 &=& Q_i\, E_{\perp r}(\tau ,\eta^u_i(\tau )) - \frac{\partial{\cal
 W}(\eta^u_1(\tau ),...,\eta^u_N(\tau
))}{\partial\eta^r_i}.
 \label{4.38}
 \eea

\bigskip

The first of Eqs.(\ref{4.36}) can be inverted to get

\bea
 \check{\kappa}_{ir}(\tau )& =&
 \Big(\frac{\h_{F\,rs}\, m_ic\,\Big(\dot{\eta}^s_i(\tau )
 + n_F^s\Big)}{\sqrt{\Big(1 + n_F\Big)^2 - \h_{F\,uv}\,
 \Big(\dot{\eta}^u_i(\tau ) + n_F^u\Big)\, \Big(\dot{\eta}^v_i(\tau )
 + n_F^v\Big)}}\Big)(\tau ,\eta^u_i(\tau ))  + \nonumber \\
 &+& {{Q_i}\over c}\, A_{\perp\,r}(\tau ,\eta^u_i(\tau )).
 \label{4.39}
  \eea

\bigskip
See the next Subsection  for its expression in a nearly
non-relativistic frame.

\bigskip

In the general case to evaluate the integral in Eq.(\ref{4.39}) we
must regularize the function $t^{rs}(\sigma^u) = \frac{1}{
\Big(\sum_u\, (\sigma^u)^2\Big)^3/2 }\,\left(\delta^{rs} - 3\,
\frac{\sigma^r\,\sigma^s}{ \Big(\sum_u\, (\sigma^u)^2\Big)
}\right)$, which is singular at $\sigma^u = 0$. By considering it as
a distribution, we must give a prescription to define the integral
$\int d^3\sigma\, t^{rs}(\sigma^u)\, f(\sigma^u)$, where
$f(\sigma^u)$ is a test function. Following Ref. \cite{27}, we
consider the sphere ${\cal S}_R$ centered in the origin and defined
by the relation $\sqrt{\sum_u\, (\sigma^u)^2} < R$ and the space
$\Omega_R$ external to it of the points such that $\sqrt{\sum_u\,
(\sigma^u)^2} \ge R$. The integral is written in the form

\beq
 \int d^3\sigma\,t^{rs}(\sigma^u)\,f(\sigma^u) =
\int_{{\cal S}_R} d^3\sigma\,t^{rs}(\sigma^u)\,f(\sigma^u)+
\int_{\Omega_R} d^3\sigma\,t^{rs}(\sigma^u)\,f(\sigma^u).
 \label{4.40}
 \eeq

The first term, containing the singularity, can be shown to have the
expression

\beq
 \lim_{R\rightarrow 0}\,\int_{{\cal S}_R}
d^3\sigma\,t^{rs}(\sigma^u)\,f(\sigma^u) =
\frac{4\pi}{3}\,\delta^{rs}\,f(0).
 \label{4.41}
 \eeq

Regarding the second term in Eq.(\ref{4.40}) we can define a
distribution $\overline{t}^{rs}(\sigma^u)$ such that the following
integral

\beq
 \lim_{R\rightarrow 0}\int_{\Omega_R}
d^3\sigma\,t^{rs}(\sigma^u)\,f(\sigma^u) = \int
d^3\sigma\,\overline{t}^{rs}(\sigma^u)\,f(\sigma^u)
 \label{4.42}
 \eeq

\noindent has no singularity in the origin. As a consequence we get

\beq
 t^{rs}(\sigma^u) = \frac{4\pi}{3}\, \delta^{rs}\, \delta^3(\sigma^u)
 + \overline{t}^{rs}(\sigma^u).
 \label{4.43}
 \eeq

\bigskip

Therefore we get

 \bea
&&{\cal W}(\eta^u_1(\tau ),...,\eta^u_N(\tau ))=\nonumber\\
&&\nonumber\\
&=&  \sum_{i\neq j}\, \int d^3\sigma\, \frac{\h_{F\,rs}(\tau
,\sigma^u)\, \Big(1 + n_F(\tau
,\sigma^u)\Big)}{2\,\sqrt{\gamma_F(\tau ,\sigma^u)}}\nonumber \\
 &&\left( \frac{1}{4\pi}\, \frac{\partial}{\partial\sigma^r}\,
\frac{Q_i}{ \sqrt{\sum_u\, (\sigma^u - \eta_i^u(\tau))^2} }\right)\,
\left( \frac{1}{4\pi}\, \frac{\partial}{\partial\sigma^s}\,
\frac{Q_j}{ \sqrt{\sum_u\, (\sigma^u - \eta_j^u(\tau))^2}
}\right) +\nonumber\\
&&\nonumber\\
&+&\int d^3\sigma\, \left[ \frac{\h_{F\,rs}\, \Big(1 + n_F\Big)}{
\sqrt{\gamma_F}}\,\pi^r_\perp + n_F^r\, F_{rs}\right](\tau
,\sigma^u)\, \left( \frac{1}{4\pi}\, \sum_j\,
\frac{\partial}{\partial\sigma^s}\, \frac{Q_j}{ \sqrt{\sum_u\,
(\sigma^u - \eta_j^u(\tau ))^2} }\right).
\nonumber \\
 &&{}
 \label{4.44}
  \eea

After some integrations by parts we get

\bea
&&{\cal W}(\eta^u_1(\tau ),...,\eta^u_N(\tau ))=\nonumber\\
&&\nonumber\\
&=&\sum_{i\neq j}\, \int d^3\sigma\, \frac{\h_{F\,rs}(\tau
,\sigma^u)\, \Big(1 + n_F(\tau ,\sigma^u)\Big)}{2\,
\sqrt{\gamma_F(\tau ,\sigma^u)}}\, \left(\frac{1}{16\pi^2}\,
\frac{Q_i\,Q_j}{ \sqrt{\sum_u\, (\sigma^u - \eta_i^u(\tau))^2}
}\right)\,t^{rs}(\sigma^u - \eta^u_j(\tau )) -\nonumber\\
&&\nonumber\\
&+&\sum_{i\neq j}\,\int d^3\sigma\,
\frac{\partial}{\partial\sigma^s}\, \left(\frac{\h_{F\,rs}(\tau
,\sigma^u)\, \Big(1 + n_F(\tau ,\sigma^u)\Big)}{2\,
\sqrt{\gamma_F(\tau ,\sigma^u)}}\right)\nonumber \\
 &&\left(\frac{1}{4\pi}\, \frac{Q_i\,Q_j}{ \sqrt{\sum_u\, (\sigma^u -
\eta_i^u(\tau))^2} }\right)\, \left(\frac{1}{4\pi}\, \frac{\sigma^r
- \eta^r_j(\tau )}
{ \Big(\sum_u\, (\sigma^u - \eta_j^u(\tau))^2\Big)^3/2}\right) -\nonumber\\
&&\nonumber\\
&-& \int d^3\sigma\, \left(\frac{1}{4\pi}\, \sum_j\, \frac{Q_j}{
\sqrt{\sum_u\, (\sigma^u - \eta_j^u(\tau))^2} }\right) \,
 \frac{\partial}{\partial\sigma^s}\, \left[ \frac{\h_{F\,rs}\,
  \Big(1 + n_F\Big)}{ \sqrt{\gamma_F}}\, \pi^r_\perp + n_F^r\,
F_{rs}\right](\tau ,\sigma^u),\nonumber \\
 &&{}
 \label{4.45}
 \eea

\noindent and then we can get the following form

 \bea
&&{\cal W}(\eta^u_1(\tau ),...,\eta^u_N(\tau )) =\nonumber\\
&&\nonumber\\
&=&\sum_{i\neq j}\, \frac{1}{12\pi}\, \sum_r\, \left(
\frac{\h_{F\,rr}(\tau ,\eta^u_j(\tau ))\, \Big(1 + n_F(\tau
,\eta_j^u(\tau ))\Big)}{2\, \sqrt{\gamma_F(\tau ,\eta^u_j(\tau ))}}
\right)\, \frac{Q_i\,Q_j}{ \sqrt{\sum_u\,
(\eta^u_j(\tau ) - \eta_i^u(\tau))^2} }+\nonumber\\
&&\nonumber\\
&+&\sum_{i\neq j}\, \int d^3\sigma\, \left( \frac{1}{4\pi}\,
\frac{Q_i\,Q_j}{ \sqrt{\sum_u\, (\sigma^u - \eta_i^u(\tau))^2}}
\right) \, \Big[ \frac{\h_{F\,rs}(\tau ,\sigma^u)\, \Big(1 +
n_F(\tau ,\sigma^u)\Big)}{2\, \sqrt{\gamma_F(\tau ,\sigma^u)}}\,
\overline{t}^{rs}(\sigma^u - \eta^u_j(\tau )) -\nonumber \\
 &+& \frac{1}{4\pi}\, \frac{\sigma^r - \eta^r_j(\tau
)}{ \Big(\sum_u\, (\sigma^u - \eta^u_j(\tau))^2\Big)^3/2 }\,
\frac{\partial}{\partial\sigma^s}\, \left(\frac{\h_{F\,rs}(\tau
,\sigma^u)\, \Big(1 + n_F(\tau ,\sigma^u)\Big)}{2\,
\sqrt{\gamma_F(\tau ,\sigma^u)}}\right)\, \Big] -\nonumber\\
&&\nonumber\\
&-&\, \int d^3\sigma\, \left(\frac{1}{4\pi}\, \sum_j\, \frac{Q_j}{
\sqrt{\sum_u\, (\sigma^u - \eta_j^u(\tau))^2} }\right) \,
 \frac{\partial}{\partial\sigma^s}\, \left[ \frac{\h_{F\,rs}\,
  \Big(1 + n_F\Big)}{ \sqrt{\gamma_F}}\, \pi^r_\perp + n_F^r\,
F_{rs}\right](\tau ,\sigma^u),\nonumber \\
 &&{}
 \label{4.46}
 \eea

\noindent which can be checked to be explicitly symmetric in the
exchange of ${\vec \eta}_i$ with ${\vec \eta}_j$.

\bigskip

Finally the Hamilton equations for the transverse electro-magnetic
fields $A_{\perp r}$ and $\pi^r_{\perp}$ in the radiation gauge
implied by the Dirac Hamiltonian (\ref{4.35}) are

\begin{eqnarray*}
 \partial_{\tau}\, A_{\perp\, r}(\tau ,\vec \sigma ) &\cir& \{
 A_{\perp\, r}(\tau ,\vec \sigma), H_{DF} \} =\nonumber \\
 &=& \delta_{rn}\, P^{nu}_{\perp}(\vec \sigma)\, \Big[
 {{(1 + n)\, {}^3e_{(a)u}\, {}^3e_{(a)v}}\over {{}^3e}}\, \Big(\pi^v_{\perp} -
 \delta^{vm}\, \sum_i\, Q_i\, \eta_i\, {{\partial\,
 c(\vec \sigma, {\vec \eta}_i(\tau))}\over {\partial\, \sigma^m}}
 \Big) +\nonumber \\
 &+&  {\bar n}_{(a)}\,  {}^3e^v_{(a)}\, F_{vu}
 \Big](\tau ,\vec \sigma),\nonumber \\
 &&{}\nonumber \\
 \partial_{\tau}\, \pi^r_{\perp}(\tau ,\vec \sigma) &\cir& \{
 \pi^r_{\perp}(\tau ,\vec \sigma), H_{DF} \} =\nonumber \\
 &=& P^{rn}_{\perp}(\vec \sigma)\, \delta_{nm}\, \Big(
 \sum_i\, \eta_i\, Q_i\, \delta^3(\vec \sigma, {\vec
 \eta}_i(\tau))\, {}^3e^m_{(a)}(\tau ,{\vec \eta}_i(\tau))
 \nonumber \\
 &&\Big[{{(1 + n)\, {}^3e^s_{(a)}\, {\check \kappa}_{is}(\tau)}\over
 {\sqrt{m_i^2\, c^2 +
 {}^3e^r_{(a)}\, \Big({\check \kappa}_{ir}(\tau ) - {{Q_i}\over c}\,
 A_{\perp\, r}\Big)\, {}^3e^s_{(a)}\, \Big({\check \kappa}_{is}(\tau ) -
 {{Q_i}\over c}\, A_{\perp \, s}\Big)}}} -\nonumber \\
  &-& {\bar n}_{(a)}\Big](\tau ,{\vec \eta}_i(\tau)) +
 \end{eqnarray*}

  \bea
 &+&\Big[(1 + n)\, \Big({}^3e\, {}^3e^s_{(a)}\, {}^3e^v_{(b)}\,
 ({}^3e^r_{(a)}\, {}^3e^m_{(b)} - {}^3e^m_{(a)}\, {}^3e^r_{(b)})\,
 \partial_r\, F_{sv} +\nonumber \\
 &+& \partial_r\, \Big[{}^3e\,  {}^3e^s_{(a)}\, {}^3e^v_{(b)}\,
 ({}^3e^r_{(a)}\, {}^3e^m_{(b)} - {}^3e^m_{(a)}\, {}^3e^r_{(b)})\,\Big]\,
 F_{sv} \Big) +\nonumber \\
 &+& \partial_r\, n\, {}^3e\, {}^3e^s_{(a)}\, {}^3e^v_{(b)}\, ({}^3e^r_{(a)}\,
 {}^3e^m_{(b)} - {}^3e^m_{(a)}\, {}^3e^r_{(b)})\,
 F_{sv} +\nonumber \\
 &+& {\bar n}_{(a)}\, \Big({}^3e^r_{(a)}\, \partial_r\,
 \pi_{\perp}^m + \partial_r\, {}^3e^r_{(a)}\, \pi^m_{\perp} -
 \partial_r\, {}^3e^m_{(a)}\, \pi^r_{\perp} +\nonumber \\
 &+& (\partial_r\, {}^3e^r_{(a)}\, \delta^{mt} - \partial_r\,
 {}^3e^m_{(a)}\, \delta^{rt})\, \sum_i\, \eta_i\, Q_i\,
  {{\partial\, c(\vec \sigma, {\vec \eta}_i(\tau)))}\over {\partial\, \sigma^t}}
 +\nonumber \\
 &+& ({}^3e^r_{(a)}\, \delta^{mt} - {}^3e^m_{(a)}\, \delta^{rt})\,
 \sum_i\, \eta_i\, Q_i\,  {{\partial^2\, c(\vec \sigma, {\vec \eta}_i(\tau)))}
 \over {\partial\, \sigma^t\, \partial\, \sigma^r}} \Big) +\nonumber \\
 &+& \partial_r\, {\bar n}_{(a)}\, ({}^3e^r_{(a)}\, \delta^{mt} -
 {}^3e^m_{(a)}\, \delta^{rt})\, \sum_i\, \eta_i\, Q_i\,
  {{\partial\, c(\vec \sigma, {\vec \eta}_i(\tau)))}\over {\partial\, \sigma^t}}
 \Big](\tau ,\vec \sigma)\, \Big).
 \label{x}
 \eea

Here $c(\sigma^u, \sigma^{{'}\, u}) = {1\over {4\pi\, \sqrt{\sum_u\,
(\sigma^u - \sigma^{{'}\, u})^2}}}$ and, following the general
relativity notation of Ref.\cite{12}, the metric has been expressed
in terms of triads ${}^3e^r_{(a)}$ and cotriads ${}^3e_{(a)r}$ on
$\Sigma_{\tau}$ as in Eq.(\ref{2.10}): $h_{F\, rs} = \sum_a\,
{}^3e_{(a)r}\, {}^3e_{(a)s}$, $h_F^{rs} = \sum_a\, {}^3e^r_{(a)}\,
{}^3e^s_{(a)}$, $\gamma_F = {}^3e$. The shift functions of
Eq.(\ref{2.4}) are replaced by ${\bar n}_{(a)} = n^r\,
{}^3e_{(a)r}$.

\subsection{On the Non-Relativistic Limit}

Let us consider the nearly non-relativistic limit of the embedding
(\ref{2.10}) given in Eqs.(\ref{2.16}). It can be done either before
or after the choice of the radiation gauge.\medskip

\medskip

Since we have $h_{rs} = \delta_{rs} + O(c^{-2})$, we can use the
vector notation of the inertial frames for the 3-vectors: $\vec V =
\{ V_r = {\tilde V}^r \}$ (since $g_{\tau\tau} = \sgn\, \Big(1 -
\sum_r\, (n^r_F)^2\Big) + O(c^{-2}) = \sgn + O(c^{-2})$, we still
have $V^r = g^{rA}\, V_A \not= {\tilde V}^r$ for 4-vectors $V_A$).
Therefore we have $\check{\vec{\kappa}}_i = \{\check{\kappa}_i^r\}
\byd \{\check{\kappa}_{ir}\}$, $\vec E = \{ E_r = {\tilde E}^r\} +
O(c^{-2})$, $\vec B = \{ B_r = {\tilde B}^r\} + O(c^{-2})$, but
$\vec{A}_\perp = \{ A_{\perp r} = {\tilde A}^r_{\perp} \not=
A^r_{\perp}\} + O(c^{-2})$.

\hfill

In these rigidly-rotating non-inertial frames the equations of
motion (\ref{4.9}) takes the form (the Newtonian functions are
$\tilde f(t) = f(\tau = c\, t)$; $\vec \Omega (c\, t)$ has the
components $\tilde \Omega (c\, t)$ defined after Eq. (\ref{2.15}))

 \bea
m_i\, \frac{d}{dt}\, \Big[ {{d\, {\vec \eta}_i(c\, t)}\over {dt}}
 &+& \vec{v}(c\, t ) + \vec{\Omega}(c\, t )\times \vec{\eta}_i(c\, t )
\Big]\,\, \on\,\,  Q_i\, \left[\vec{E} + {1\over c}\, {{d\, {\vec
\eta}_i(c\, t)}\over {dt}} \times \vec{B}\,\right](c\, t ,{\vec
\eta}_i(c\, t )) +\nonumber \\
 &+& \vec{\cal F}_i(c\, t ),\nonumber \\
 &&{}\nonumber \\
 \vec{\cal F}_i(c\, t ) &=& - m_i\, \vec{\Omega}(c\, t ) \times
\left[ {{d\, {\vec \eta}_i(c\, t)}\over {dt}} + \vec{v}(c\, t ) +
\vec{\Omega}(c\, t ) \times \vec{\eta}_i(c\, t )\right].
 \label{4.47}
 \eea
 \medskip

As a consequence the final form of the equations of motion of the
particles is

 \bea
 m_i\, {{d^2\, \vec{\eta}_i(c\, t )}\over {dt^2}} &\on& + Q_i\, \left[\vec{E} +
 {{d\, \vec{\eta}_i(c\, t )}\over {dt}} \times \vec{B}\,\right](c\, t ,{\vec \eta}_i(c\, t ))
 + \vec{\cal F}_i^{(in)}(c\, t ),\nonumber \\
 &&{}\nonumber \\
 \vec{\cal F}_i^{(in)}(c\, t ) &=& \vec{\cal F}_i(c\, t ) +
m_i\, \frac{d}{dt}\,
\Big(\vec{v}(c\, t) + \vec{\Omega}(c\, t ) \times \vec{\eta}_i(c\, t )\Big) =\nonumber \\
 &=&- m_i\, \Big[\vec{\Omega}(c\, t ) \times \Big(\vec{\Omega}(c\, t )
\times \vec{\eta}_i(c\, t )\Big) + 2\,  \vec{\Omega}(c\, t ) \times
{{d\, \vec{\eta}_i(c\, t )}\over {dt}} + {{d\, \vec{\Omega}(c\, t
)}\over {dt}} \times \vec{\eta}_i(c\, t ) +\nonumber \\
 &+& {{d\, \vec{v}(c\, t )}\over {dt}} +
\vec{\Omega}(c\, t ) \times \vec{v}(c\, t )\Big],
 \label{4.48}
 \eea

\noindent ${\vec {\cal F}}_i^{(in)}(\tau )$ is the sum of all the
inertial forces (centrifugal, Coriolis, Jacobi, the two pieces of
the linear acceleration) present in Newtonian rigid non-inertial
frames.

\hfill

The equations of motion (\ref{4.36}), (\ref{4.29}) of the particles
in the radiation gauge become

 \bea
  m_i\, {{d^2\, \vec{\eta}_i(c\, t )}\over {dt^2}} &\on& - {{\partial}\over {\partial\,
  {\vec \eta}_i}}\, {\cal W}(\vec{\eta_1}(\tau ),...,\vec{\eta}_N(\tau ))
  + Q_i\, \left[- {1\over c}\, \frac{\partial \vec{A}_\perp}{\partial\, t}
  + {1\over c}\, {{d\, \vec{\eta}_i(c\, t )}\over {dt}} \times \vec{B}\,\right](c\, t ,{\vec
\eta}_i(c\, t ))  +\nonumber \\
  &+& \vec{\cal F}_i^{(in)}(c\, t ),
 \label{4.49}
 \eea

\noindent where the non-inertial Coulomb potential takes the form
($\tau = ct$)\footnote{ In this case from Eq.(\ref{4.30}) we get
\[
a_\tau(\tau,\vec{\sigma}) = -\left[ \sum_{k}\,
\frac{Q_k}{4\pi\mid\vec{\sigma} - \vec{\eta}_k\mid} -
{{\vec{v}}\over c} \cdot \vec{A}_\perp(\tau,\vec{\sigma}) -
{{\vec{\Omega}}\over c} \times\vec{\sigma}\cdot
\vec{A}_\perp(\tau,\vec{\sigma}) \right],
\]
}

 \bea
&&{\cal W}(\vec{\eta_1}(\tau ),...,\vec{\eta}_N(\tau )) = \nonumber\\
&&\nonumber\\
 &=& + \sum_{i>j}\, \frac{Q_i\, Q_j}{4\pi\, \mid\vec{\eta}_i(\tau )
 - \vec{\eta}_j(\tau )\mid} - \sum_i\, {{Q_i}\over c}\, \left[
\vec{v}(\tau ) \cdot \vec{A}_\perp(\tau ,\vec{\eta_i}(\tau )) +
\vec{\Omega}(\tau ) \times\vec{\eta_i}(\tau )\cdot
\vec{A}_\perp(\tau ,\vec{\eta_i}(\tau
))\right].\nonumber \\
 &&{}
 \label{4.50}
 \eea

\hfill

Finally the Hamiltonian (\ref{4.35}) becomes

\bea
 \check{H}_R&=&\sum_i\, \sqrt{m_i^2\, c^2 + \Big(\check{\vec{\kappa}}_i(\tau )
 - {{Q_i}\over c}\, \vec{A}_{\perp}(\tau ,\vec{\eta}_i(\tau ))\Big)^2}
+ \sum_{i>j}\, \frac{Q_i\, Q_j}{4\pi\, c\, \mid\vec{\eta}_i(\tau ) -
\vec{\eta}_j(\tau )\mid} +\nonumber\\
 &&\nonumber\\
&+& {1\over {2 c}}\, \int d^3\sigma\, \Big( \vec{\pi}_\perp^2(\tau
,\vec{\sigma}) - \vec{A}_\perp(\tau ,\vec{\sigma}) \cdot
\left[\Delta\, \vec{A}_\perp(\tau ,\vec{\sigma})\right]\Big) +\nonumber\\
&&\nonumber\\
&-&{{\vec{v}(\tau )}\over c} \cdot
\left[\sum_i\,\check{\vec{\kappa}}_i(\tau ) - {1\over c}\, \int
d^3\sigma\, \vec{\pi}_\perp(\tau ,\vec{\sigma}) \times (\vec
\partial \times \vec{A}_\perp(\tau ,\vec{\sigma}))\right] +\nonumber\\
&&\nonumber\\
&-&{{\vec{\Omega}(\tau )}\over c} \cdot \left[\sum_i\,
\vec{\eta}_i(\tau ) \times \check{\vec{\kappa}}_i(\tau ) + \vec{\cal
J}(\tau )\right],
\nonumber \\
 &&{}\nonumber \\
  \vec{\cal J}(\tau ) &=& - {1\over c}\,  \int
d^3\sigma\, \sum_r\, \pi^r_\perp(\tau ,\vec{\sigma})\, \Big(
\vec{\sigma} \times \vec \partial\Big)\, {\tilde A}^r_\perp(\tau
,\vec{\sigma}) - \vec{A}_\perp(\tau ,\vec{\sigma}) \times
\vec{\pi}_\perp(\tau ,\vec{\sigma}),
 \label{4.51}
 \eea

\noindent where ${\vec {\cal J}}(\tau )$ is the total angular
momentum of the electro-magnetic field.
\medskip

It can be checked that this Hamiltonian generates the previous limit
of the equations of motion of the particles. In particular the first
set of Hamilton equations is

 \beq
 {1\over c}\, {{d\, \vec{\eta}_i(\tau )}\over {dt}} = \frac{\check{\vec{\kappa}}_i(\tau ) -
{{Q_i}\over c}\, \vec{A}_{\perp}(\tau ,\vec{\eta}_i(\tau
))}{\sqrt{m_i^2\, c^2 + \Big(\check{\vec{\kappa}}_i(\tau ) -
{{Q_i}\over c}\, \vec{A}_{\perp}(\tau ,\vec{\eta}_i(\tau ))\Big)^2}}
- {{\vec{v}(\tau )}\over c} -{{\vec{\Omega}(\tau )}\over c} \times
\vec{\eta}_i(\tau ).
 \label{4.52}
 \eeq

\vfill\eject

\section{The Instant Form of
Dynamics in Non-Inertial Frames and in the Inertial and Non-Inertial
Rest Frames.}

In this Section we study the problem of the separation of the
relativistic non-covariant canonical 4-center of mass of an isolated
system from the relative variables describing its dynamics. We first
recall how this problem is solved in the inertial rest-frame instant
form of dynamics \cite{1,3,4,5,8}. As said in the Introduction the
isolated system is described as a decoupled pseudo-particle
(described by the non-covariant canonical variables $\vec z$ and
$\vec h$) carrying a pole-dipole structure given by its invariant
mass and its rest spin. On each instantaneous Wigner 3-space,
centered on the inertial observer corresponding to the Fokker-Pryce
4-center of inertia, these quantities are functions of the relative
variables of the isolated system after the elimination of the
internal 3-center of mass. The double counting of the center of mass
is avoided by the presence of three pairs of second class
constraints: the rest-frame conditions, i.e. the vanishing of the
internal 3-momentum, and the vanishing of the internal
boosts.\medskip

In Subsection A we will show how to get these conditions in the
inertial rest frames starting from the embeddings (\ref{1.1}), from
the determination (\ref{3.8}) of their conjugate momenta and from
the Poincare' generators (\ref{3.17}).\medskip

In Subsection B we will extend this construction to determine the
three pairs of second class constraints in an arbitrary admissible
non-inertial frame described by the embeddings (\ref{2.1}) and
centered on an arbitrary time-like observer. Again the isolated
system can be visualized as a pole-dipole carried by the external
decoupled center of mass.

\medskip

In Subsection C we will define the special family of the {\it
non-inertial rest-frames}, centered on the inertial Fokker-Pryce
4-center of inertia, and the associated {\it non-inertial rest-frame
instant form}. They are relevant because they are the only global
non-inertial frames allowed by the equivalence principle (forbidding
the existence of global inertial frames) in canonical metric and
tetrad gravity in globally hyperbolic, asymptotically flat
(asymptotically Minkowskian) space-times without super-translations,
so to have the asymptotic ADM Poincare' group \cite{11}. Also in
this case we identify the three pairs of second class constraints
eliminating the internal 3-center of mass, visualizing the isolated
system as a pole-dipole and allowing to describe the dynamics on the
instantaneous (non-Euclidean) 3-spaces only in terms of relative
variables. Then in Subsection D we show how the Hamiltonian
description of Section IV has to be modified if we take this point
of view in the description of the isolated system. We also delineate
the analogue of this procedure for the general case of Subsection B.

\subsection{The Inertial Rest-Frame Instant Form}

As said in the Introduction every configuration of an isolated
system, with associated finite Poincare' generators $P^{\mu}$,
$J^{\mu\nu}$, identifies a unique inertial frame in an intrinsic
way: the {\it inertial rest frame} whose Euclidean instantaneous
3-spaces (the Wigner 3-spaces) are orthogonal to the conserved
4-momentum $P^{\mu}$ of the configuration. The embedding
corresponding to the inertial rest frame, centered on the
Fokker-Pryce center of inertia, is given in Eq.(\ref{1.1})
\bigskip

The generators of the external realization of the Poincare' algebra
are (following footnote 10 we use only $\epsilon_{ijk}$; $M$ and
${\vec {\bar S}}$ have vanishing Poisson brackets with $\vec z$ and
$\vec h$ and we have $\{ {\bar S}^i, {\bar S}^j \} = \delta^{im}\,
\delta^{jn}\, \epsilon_{mnk}\, {\bar S}^k$)

\bea
 P^o &=& Mc\, h^{\mu},\qquad h^{\mu} = \Big(\sqrt{1 + {\vec h}^2}; \vec
 h\Big),\nonumber \\
 J^i &=& \delta^{im}\, \epsilon_{mnk}\, \Big(z^n\, h^k + {\bar S}^k\Big),\qquad K^i = -
 \sqrt{1 + {\vec h}^2}\, z^i + {{ \delta^{in}\, \epsilon_{njk}\, {\bar
 S}^j\, h^k }\over {1 + \sqrt{1 + {\vec h}^2}}},\nonumber \\
 &&{}
 \label{5.1}
 \eea
 \bigskip

\noindent while those of the unfaithful internal realization of the
Poincare' algebra determined by the energy-momentum tensor (in
inertial frames Eqs.(\ref{3.8}) imply $T_{\perp\perp} =
T^{\tau\tau}$ and $T_{\perp r} = \delta_{rs}\, T^{\tau s}$) are

\bea
 Mc &=&  \int d^3\sigma\, T^{\tau\tau}(\tau ,\sigma^u),\qquad
  {\bar S}^r = {1\over 2}\, \delta^{rs}\, \epsilon_{suv}\, \int d^3\sigma\,
  \sigma^u\, T^{\tau v}(\tau ,\sigma^u), \nonumber \\
 &&{}\nonumber \\
 {\cal P}^r &=& \int d^3\sigma\, T^{\tau r}(\tau ,\sigma^u)
 \approx 0,\qquad {\cal K}^r = - \int d^3\sigma\, \sigma^r\,
 T^{\tau\tau}(\tau ,\sigma^u) \approx 0.
 \label{5.2}
 \eea

 \medskip

The constraints ${\vec {\cal P}} \approx 0$ are the rest-frame
conditions identifying the inertial rest frame. Having chosen the
Fokker-Pryce center of inertia as origin of the 3-coordinates, the
({\it interaction-dependent}) constraints ${\vec {\cal K}} \approx
0$ are their gauge fixing: they eliminate the internal 3-center of
mass so not to have a double counting (external, internal).
Therefore the isolated system is described by the external
non-covariant 3-center of mass $\vec z$, $\vec h$, and by an {\it
internal space} of Wigner-covariant relative variables ($M$ and
${\vec {\bar S}}$ depend only upon them).\medskip

Eqs. (\ref{5.1}) and (\ref{5.2}) are obtained in the following way.
If we put the embedding (\ref{1.1}), namely $z^{\mu}(\tau ,\sigma^u)
= Y^{\mu}(0) + h^{\mu}\, \tau + \epsilon^{\mu}_r(\vec h)\, \sigma^r
= Y^{\mu}(0) + \epsilon^{\mu}_A(\vec h)\, \sigma^A$,  in
Eq.(\ref{3.8}), we get $\rho^{\mu}(\tau ,\sigma^u) \approx h^{\mu}\,
T^{\tau\tau}(\tau ,\sigma^u) + \epsilon^{\mu}_r(\vec h)\, T^{\tau
r}(\tau ,\sigma^u) = \epsilon^{\mu}_A(\vec h)\, T^{A\tau}(\tau
,\sigma^u)$. Then the first of Eqs.(\ref{3.17}) implies $P^{\mu} =
Mc\, h^{\mu}$ if $Mc = \int d^3\sigma\, T^{\tau\tau}(\tau
,\sigma^u)$ and ${\cal P}^r = \int d^3\sigma\, T^{\tau r}(\tau
,\sigma^u) \approx 0$.\medskip

The second of Eqs.(\ref{3.17}) gives $J^{\mu\nu} = \Big(Y^{\mu}(o)\,
\epsilon^{\nu}_A(\vec h) - Y^{\nu}(0)\, \epsilon^{\mu}_A(\vec
h)\Big)\, \int d^3\sigma\, T^{A\tau}(\tau ,\sigma^u) +
\epsilon^{\mu}_A(\vec h)\, \epsilon^{\nu}_B(\vec h)\, S^{AB}$ with
$S^{AB} = \int d^3\sigma\, \Big(\sigma^A\, T^{B\tau} - \sigma^B\,
T^{A\tau}\Big)(\tau ,\sigma^u)$. By using ${\cal P}^r \approx 0$ we
get $J^{\mu\nu} \approx Mc\, \Big(Y^{\mu}(0)\, h^{\nu} -
Y^{\nu}(0)\, h^{\mu}\Big) + \epsilon^{\mu}_A(\vec h)\,
\epsilon^{\nu}_B(\vec h)\, S^{AB}$ with $S^{\tau r} \approx \int
d^3\sigma\, \sigma^r\, T^{\tau\tau}(\tau ,\sigma^u)$ and $S^{rs} =
\int d^3\sigma\, \Big(\sigma^r\, T^{s\tau} - \sigma^s\,
T^{r\tau}\Big)(\tau ,\sigma^u)$. Then, by using the expression of
the Fokker-Pryce 4-center of inertia given in Eq.(2.20) of
Ref.\cite{8}, i.e. $Y^{\mu}(\tau ) = Y^{\mu}(0) + h^{\mu}\, \tau$
with $Y^{\mu}(0) = \Big(\sqrt{1 + {\vec h}^2}\, {{\vec h \cdot \vec
z}\over {Mc}};\,\, {{\vec z}\over {Mc}} + {{\vec h \cdot \vec
z}\over {Mc}}\, \vec h + {{{\vec V}_S}\over {Mc\, (1 + \sqrt{1 +
{\vec h}^2})}}\Big)$, as a function of $\tau$, $\vec z$, $\vec h$,
$Mc$ and of  ${\vec {\bar S}}$, and the expression of
$\epsilon^{\mu}_A(\vec h)$ given after Eq.(\ref{1.1}), we get:

a) $J^{ij} = z^i\, h^j - z^j\, h^i + \delta^{ir}\, \delta^{js}\,
\epsilon_{rsk}\, \int d^3\sigma\, \sigma^r\, T^{s\tau}(\tau
,\sigma^u)$, which coincides with Eq.(\ref{5.1}) if ${\vec {\bar
S}}$ has the expression given in Eq.(\ref{5.2});

b) $J^{oi} = - \sqrt{1 + {\vec h}^2}\, z^i + \delta^{in}\,
\epsilon_{njk}\, {\bar S}^j\, h^k + \epsilon^o_{\tau}(\vec h)\,
\epsilon^i_r(\vec h)\, S^{\tau r}$, which coincides with
Eq.(\ref{5.1}) if ${\cal K}^r = - S^{\tau r} \approx 0$ as in
Eqs.(\ref{5.2}).

Therefore we have $S^{AB} \approx (\delta^A_r\, \delta^B_{\tau} -
\delta^A_{\tau}\, \delta^B_r)\, {\cal K}^r + \delta^{Ar}\,
\delta^{Bs}\, \epsilon_{rsk}\, {\bar S}^k \approx \delta^{Ar}\,
\delta^{Bs}\, \epsilon_{rsk}\, {\bar S}^k $.

\bigskip

As shown in Ref.\cite{8}, the restriction of the embedding
$z^{\mu}(\tau ,\sigma^u )$ to the Wigner 3-spaces (\ref{1.1})
implies the replacement of the Dirac Hamiltonian (\ref{3.13}) with
the new one

\beq
 H_{D\, W} = Mc  + \int
 d^3\sigma\, \Big(\mu\, \pi^{\tau} - A_{\tau}\, \Gamma\Big)(\tau
 ,\sigma^u).
 \label{5.3}
 \eeq

Therefore, consistently with Eqs.(\ref{5.2}),  the effective
Hamiltonian is the invariant mass of the isolated system, whose
conserved rest spin is ${\vec {\bar S}}$. As already said, the three
pairs of second class constraints ${\vec {\cal P}} \approx 0$,
${\vec {\cal K}} \approx 0$, eliminate the internal 3-center of
mass.\medskip

\bigskip

As shown in Refs.\cite{8,9}, in the rest-frame instant form it is
possible to restrict the description of N charged positive-energy
particles plus the electro-magnetic field to the radiation gauge
(see next Section for the non-inertial case), where all the
electro-magnetic quantities are transverse. The mutual Coulomb
interaction among the particles appears in this gauge, the
Hamiltonian (\ref{5.3}) reduces to $Mc$ and we get the following
form of the internal Poincare' generators (\ref{5.2}) \footnote{In
this equation we use the notation ${\vec \kappa}_i(\tau )$ for the
Coulomb-dressed momenta ${\check {\vec \kappa}}_i(\tau ) = {\vec
\kappa}_i(\tau ) - {{\partial\, \eta_{em}(\tau ,{\vec \eta}_i(\tau
))}\over {\partial\, {\vec \eta}_i}}$ belonging to the
Shanmugadhasan canonical basis defined in Eqs.(\ref{4.26}).}

\begin{eqnarray*}
\mathcal{E}_{(int)} &=& \mathcal{P}^{\tau }_{(int)}\, c = M\, c^2 =
c\, \int d^3\sigma\, T^{\tau\tau}(\tau ,\vec \sigma ) =  \nonumber \\
&=& c\, \sum_{i=1}^{N}\, \sqrt{ m_{i}^{2}\, c^2 + \Big({\vec{
\kappa}} _i(\tau ) - {\frac{{Q_i}}{c}}\, {\vec{A }}_{\perp }(\tau
,\vec{\eta} _i(\tau ))\Big)^2} +  \nonumber \\
&+&\sum_{i\neq j}\, \frac{Q_{i}\, Q_{j}}{4\pi\, \mid
\vec{\eta}_{i}(\tau ) - \vec{\eta} _{j}(\tau )\mid } +
{\frac{1}{2}}\, \int d^{3}\sigma \, [{\vec{ \pi }}_{\perp }^{2} +
{\vec{B}}^{2}](\tau ,\vec{\sigma}) =
 \end{eqnarray*}

 \begin{eqnarray*}
 &=&c\, \sum_{i=1}^N\, \Big(\sqrt{m^2_i\, c^2 + {\vec
\kappa}^2_i(\tau )} - { \ \frac{{Q_i}}{c}}\, {\frac{{{\vec
\kappa}_i(\tau ) \cdot {\vec A} _{\perp}(\tau , {\vec \eta}_i(\tau
))}}{\sqrt{m^2_i\, c^2 + {\vec \kappa}
^2_i(\tau )}}} \Big) +  \nonumber \\
&+&\sum_{i\neq j}\, \frac{Q_{i}\, Q_{j}}{4\pi\, \mid
\vec{\eta}_{i}(\tau ) - \vec{\eta} _{j}(\tau )\mid } +
{\frac{1}{2}}\, \int d^{3}\sigma \, [{\vec{ \pi }}_{\perp }^{2} +
{\vec{B}}^{2}](\tau ,\vec{\sigma}),
 \end{eqnarray*}

\begin{eqnarray*}
 \mathcal{\vec{P}}_{(int)} &=& \int d^3\sigma\, T^{r\tau}(\tau ,\vec
\sigma ) = \sum_{i=1}^N\, {\vec{\kappa}}_i(\tau ) + {\frac{1}{c}}\,
\int d^{3}\sigma\, \lbrack {\vec{\pi}}_{\perp } \times
{\vec{B}}](\tau ,\vec{\sigma}) \approx 0,  \nonumber \\
&&{}  \nonumber \\
\mathcal{J}_{(int)}^r &=& {\bar S}^r = {\frac{1}{2}}\, \delta^{rs}\,
\epsilon_{suv}\, \int d^3\sigma\, \sigma^u\, T^{v\tau}(\tau
,\vec \sigma ) =  \nonumber \\
&=&\sum_{i=1}^{N}\,\Big(\vec{\eta}_{i}(\tau )\times
{\vec{\kappa}}_{i}(\tau ) \Big)^{r} + {\frac{1}{c}}\, \int
d^{3}\sigma (\vec{\sigma}\times \,\Big([{ \vec{\pi}}_{\perp }{\
\times }{\vec{B}}]\Big)^{r}(\tau ,\vec{\sigma}),
 \end{eqnarray*}

\bea
 \mathcal{K}_{(int)}^{r} &=&{\bar{S}}^{\tau r} = -
{\bar{S}}^{r\tau } = - \int d^3\sigma\, \sigma^r\, T^{\tau\tau}(\tau
,\vec \sigma ) =\nonumber \\
 &=& - \sum_{i=1}^{N}\, \eta^r_{i}(\tau )\, \Big( \sqrt{m^2_i\, c^2 +
{\vec \kappa}^2_i(\tau )} - {\frac{{Q_i}}{c}}\, {\frac{{{\vec
\kappa}_i(\tau ) \cdot {\vec A}_{\perp}(\tau , {\vec \eta}_i(\tau
))}}{\sqrt{m^2_i\, c^2 + {\
\vec \kappa}^2_i(\tau )}}} \Big) +  \nonumber \\
 &+& {\frac{1}{c}}\, \sum_{i=1}^{N}\, \sum_{j\not=i}^{1..N}\,
Q_{i}\, Q_{j}\, \Big[\int d^3\sigma\, {1\over {4\pi\, |\vec \sigma -
{\vec \eta}_j(\tau )|}}\, {{\partial}\over {\partial\, \sigma^r}}\,
{1\over {4\pi\, |\vec \sigma - {\vec \eta}_i(\tau )|}} +\nonumber \\
 &+&{{\eta^r_j(\tau )}\over {4\pi\, |{\vec \eta}_i(\tau ) - {\vec
\eta}_j(\tau )|}} \Big] -  \nonumber \\
 &-& {1\over c}\, \sum_{i=1}^N\, Q_{i}\, \int d^{3}\sigma\, {{{\pi}_{\perp }^{r}(\tau
,\vec{\sigma})} \over {4\pi\, |\vec{\sigma} - {\
\vec{\eta}}_{i}(\tau )|}} - {\frac{1}{2c}}\, \int d^{3}\sigma\,
\sigma ^{r}\, ({{\vec{\pi}}}_{\perp }^{2} + {{\vec{B} }} ^{2})(\tau
,\vec{\sigma}) \approx 0.
  \label{5.4}
\end{eqnarray}

\bigskip

Note that, as required by the Poincare' algebra in an instant form
of dynamics, there are interaction terms both in the internal energy
and in the internal Lorentz boosts, but not in the 3-momentum and in
the angular momentum.

\bigskip

As shown in Ref.\cite{8}, we can reconstruct the original gauge
potential ${\tilde A}_{\mu}(x)$ in the radiation gauge. It has the
following form

\beq
 {\tilde A}^{\mu}(Y^{\alpha}(\tau ) + \epsilon^{\alpha}_r(\vec h)\, \sigma^r)
= {{P^{\mu}}\over {Mc}}\, \sum_i\, {{Q_i}\over {|\vec \sigma - {\vec
\eta}_i(\tau )|}} - \epsilon^{\mu}_r(\vec h)\, A^r_{\perp}(\tau
,\sigma^u).
 \label{5.5}
 \eeq

\subsection{Amissible Non-Inertial Frames}

Let us now see whether in an arbitrary admissible non-inertial
frame, centered on an arbitrary non-inertial observer and described
by the embeddings (\ref{2.1}), we can arrive at the same picture of
an isolated system as a decoupled external canonical non-covariant
center of mass $\vec z$, $\vec h$, carrying a pole-dipole structure,
with the external Poincare' generators given by expressions like
Eqs.(\ref{5.1}) and with the dynamics described by suitable relative
variables after an appropriate elimination of the internal 3-center
of mass inside the instantaneous 3-spaces. If this is possible,
there will be a new expression for the internal invariant mass $M$,
a new effective spin ${\vec {\tilde S}}$ (supposed to satisfy the
Poisson brackets of an angular momentum and such that $J^i =
\delta^{im}\, \epsilon_{mnk}\, \Big(z^n\, h^k + {\tilde S}^k\Big)$)
and a new form of the three pairs of second class constraints
replacing the expressions given in Eqs.(\ref{5.2}) for the case of
the inertial rest frame centered on the Fokker-Pryce center of
inertia.
\bigskip

Now the embeddings (\ref{2.1}) imply the form (\ref{3.8}) for the
conjugate momenta $\rho^{\mu}(\tau , \sigma^u)$. Therefore  we must
evaluate the Poincare' generators (\ref{3.17}) by using
Eqs.(\ref{2.1}) and (\ref{3.8}). By equating the resulting
expressions with Eqs.(\ref{5.1}) we will find the new expression of
the invariant mass, of the effective spin and of the second class
constraints.
\medskip

Since the embedding (\ref{2.1}) depend on the asymptotic tetrads
$\epsilon^{\mu}_A$, we must express them in terms of the tetrads
$\epsilon^{\mu}_A(\vec h)$ determined by $P^{\mu}$ (whose expression
is given after Eq.(\ref{1.1})): $\epsilon^{\mu}_A =
\Lambda_A{}^B(\vec h)\, \epsilon^{\mu}_B(\vec h)$ with $\Lambda
(\vec h)$ a Lorentz matrix.\medskip

Then, by using Eqs.(\ref{2.1}), (\ref{3.8}) and (\ref{5.1})  the
first of Eqs.(\ref{3.17}) becomes

\bea
 P^{\mu} &=& Mc\, h^{\mu} = Mc\, \epsilon^{\mu}_{\tau}(\vec h)
 \approx \epsilon^{\mu}_A\, {\hat {\cal
 P}}^A = {\hat {\cal P}}^A\, \Lambda_A{}^B(\vec h)\,
 \epsilon^{\mu}_B(\vec h) =\nonumber \\
 &=& {\hat {\cal P}}^A\, \Big[\Lambda_A{}^{\tau}(\vec h)\, h^{\mu} +
 \Lambda_A{}^r(\vec h)\, \epsilon^{\mu}_r(\vec h)\Big],\nonumber \\
  &&{}\nonumber \\
  {\hat {\cal P}}^A &=& \int d^3\sigma\, \sqrt{\gamma (\tau ,\sigma^u)}\,
  \Big[T_{\perp\perp}\, l^A - T_{\perp s}\, h^{sr}\, \partial_r\,
  F^A\Big](\tau ,\sigma^u),
 \label{5.6}
 \eea

\noindent with $l^A(\tau ,\sigma^u)$ given in Eq.(\ref{2.7}).
\medskip

Therefore the invariant mass $M$ and the three constraints ${\tilde
{\cal P}}^r \approx 0$ replacing the rest-frame conditions are

\beq
 Mc  \approx {\hat {\cal P}}^A\, \Lambda_A{}^{\tau}(\vec h),\qquad
 {\tilde {\cal P}}^r = {\hat {\cal P}}^A\, \Lambda_A{}^r(\vec h) \approx
 0,\qquad \Rightarrow\,\, {\hat {\cal P}}^A \approx Mc\,
 \Lambda_{\tau}{}^A(\vec h).
 \label{5.7}
 \eeq

\bigskip

If we define

\bea
  {\hat S}^{AB} &=& \int d^3\sigma\, \sqrt{\gamma (\tau ,\sigma^u)}\,
 \Big[\Big(f^A(\tau ) + F^A(\tau ,\sigma^u)\Big)\, \Big(T_{\perp\perp}\, l^B -
 T_{\perp s}\, h^{sr}\, \partial_r\, F^B\Big)(\tau ,\sigma^u) -\nonumber \\
 &-& \Big(f^B(\tau ) + F^B(\tau ,\sigma^u)\Big)\, \Big(T_{\perp\perp}\,
l^A - T_{\perp s}\, h^{sr}\, \partial_r\, F^A\Big)
 (\tau ,\sigma^u)\Big] =\nonumber \\
  &=& f^A(\tau )\, {\hat {\cal P}}^B  - f^B(\tau )\, {\hat {\cal P}}^A
  + {\hat {\cal S}}^{CD}\, \Lambda_C{}^A(\vec h)\, \Lambda_D{}^B(\vec h),\nonumber \\
 &&{}\nonumber \\
 {\hat {\cal S}}^{AB} &=& \int d^3\sigma\, \sqrt{\gamma (\tau ,\sigma^u)}\,
 \Big[F^C(\tau ,\sigma^u)\, \Big(T_{\perp\perp}\, l^D -
 T_{\perp s}\, h^{sr}\, \partial_r\, F^D\Big)(\tau ,\sigma^u) -\nonumber \\
 &-& F^D(\tau ,\sigma^u)\, \Big(T_{\perp\perp}\,
l^C - T_{\perp s}\, h^{sr}\, \partial_r\, F^C\Big) (\tau
,\sigma^u)\Big]\, \Lambda_C{}^A(\vec h)\, \Lambda_D{}^B(\vec h) =\nonumber \\
  &{\buildrel {def}\over =}&
 (\delta^A_r\, \delta^B_{\tau} - \delta^A_{\tau}\,
 \delta^B_r)\, {\hat {\cal K}}^r + \delta^{Ar}\, \delta^{Bs}\,
 \epsilon_{rsk}\, {\hat S}^k,
 \label{5.8}
 \eea

\noindent then, by using Eq.(\ref{5.7}), the second of
Eqs.(\ref{3.17}) becomes

\bea
 J^{\mu\nu} &\approx& (x^{\mu}_o\, \epsilon^{\nu}_A - x^{\nu}_o\,
 \epsilon^{\mu}_A)\, {\hat {\cal P}}^A + \epsilon^{\mu}_A\, \epsilon^{\nu}_B
 \, {\hat S}^{AB} =\nonumber \\
 &=& {\hat {\cal P}}^B\, \Lambda_B{}^A(\vec h)\, \Big(x^{\mu}_o\,
 \epsilon^{\nu}_A(\vec h) - x^{\nu}_o\, \epsilon^{\mu}_A(\vec h)\Big) +
  {\hat S}^{CD}\, \Lambda_C{}^A(\vec h)\,
 \Lambda_D{}^B(\vec h)\, \epsilon^{\mu}_A(\vec h)\,
 \epsilon^{\nu}_B(\vec h) = \nonumber \\
 &&{}\nonumber \\
 &=& {\hat {\cal P}}^A\, \Lambda_A{}^D(\vec h)\, \Big[\Big(x^{\mu}_o +
 f^B(\tau)\, \Lambda_B{}^C(\vec h)\, \epsilon^{\mu}_C(\vec h)\Big)\,
 \epsilon^{\nu}_D(\vec h) -\nonumber \\
 &-& \Big(x^{\nu}_o + f^B(\tau)\, \Lambda_B{}^C(\vec h)\, \epsilon^{\nu}_C(\vec h)
 \Big)\, \epsilon^{\mu}_D(\vec h)\Big] + \epsilon^{\mu}_A(\vec h)\,
 \epsilon^{\nu}_B(\vec h)\, {\hat {\cal S}}^{AB} \approx\nonumber \\
 &\approx& Mc\, \Big[\Big(x^{\mu}_o + f^B(\tau)\, \Lambda_B{}^C(\vec h)\,
 \epsilon^{\mu}_C(\vec h)\Big)\, h^{\nu} -\nonumber \\
 &-& \Big(x^{\nu}_o + f^B(\tau)\, \Lambda_B{}^C(\vec h)\, \epsilon^{\nu}_C(\vec h)
 \Big)\, h^{\mu}\Big] + \epsilon^{\mu}_A(\vec h)\,
 \epsilon^{\nu}_B(\vec h)\, {\hat {\cal S}}^{AB}.
 \label{5.9}
 \eea
\medskip

After some algebra Eqs.(\ref{5.1}) and (\ref{5.9}) imply

\bea
 J^{ij} &=& z^i\, h^j - z^j\, h^i + \delta^{iu}\, \delta^{jv}\,
 \epsilon_{uvk}\, {\tilde S}^k \approx\nonumber \\
 &&{}\nonumber \\
 &\approx& Mc\, \Big[\Big(x^i_o + f^B(\tau )\, \Lambda_B{}^C(\vec h)\, \epsilon^i_C(\vec h)
 + {1\over {Mc}}\, \Big[\epsilon^i_r(\vec h)\, {\tilde {\cal K}}^r +
 {{\delta^{im}\, \epsilon_{mnk}\, h^n\, {\hat S}^k}\over {1 + \sqrt{1 +
 {\vec h}^2}}}\Big]\Big)\, h^j -\nonumber \\
 &-& \Big(x^j_o + f^B(\tau )\, \Lambda_B{}^C(\vec h)\, \epsilon^j_C(\vec h)
 + {1\over {Mc}}\, \Big[\epsilon^j_r(\vec h)\, {\hat {\cal K}}^r +
 {{\delta^{jm}\, \epsilon_{mnk}\, h^n\, {\hat S}^k}\over {1 + \sqrt{1 +
 {\vec h}^2}}}\Big]\Big)\, h^i \Big] +\nonumber \\
 &+& \delta^{im}\, \delta^{jn}\, \epsilon_{mnk}\, {\hat S}^k
 =\nonumber \\
 &&{}\nonumber \\
 &{\buildrel {def}\over =}& X^i\, h^j - X^j\, h^i + \delta^{iu}\, \delta^{jv}\,
 \epsilon_{uvk}\, {\hat S}^k,
 \label{5.10}
 \eea

 \bea
   J^{oi} &=& - \sqrt{1 + {\vec h}^2}\, z^i - {{\delta^{im}\, \epsilon_{mnk}\,
 h^n\, {\tilde S}^k}\over {1 + \sqrt{1 + {\vec h}^2}}} \approx \nonumber \\
 &\approx& Mc\, \Big[x^o_o + f^B(\tau )\, \Lambda_B{}^C(\vec h)\, \epsilon^o_C(\vec h)
 + {{\sum_r\, h^r\, {\hat {\cal K}}^r}\over {Mc}} \Big]\, h^i -\nonumber \\
 &-& \sqrt{1 + {\vec h}^2}\, \Big[x^i_o + f^B(\tau )\, \Lambda_B{}^C(\vec h)\,
 \epsilon^i_C(\vec h) + {1\over {Mc}}\, \Big(\epsilon^i_r(\vec h)\, {\hat {\cal K}}^r +
 {{\delta^{im}\, \epsilon_{mnk}\, h^n\, {\hat S}^k}\over {1 + \sqrt{1 +
 {\vec h}^2}}}\Big)\Big] -\nonumber \\
 &-& {{\delta^{im}\, \epsilon_{mnk}\, h^n\, {\hat S}^k}\over {1 + \sqrt{1 + {\vec h}^2}}}
 =\nonumber \\
 &&{}\nonumber \\
  &{\buildrel {def}\over =}& X^o\, h^i - \sqrt{1 + {\vec h}^2}\, X^i
  - {{\delta^{im}\, \epsilon_{mnk}\, h^n\, {\hat S}^k}\over {1 + \sqrt{1 + {\vec
  h}^2}}},
 \label{5.11}
 \eea

\noindent where in the last lines we introduced the definition of
the quantities $X^o$ and $X^i$.\medskip

This implies the reformulation of the isolated system as an external
center of mass $\vec z$, $\vec h$, plus a pole-dipole structure $M$
and ${\vec {\tilde S}}$.
\medskip

If we solve Eq.(\ref{5.11}) in $\vec z$,  we get $\vec z = \vec X -
X^o\, {{\vec h}\over {\sqrt{1 + {\vec h}^2}}} - {{({\vec {\hat S}} -
{\vec {\tilde S}}) \times \vec h}\over {\sqrt{1 + {\vec h}^2}\, (1 +
\sqrt{1 + {\vec h}^2})}}$ (we use a vector notation). If we put this
expression in Eq.(\ref{5.10}), we get the following equation:
$[({\vec {\hat S}} - {\vec {\tilde S}}) \times \vec h] \times \vec h
= \sqrt{1 + {\vec h}^2}\, (1 + \sqrt{1 + {\vec h}^2})\, ({\vec {\hat
S}} - {\vec {\tilde S}})$. It implies $({\vec {\hat S}} - {\vec
{\tilde S}}) \cdot \vec h = 0$ and then we get

\beq
 {\tilde S}^r \approx {\hat S}^r,
 \label{5.12}
 \eeq

\noindent namely the effective spin ${\vec {\tilde S}}$ is given by
${\hat S}^{rs}$ of Eqs.(\ref{5.8}).\medskip

By using Eq.(\ref{5.12}) inside Eq.(\ref{5.11}) we get three
constraints, eliminating the internal 3-center of mass and allowing
to re-express the dynamics inside the instantaneous 3-spaces only in
terms of relative variables, which are

\bea
 Mc&& \Big[x^o_o + f^B(\tau )\, \Lambda_B{}^C(\vec h)\, \epsilon^o_C(\vec h)
 + {{\sum_r\, h^r\, {\hat {\cal K}}^r}\over {Mc}} \Big]\, h^i -
  \sqrt{1 + {\vec h}^2}\, \Big[x^i_o - z^i +\nonumber \\
  &+& f^B(\tau )\, \Lambda_B{}^C(\vec h)\,
 \epsilon^i_C(\vec h) + {1\over {Mc}}\, \Big(\epsilon^i_r(\vec h)\, {\hat {\cal K}}^r +
 {{\delta^{im}\, \epsilon_{mnk}\, h^n\, {\hat S}^k}\over {1 + \sqrt{1 +
 {\vec h}^2}}}\Big)\Big] \approx 0,\nonumber \\
 &&{}\nonumber \\
 &&\Downarrow\nonumber \\
 &&{}\nonumber \\
 {\hat {\cal K}}^r &\approx& Mc\, h^r\, \Big(x^o_o + f^B(\tau )\,
 \Lambda_B{}^C(\vec h)\, \epsilon^o_C(\vec h) -
  {{\sum_u\, h^u\, \Big(x^u_o - z^u + f^B(\tau )\, \Lambda_B{}^C(\vec h)\,
 \epsilon^u_C(\vec h)\Big)}\over {1 + \sqrt{1 + {\vec h}^2}}} \Big)
 -\nonumber \\
 &-& \Big(x^r_o - z^r + f^B(\tau )\, \Lambda_B{}^C(\vec h)\,
 \epsilon^r_C(\vec h) + {{\delta^{rm}\, \epsilon_{mnk}\, h^n\, {\hat S}^k}\over
 {Mc\, (1 + \sqrt{1 + {\vec h}^2})}}\Big).
 \label{5.13}
 \eea

They replace the constraints ${\cal K}^r \approx 0$ of Subsection A.

Now we have ${\hat {\cal S}}^{AB} \approx \delta^{Ar}\,
\delta^{Bs}\, \epsilon_{rsk}\, {\hat S}^k + (\delta^A_r\,
\delta^B_{\tau} - \delta^A_{\tau}\, \delta^B_r)\, {\hat {\cal
K}}^r$.

\bigskip

Let us remark that that if we put $\Lambda_A{}^B(\vec h) =
\delta^B_A$ and $x^{\mu}_o + f^B(\tau)\, \Lambda_B{}^C(\vec h)\,
\epsilon^{\mu}_C(\vec h) = Y^{\mu}(0) + h^{\mu}\, \tau$, then we
recover the results of Subsection A for the inertial rest frame
centered on the Fokker-Pryce inertial observer.\medskip

Instead the conditions $\Lambda_A{}^B(\vec h) = \delta^B_A$ and
$f^B(\tau)\, \Lambda_B{}^C(\vec h)\, \epsilon^{\mu}_C(\vec h) =
h^{\mu}\, \tau$, identifying the inertial rest frame centered on the
inertial observer $x^{\mu}_o + h^{\mu}\, \tau$, have the constraints
${\cal K}^r \approx 0$ replaced by Eqs.(\ref{5.13}).

\bigskip

Equations of the type (\ref{5.7}), (\ref{5.12}) and (\ref{5.13})
holds not only for admissible embeddings with pure differential
rotations like the ones of Eq.(\ref{2.14}), but also for the
admissible embeddings with pure linear acceleration. If in
Eq.(\ref{2.1}) we put $F^{\tau}(\tau ,\sigma^u) = 0$, $F^r(\tau
,\sigma^u) = \sigma^r$, so that the embedding becomes $z^{\mu}(\tau
,\sigma^u) = x_o^{\mu} + \epsilon^{\mu}_{\tau}\,f^{\tau}(\tau ) +
\epsilon^{\mu}_r\, \Big(f^r(\tau ) + \sigma^r\Big)$, the
instantaneous 3-spaces are space-like hyper-planes orthogonal to
$l^{\mu} = \epsilon^{\mu}_{\tau}$ and we get $h_{rs} = \delta_{rs}$,
$1 + n(\tau ) = {\dot f}^{\tau}(\tau )$, $n_r(\tau ) = \delta_{rs}\,
{\dot f}^s(\tau )$. In the case of Eq.(\ref{2.13}), i.e. $f^r(\tau )
= 0$ and $f^{\tau}(\tau ) = f(\tau )$, we get $1 + n(\tau ) = \dot
f(\tau )$, $n_r = 0$. If $f^{\tau}(\tau ) = \tau$ and $f^r(\tau ) =
a^r = const.$, we have inertial frames centered on inertial
observers: changing $a^r$ we change the inertial observer origin of
the 3-coordinates $\sigma^r$.

\bigskip

Let us remark that the final Dirac Hamiltonian (\ref{4.35}) does not
coincide with $Mc$ due to the presence of the inertial potentials
$g_{AB}(\tau ,\sigma^u)$.

\subsection{The Non-Inertial Rest Frames}

The family of non-inertial rest frames for an isolated system
consists of all the admissible 3+1 splittings of Minkowski
space-time whose instantaneous 3-spaces $\Sigma_{\tau}$ tend to
space-like hyper-planes orthogonal to the conserved 4-momentum of
the isolated system at spatial infinity. Therefore they tend to the
Wigner 3-spaces (\ref{1.1}) of the inertial rest frame
asymptotically.\bigskip

These non-inertial frames can be centered on the external
Fokker-Pryce center of inertia like the inertial ones and  are
described by the following embeddings

\bea
 &&z^{\mu}(\tau ,\sigma^u )\approx z^{\mu}_F(\tau ,\sigma^u ) =
 Y^{\mu}(\tau ) + u^{\mu}(\vec h)\, g(\tau ,\sigma^u ) + \epsilon^{\mu}_r(\vec
 h)\, [\sigma^r + g^r(\tau ,\sigma^u )],\nonumber \\
&&\nonumber\\
&&\qquad{\rightarrow_{|\vec \sigma |\, \rightarrow \infty}}
 z^{\mu}_W(\tau ,\sigma^u ) = Y^{\mu}(\tau ) +
 \epsilon^{\mu}_r(\vec h)\, \sigma^r,\qquad
x^{\mu}(\tau ) = z^{\mu}_F(\tau , 0^u), \nonumber \\
 &&{}\nonumber \\
&&g(\tau ,0^u) = g^r(\tau ,0^u) = 0,\qquad  g(\tau ,\sigma^u )\,
\rightarrow_{|\vec \sigma | \rightarrow
 \infty}\, 0,\qquad g^r(\tau ,\sigma^u )\, \rightarrow_{|\vec \sigma | \rightarrow
 \infty}\, 0.
 \label{5.14}
 \eea

These embeddings  are a special case of Eqs.(\ref{4.1}) with
$x^{\mu}(\tau ) = Y^{\mu}(\tau )$ and $F^{\mu}(\tau ,\sigma^u ) =
\epsilon_{\tau}^{\mu}(\vec h)\, g(\tau ,\sigma^u ) +
\epsilon^{\mu}_r(\vec h)\, [\sigma^r + g^r(\tau ,\sigma^u )]\,\,$,
$\epsilon^{\mu}_{\tau}(\vec h) = h^{\mu} = {\dot Y}^{\mu}(\tau )$.

\bigskip

For the induced metric we have

\begin{eqnarray*}
 z^{\mu}_{\tau}(\tau ,\sigma^u ) &\approx& z^{\mu}_{F\,
 \tau}(\tau ,\sigma^u ) = h^{\mu}\, [1 +
 \partial_{\tau}\, g(\tau ,\sigma^u )] + \epsilon^{\mu}_r(\vec
 h)\, \partial_{\tau}\, g^r(\tau ,\sigma^u ),\nonumber \\
&&\nonumber\\
z^{\mu}_r(\tau ,\sigma^u ) &\approx& z^{\mu}_{F\, r}(\tau ,
 \sigma^u ) = h^{\mu}\, \partial_r\, g(\tau ,\sigma^u ) +
 \epsilon^{\mu}_s(\vec h)\, [\delta^s_r + \partial_r\, g^s(\tau
 ,\sigma^u )],\nonumber \\
 &&{}\nonumber \\
 \sgn\,g_{F\, \tau\tau}(\tau ,\sigma^u ) &=& [1 + \partial_{\tau}\, g(\tau
 ,\sigma^u )]^2 - \sum_r\, [\partial_{\tau}\, g^r(\tau ,\sigma^u
 )]^2 =\nonumber \\
 &=& \Big[(1 + n_F)^2 - \h_F^{rs}\, n_{F\, r}\, n_{F\, s}\Big](\tau ,\sigma^u),
 \nonumber \\
&&\nonumber\\
 \sgn\,g_{F\, \tau u}(\tau ,\sigma^u ) &=& [1 + \partial_{\tau}\,
g(\tau ,\sigma^u )]\, \partial_u\, g(\tau ,\sigma^u ) - \sum_r\,
\partial_{\tau}\, g^r(\tau ,\sigma^u )\, [\delta^r_u +
 \partial_u\, g^r(\tau ,\sigma^u )] =\nonumber \\
 &=& \Big([1 + \partial_{\tau}\, g]\, \partial_u\, g - \partial_{\tau}\, g^u
 - \sum_r\, \partial_{\tau}\, g^r\, \partial_u\, g^r\Big)(\tau
 ,\sigma^u ) = - n_{F\, u}(\tau ,\sigma^u),
 \end{eqnarray*}

\bea
\sgn\,g_{F\, uv}(\tau ,\sigma^u ) &=& - \h_{F\, uv}(\tau ,\sigma^u) =\nonumber \\
 &=& \partial_u\, g(\tau ,\sigma^u
 )\, \partial_v\, g(\tau ,\sigma^u ) - \sum_r\, [\delta^r_u +
 \partial_u\, g^r(\tau ,\sigma^u )]\, [\delta^r_v + \partial_v\,
 g^r(\tau ,\sigma^u )] =\nonumber \\
 &=& - \delta_{uv} + \Big(\partial_u\, g\, \partial_v\, g - (\partial_u\,
 g^v + \partial_v\, g^u) - \sum_r\, \partial_u\, g^r\, \partial_v\,
 g^r\Big)(\tau ,\sigma^u ),\nonumber \\
 &&{}
 \label{5.15}
 \eea

\bigskip

The admissibility conditions  of Eqs.(\ref{2.9}), plus the
requirement $1 + n_F(\tau ,\sigma^u) > 0$, can be written as
restrictions on the functions $g(\tau ,\sigma^u)$ and $g^r(\tau
,\sigma^u)$.

\bigskip

The unit normal $l_F^\mu(\tau ,\sigma^u )$ and the tangent 4-vectors
$z^\mu_{F\,r}(\tau ,\sigma^u )$ to the instantaneous 3-spaces
$\Sigma_{\tau }$ can be projected on the asymptotic tetrad
 $h^\mu = \epsilon^{\mu}_{\tau}(\vec h)$, $\epsilon^\mu_r(\vec h)$

\bea
 z^\mu_{F\,r}(\tau ,\sigma^u)&=& \Big[\partial_r\, g\, h^\mu +
 \partial_r\, g^s\,\epsilon^\mu_s(\vec h)\Big](\tau ,\sigma^u)\nonumber\\
&&\nonumber\\
 l^{\mu}_F(\tau ,\sigma^u) &=& \Big[{1\over {\sqrt{\gamma}}}\,
\epsilon^{\mu}{}_{\alpha\beta\gamma}\, z^{\alpha}_{F1}\,
z^{\beta}_{F2}\, z^{\gamma}_{F3}\Big](\tau ,\sigma^u) = \nonumber \\
 &=&{1\over {\sqrt{\gamma (\tau ,\sigma^u)}}}\, \Big[det\, (\delta^s_r
 + \partial_r\, g^s)\, h^{\mu} -\nonumber \\
 &-& \delta^{ra}\, \epsilon_{asu}\, \epsilon_{vwt}\,
\partial_v\, g\,\, \partial_w\, g^s\,\, \partial_t\, g^u\,
\epsilon^{\mu}_r(\vec h) \Big](\tau ,\sigma^u),\nonumber \\
 &&{}\nonumber \\
 1 + n_F(\tau ,\sigma^u) &=& \sgn\, z^{\mu}_{\tau}(\tau ,\sigma^u)\,
 l_{F\mu}(\tau ,\sigma^u) =\nonumber \\
 &=& {1\over {\sqrt{\gamma (\tau ,\sigma^u)}}}\, \Big[(1 +
 \partial_{\tau}\, g\, det\, (\delta^s_r + \partial_r\, g^s)
 -\nonumber \\
 &-& \partial_{\tau}\, g^r\, \epsilon_{rsu}\, \epsilon_{vwt}\,
\partial_v\, g\,\, \partial_w\, g^s\,\, \partial_t\, g^u
\Big](\tau ,\sigma^u),\nonumber \\
 &&{}\nonumber \\
  l^2_F(\tau ,\sigma^u) = \sgn,\,\, \Rightarrow\,\,
 \gamma_F(\tau ,\sigma^u) &=& \Big[\Big(det\, (\delta^s_r +
 \partial_r\, g^s)\Big)^2 -\nonumber \\
 &-& 2\, \epsilon_{vwt}\,
\partial_v\, g\,\, \partial_w\, g^s\,\, \partial_t\, g^u\,
\epsilon_{hmn}\, \partial_h\, g\,\, \partial_m\, g^s\,\,
\partial_n\, g^u \Big](\tau ,\sigma^u).\nonumber \\
 &&{}
 \label{5.16}
 \eea

 \bigskip

To define the non-inertial rest-frame instant form we must find the
form of the internal Poincare' generators replacing the ones of the
inertial rest-frame one, given in Eqs.(\ref{5.2}).\medskip

Eq.(\ref{3.8}) and  the first of Eqs.(\ref{3.17}) imply

\bea
 P^{\mu} &=& Mc\, h^{\mu} =  \int d^3\sigma\,
 \rho^{\mu}(\tau ,\sigma^u ) \approx\nonumber \\
 &\approx& h^{\mu}\, \int d^3\sigma\, \sqrt{\gamma (\tau ,\sigma^u)}\,
 \Big({{det\, (\delta^s_r + \partial_r\, g^s)}\over {\sqrt{\gamma_F}}}\,
 T_{F\, \perp\perp} -\nonumber \\
 &-& \partial_r\, g\, h_F^{rs}\,
 T_{F\, \perp s}\Big)(\tau , \sigma^u ) +\nonumber \\
 &+&  \epsilon^{\mu}_u(\vec h)\, \int d^3\sigma\, \Big(
 - {{\delta^{ua}\, \epsilon_{asr}\, \epsilon_{vwt}\,
\partial_v\, g\,\, \partial_w\, g^s\,\, \partial_t\, g^r}
\over {\sqrt{\gamma_F}}}\, T_{F\, \perp\perp} -\nonumber \\
 &-& (\delta^u_r + \partial_r\, g^u)\,
 h_F^{rs}\, T_{F\, \perp s}\Big)(\tau ,\sigma^u ) =\nonumber \\
 &{\buildrel {def}\over =}& \int d^3\sigma\, {\cal T}^{\mu}_F(\tau ,\sigma^u),
 \label{5.17}
 \eea

\noindent so that the internal mass and the rest-frame conditions
become (Eqs.(\ref{5.2}) are recovered for the inertial rest frame)

\bea
 Mc &=& \int d^3\sigma\, \Big({{det\, (\delta^s_r + \partial_r\,
 g^s)}\over {\sqrt{\gamma}}}\, T_{F\, \perp\perp} -
 \partial_r\, g\, h_F^{rs}\, T_{F\, \perp s}\Big)(\tau ,
 \sigma^u ),\nonumber \\
  &&{}\nonumber \\
  {\hat {\cal P}}^u &=& \int d^3\sigma\, \Big(
 - {{\delta^{ua}\, \epsilon_{asr}\, \epsilon_{vwt}\,
\partial_v\, g\,\, \partial_w\, g^s\,\, \partial_t\, g^r}
\over {\sqrt{\gamma_F}}}\, T_{F\, \perp\perp} -\nonumber \\
 &-& (\delta^u_r + \partial_r\, g^u)\,
 h_F^{rs}\, T_{F\, \perp s}\Big)(\tau ,\sigma^u )
 \approx 0.\nonumber \\
 &&{}
 \label{5.18}
 \eea

\bigskip

By using Eqs.(\ref{3.17}) for the angular momentum  we get
$J^{\mu\nu} \approx \int d^3\sigma\, \Big(z^{\mu}_F\, \rho_F^{\nu} -
 z_F^{\nu}\, \rho_F^{\mu}\Big)(\tau ,\sigma^u )$ with
 $\rho^{\mu}_F(\tau ,\sigma^u) = \Big[\sqrt{\gamma_F}\, \Big(T_{\perp\perp}\,
 l^{\mu}_F - T_{\perp s}\, h_F^{sr}\, z^{\mu}_{Fr}\Big)\Big](\tau
 ,\sigma^u)$, where $z^{\mu}_F$, $z^{\mu}_{Fr}$ and $l^{\mu}_F$ are
given in Eqs.(\ref{5.14}), (\ref{5.15}) and (\ref{5.16})
respectively. The description of the isolated system as a
pole-dipole carried by the external center of mass $\vec z$ requires
that we must identify the previous $J^{ij}$ and $J^{oi}$ with the
expressions like the ones given in Eqs.(\ref{5.1}), now functions of
$\vec z$, $\vec h$, $Mc$ of Eq.(\ref{5.18}) and of an effective spin
${\vec {\tilde S}}$. This identification will allow to find the
effective spin ${\vec {\tilde S}}$ and three constraints ${\tilde
{\cal K}}^r \approx 0$ eliminating the internal 3-center of mass: in
the limit of the inertial rest frame they must reproduce the
quantities in Eqs.(\ref{5.2}).\medskip

By using Eqs.(\ref{5.18}) this procedure implies (${\hat {\cal
K}}^r$ and ${\hat {\cal S}}^r$ are the analogue of the quantities
defined in Eqs.(\ref{5.8}) for the embedding (\ref{5.14}))

\begin{eqnarray*}
 J^{\mu\nu} &\approx& \int d^3\sigma\, \Big(z^{\mu}_F\, \rho_F^{\nu} -
 z_F^{\nu}\, \rho_F^{\mu}\Big)(\tau ,\sigma^u ) =\nonumber \\
 &=& Mc\, \Big(Y^{\mu}(0)\, h^{\nu} - Y^{\nu}(0)\, h^{\mu}\Big)
 + {\hat {\cal P}}^u\, \Big(Y^{\mu}(0)\, \epsilon^{\nu}_u(\vec h) -
 Y^{\nu}(0)\, \epsilon^{\mu}_u(\vec h)\Big) +\nonumber \\
 &+& \Big(\tau\, {\hat {\cal P}}^u + {\hat {\cal K}}^u\Big)\, \Big(h^{\mu}\,
 \epsilon^{\nu}_u(\vec h) - h^{\nu}\, \epsilon^{\mu}_u(\vec h)\Big)
  + \delta^{un}\, \epsilon_{nvr}\, {\hat S}^r\, \epsilon^{\mu}_u(\vec h)\,
 \epsilon^{\nu}_v(\vec h) \approx\nonumber \\
 &\approx& Mc\, \Big(Y^{\mu}(0)\, h^{\nu} - Y^{\nu}(0)\, h^{\mu}\Big)
 + {\hat {\cal K}}^u\, \Big(h^{\mu}\, \epsilon^{\nu}_u(\vec h) - h^{\nu}\,
 \epsilon^{\mu}_u(\vec h)\Big) +\nonumber \\
 &+& \delta^{un}\, \epsilon_{nvr}\, {\hat S}^r\, \epsilon^{\mu}_u(\vec h)\,
 \epsilon^{\nu}_v(\vec h),\nonumber \\
 &&{}\nonumber \\
 &&so\, that\, we\, get
 \end{eqnarray*}

 \bea
 J^{ij} &=& z^i\, h^j - z^j\, h^i + \delta^{iu}\, \delta^{jv}\,
 \epsilon_{uvk}\, {\tilde S}^k \approx\nonumber \\
 &\approx& Mc\, \Big(Y^i(0)\, h^j - Y^j(0)\, h^i\Big)
 + {\hat {\cal K}}^u\, \Big(h^i\, \epsilon^j_u(\vec h) - h^j\,
 \epsilon^i_u(\vec h)\Big) +\nonumber \\
 &+& \delta^{un}\, \epsilon_{nvr}\, {\hat S}^r\, \epsilon^i_u(\vec h)\,
 \epsilon^j_v(\vec h),\nonumber \\
 &&{}\nonumber \\
 J^{oi} &=& - \sqrt{1 + {\vec h}^2}\, z^i + {{\delta^{in}\, \epsilon_{njk}\,
 {\tilde S}^j\, h^k}\over {1 + \sqrt{1 + {\vec h}^2}}} \approx \nonumber \\
 &\approx& Mc\, \Big(Y^o(0)\, h^i - Y^i(0)\, h^o\Big)
 + {\hat {\cal K}}^u\, \Big(h^o\, \epsilon^i_u(\vec h) - h^i\,
 \epsilon^o_u(\vec h)\Big) +\nonumber \\
 &+& \delta^{un}\, \delta^{vm}\, \epsilon_{nmr}\, {\hat S}^r\, \epsilon^o_u(\vec h)\,
 \epsilon^i_v(\vec h).
 \label{5.19}
 \eea

\bigskip

As a consequence, by using the expression of $Y^{\mu}(0)$ given
after Eq.(\ref{5.2}), the constraints eliminating the 3-center of
mass and the effective spin are

\bea
 {\hat {\cal K}}^u &=& \int d^3\sigma\, \Big(g\,
 \Big[\delta^{ur}\, \partial_r\, g\, T_{F\, \perp\perp}
 - (\delta^u_r + \partial_r\, g^u)\, h_F^{rs}\, T_{F\, \perp s}\Big]
 -\nonumber \\
 &-& (\sigma^u + g^u)\, \Big[{{det\, (\delta^s_r + \partial_r\, g^s)}\over
 {\sqrt{\gamma}}}\, T_{F\, \perp\perp} - \partial_r\, g\,
 h_F^{rs}\, T_{F\, \perp s}\Big]\Big)(\tau ,\sigma^u
 ) \approx 0,\nonumber \\
 &&{}\nonumber \\
 {\tilde S}^r &\approx& {\hat S}^r
 = {1\over 2}\, \delta^{rn}\, \epsilon_{nuv}\,
 \int d^3\sigma\, \Big((\sigma^u + g^u)\, \Big[\delta^{vm}\,
 \partial_m\, g\, T_{F\, \perp\perp} - (\delta^v_r + \partial_r\, g^v)\,
 h_F^{rs}\, T_{F\, \perp s}\Big] -\nonumber \\
 &-&(\sigma^v + g^v)\, \Big[\delta^{um}\, \partial_m\, g\, T_{F\,
 \perp\perp} - (\delta^u_r + \partial_r\, g^u)\, h_F^{rs}\,
 T_{F\, \perp s}\Big] \Big)(\tau ,\sigma^u ).
 \label{5.20}
 \eea

\noindent and these formulas allow to recover Eqs.(\ref{5.2}) of the
inertial rest frame. \medskip

Therefore the non-inertial rest-frame instant form of dynamics is
well defined.

\subsection{The Hamiltonian of the Non-Inertial Rest-Frame Instant
Form}

We have now to find which is the effective Hamiltonian of the
non-inertial rest-frame instant form replacing $Mc$ of the inertial
rest-frame one. The gauge fixing (\ref{5.20}) is a special case of
Eqs.(\ref{4.1}), whose  final Dirac Hamiltonian is given in
Eq.(\ref{4.4}) [or in Eq.(\ref{4.35}) in the radiation gauge].
\medskip

To be able to impose this gauge fixing, let us put $F^{\mu}(\tau
,\sigma^u ) = h^{\mu}\, g(\tau ,\sigma^u ) + \epsilon^{\mu}_r(\vec
h)\, [\sigma^r + g^r(\tau ,\sigma^u )]$ in Eq.(\ref{4.1}), but let
us leave $x^{\mu}(\tau )$ as an arbitrary time-like observer to be
restricted to $Y^{\mu}(\tau )$ at the end. We will only assume that
$x^{\mu}(\tau )$ is canonically conjugate with $P^{\mu} = \int
d^3\sigma\, \rho^{\mu}(\tau , \sigma^u )$, $\{ x^{\mu}(\tau ),
P^{\nu} \} = - \sgn\, \eta^{\mu\nu}$.
\medskip

Due to the dependence of $F^{\mu}(\tau ,\sigma^u)$ and of
$Y^{\mu}(\tau)$ on $\vec h = \vec P/\sqrt{\sgn\, P^2}$ we must
develop a different procedure for the identification of the Dirac
Hamiltonian.

\medskip

In this case the constraints (\ref{3.10}) can be rewritten in the
following form (${\cal T}^{\mu}_F(\tau ,\sigma^u)$ is defined in
Eq.(\ref{5.17}))

 \bea
 {\cal H}^{\mu}(\tau ,\sigma^u ) &=& {\tilde {\cal H}}^{\mu}(\tau ,
 \sigma^u ) + \delta^3(\sigma^u )\, \int d^3\sigma_1\, {\cal
 H}^{\mu}(\tau, {\sigma}^u_1) \approx 0,\nonumber \\
   with \qquad &&\int d^3\sigma\,{\tilde {\cal H}}^{\mu}(\tau ,\sigma^u )
  \equiv 0,\nonumber \\
  &&{}\nonumber \\
  &&\Downarrow\nonumber \\
  &&{}\nonumber \\
  \rho^{\mu}(\tau ,\sigma^u ) &\approx& P^{\mu}\, \delta^3(
 \sigma^u ) + \Big[{\cal T}_F^{\mu}(\tau ,\sigma^u ) - \delta^3(\sigma^u )\,
 {\cal R}_F^{\mu}(\tau )\Big] =\nonumber\\
 &=& \delta^3(\sigma^u )\, H^\mu(\tau) + {\cal T}_F^{\mu}(\tau
,\sigma^u ),\nonumber \\
 &&{}\nonumber \\
 &&H^\mu(\tau) = P^\mu - {\cal R}_F^\mu(\tau ) \approx 0,\qquad {\cal
R}_F(\tau ) {\buildrel {def}\over =} \int d^3\sigma\, {\cal
T}^{\mu}_F(\tau ,\sigma^u).
 \nonumber \\
 &&{}
  \label{5.21}
 \eea

In this way the original canonical variables $z^{\mu}(\tau ,\vec
\sigma )$, $\rho^{\mu}(\tau ,\vec \sigma )$ are replaced by the
observer $x^{\mu}(\tau )$, $P^{\mu}$ and by relative variables with
respect to it.\medskip

From Eq.(\ref{5.14}) we get:\hfill\break

a) the gauge fixing to the constraints ${\tilde {\cal H}}^{\mu}(\tau
,\sigma^u ) \approx 0$ is

 \beq
 \psi^{\mu}_r(\tau ,\sigma^u ) = {{\partial\, \chi^{\mu}(\tau ,
 \sigma^u )}\over {\partial\, \sigma^r}} =
 \Big(z^{\mu}_r - \epsilon^{\mu}_s(\vec h)\, \Big[\delta^s_r +
 {{\partial\, g^s}\over {\partial\, \sigma^r}} - u^{\mu}(\vec h)\,
 {{\partial\, g}\over {\partial\, \sigma^r}}\Big]\Big)(\tau ,\sigma^u )
 \approx 0;
 \label{5.22}
 \eeq

b) the gauge fixing to the constraints $H^\mu(\tau) = P^\mu - {\cal
R}_F^\mu \approx 0$ is $\chi^{\mu}(\tau ,0) = z^{\mu}(\tau ,0) -
Y^{\mu}(\tau ) = x^{\mu}(\tau) - Y^{\mu}(\tau) \approx 0$.\medskip

The gauge fixing (\ref{5.22}) has the following Poisson brackets
with the collective variables $x^{\mu}(\tau )$, $P^{\mu}$

\bea
 &&\{ P^{\mu}, \psi^{\nu}_r(\tau, \sigma^u ) \} = 0,
 \nonumber \\
 && \{ x^{\mu}(\tau ), \psi^{\nu}_r(\tau ,\sigma^u ) \} = -
 {{\partial\, \epsilon^{\nu}_s(\vec h)}\over {\partial\, P_{\mu}}}\,
 \Big(\delta^s_r + {{\partial\, g^s(\tau ,\sigma^u )}\over {\partial\,
 \sigma^r}} \Big)- {{\partial\, \epsilon^{\nu}_{\tau}(\vec h)}\over {\partial\, P_{\mu}}}\,
 {{\partial\, g(\tau ,\sigma^u )}\over {\partial\, \sigma^r}}
 \not= 0.
 \label{5.23}
  \eea

Therefore $x^{\mu}(\tau )$ is no more a canonical variable after the
gauge fixing $\psi^{\mu}_r(\tau ,\sigma^u ) \approx 0$.
\bigskip

By introducing the notation ($\epsilon^A_\mu = \eta^{AB}\,
\epsilon_{B\mu}\, \Rightarrow\, \epsilon^\tau_\mu(\vec h) = \sgn\,
h_\mu$)

\beq
 {\cal T}_F^\mu(\tau ,\sigma^u ) \byd\, h^\mu\,
 {\cal T}_F^{\tau}(\tau ,\sigma^u ) + \epsilon^\mu_r(\vec h)\, {\cal
T}^r_F(\tau ,\sigma^u ),\qquad
 \Rightarrow {\cal T}^A_F(\tau ,\sigma^u ) = \epsilon^A_{\mu}(\vec h)\, {\cal
T}^{\mu}_F(\tau ,\sigma^u ),
 \label{5.24}
 \eeq

\noindent the angular momentum generator of Eq.(\ref{3.17}) takes
the form

\bea
 J^{\mu\nu} &=& x^{\mu}(\tau )\, P^{\nu} - x^{\nu}(\tau)\, P^{\mu} +
 S^{\mu\nu},\nonumber \\
 &&{}\nonumber \\
 S^{\mu\nu} &\approx& \epsilon^{\mu}_r(\vec h)\,
 \epsilon^{\nu}_s(\vec h)\, \int d^3\sigma\, \Big[(\sigma^r + g^r)\, {\cal T}^s -
 (\sigma^s + g^s)\, {\cal T}^r\Big](\tau ,\sigma^u ) +\nonumber \\
&&\nonumber\\
&+& \Big(\epsilon^{\mu}_r(\vec h)\, \epsilon^{\nu}_{\tau}(\vec h) -
 \epsilon^{\nu}_r(\vec h)\, \epsilon^{\mu}_{\tau}(\vec h)\Big)\,
 \int d^3\sigma\, \Big[(\sigma^r + g^r)\, {\cal T}^{\tau} + g\, {\cal T}^r\Big](\tau
 ,\sigma^u ) =\nonumber \\
&&\nonumber\\
&=& \epsilon^{\mu}_A(\vec h)\, \epsilon^{\nu}_B(\vec h)\,
 S^{AB},\nonumber \\
 &&{}\nonumber \\
 S^{rs} &=& \int d^3\sigma\, \Big[(\sigma^r + g^r)\, {\cal T}^s -
 (\sigma^s + g^s)\, {\cal T}^r\Big](\tau ,\sigma^u )\byd
 \delta^{rn}\, \epsilon_{nsu}\,{\cal J}^u,\nonumber \\
 S^{\tau r} &=& - S^{r\tau} = -  \int d^3\sigma\, \Big[(\sigma^r +
 g^r)\, {\cal T}^{\tau} + g\, {\cal T}^r\Big](\tau ,\sigma^u )\byd{\cal
 K}^r,
 \label{5.25}
 \eea

\noindent where only the constraints $\tilde{\cal H}^\mu(\tau
,\sigma^u) \approx 0$ have been used.

 \medskip

 Since we have

 \bea
 \{ x^{\mu}(\tau ), S^{\alpha\beta} \} &=& 0,\nonumber \\
 &&{}\nonumber \\
 \{ {{\partial\, z^{\mu}(\tau ,\sigma^u )}\over
 {\partial\, \sigma^r}}, S^{\alpha\beta} \} &=& \Big({{\partial\, z^{\beta}}\over
 {\partial\, \sigma^r}}\,\eta^{\mu\alpha} - {{\partial\, z^{\alpha}}\over
 {\partial\, \sigma^r}}\,\eta^{\mu\beta} \Big)(\tau ,\sigma^u )
 \approx\nonumber \\
 &\approx& \Big(\Big[\epsilon^{\beta}_s(\vec h)\, (\delta^s_r + {{\partial\,
 g^s}\over {\partial^r}}) + h^{\beta}\, {{\partial\, g}\over {\partial\,
 \sigma^r}}\Big]\, \eta^{\mu\alpha}\Big)(\tau ,\sigma^u ),
 \label{5.26}
 \eea

\noindent after the gauge fixing the new canonical variable for the
observer becomes

\beq
 {\tilde x}^{\mu}(\tau ) = x^{\mu}(\tau ) - {1\over 2}\,
 \epsilon_{\sigma\, A}(\vec h)\, {{\partial\, \epsilon^A_{\rho}(\vec
 h)}\over {\partial\, P_{\mu}}}\, S^{\sigma\rho},\qquad
 \{ {\tilde x}^{\mu}(\tau ), \psi^{\nu}_r(\tau ,\vec \sigma )\} =
 0.
 \label{5.27}
 \eeq
 \medskip

If we eliminate the relative variables by going to Dirac brackets
with respect to the second class constrainta ${\tilde {\cal
H}}^{\mu}(\tau ,\sigma^u ) \approx 0$, $\psi^{\mu}_r(\tau , \sigma^u
) \approx 0$, the canonical variables $z^{\mu}(\tau ,\sigma^u )$,
$\rho_{\mu}(\tau ,\sigma^u )$ are reduced to the canonical variables
${\tilde x}^{\mu}(\tau )$, $P^{\mu}$.\medskip

By defining ${\cal R}_F(\tau ) = \sgn\, h_{\mu}\, {\cal
R}^{\mu}_F(\tau ) \approx Mc = \sqrt{\sgn\, P^2}$, the remaining
constraints are

\bea
 H^\mu(\tau)&=&h^\mu\, \Big(\sqrt{\sgn\,P^2} - {\cal
R}_f(\tau)\Big) + \epsilon^\mu_r(\vec{h})\, {\hat {\cal P}}^r,
\nonumber\\
 &&{}\nonumber \\
 or&& \sgn\,h^\mu\, H_\mu(\tau) = \sqrt{\sgn\,P^2} - {\cal
R}_F(\tau)\approx 0,\qquad
 \epsilon^r_\mu(\vec{h})H^\mu(\tau) = {\hat {\cal P}}^r \approx 0.
 \nonumber \\
 &&{}
 \label{5.28}
  \eea
\bigskip

Like in Eqs.(\ref{4.1}) and (\ref{4.2}), after this reduction the
Dirac multiplier $\lambda^{\mu}(\tau ,\sigma^u )$ in the Dirac
Hamiltonian (\ref{3.16}) becomes

\bea
 \lambda_\mu(\tau,\sigma^u)&=& \sgn\, h_\mu\,
 \Big(\lambda_\tau(\tau) - \frac{\partial
g(\tau,\sigma^u)}{\partial\tau}\Big) + \sgn\,
\epsilon_{\mu\,r}(\vec{h})\, \Big(\lambda^r(\tau) - \frac{\partial
g^r(\tau,\sigma^u)}{\partial\tau}\Big)\on \nonumber \\
 &\cir& - \sgn\,\frac{\partial z^\mu_F(\tau ,\sigma^u)}{\partial\tau}
 \label{5.29}
 \eea

At this stage the Dirac Hamiltonian depends only on the residual
Dirac multipliers $\lambda_{\tau}(\tau )$ and $\vec \lambda (\tau )$

 \beq
 H_D = \lambda_{\tau}(\tau )\, (\sqrt{\sgn\,P^2} - {\cal R}_F)
 - \vec \lambda (\tau ) \cdot {\hat {\vec {\cal P}}} +
 \int d^3\sigma\, \Big({{\partial\, g^r}\over {\partial\, \tau}}\,
 {\cal T}_{F\, r} + \frac{\partial g}{\partial\tau}\,
 {\cal T}_{F\, \tau}\Big)(\tau ,\sigma^u ),
 \label{5.30}
 \eeq

\noindent where we introduced the notation ${\cal T}_{F\, A}(\tau
,\sigma^u ) \byd \sgn\,\epsilon_{\mu\,A}(\vec h)\, {\cal
T}_F^{\mu}(\tau ,\sigma^u )$ so that ${\cal T}_F^\tau = {\cal
T}_{F\, \tau},\qquad {\cal T}_F^r = - \sgn\, {\cal T}_{F\, r}$.
\bigskip

To implement the gauge fixing $x^{\mu}(\tau) - Y^{\mu}(\tau) \approx
0$ requires two other steps:\medskip

1) Firstly we impose the gauge fixing  ${\tilde x}^{\mu}(\tau )\,
h_{\mu} = \sgn\, \tau$. It implies $\lambda_{\tau}(\tau ) = -1$ and
$\sqrt{\sgn\,P^2} = Mc \equiv  {\cal R}_F$. The Dirac Hamiltonian
becomes

 \bea
 H_{F\, D} &=& {\cal M}\, c - \vec \lambda (\tau ) \cdot {\hat {\vec {\cal P}}}+
 \int d^3\sigma\, \Big[\mu\, \pi^\tau - A_\tau\,\Gamma\Big](\tau, \sigma^u )
,\nonumber \\
 &&{}\nonumber \\
 {\cal M}\, c &=& Mc + \int d^3\sigma\, \Big({{\partial\,
 g^r}\over {\partial\, \tau}}\,
 {\cal T}_{F\, r} + \frac{\partial g}{\partial\tau}\,
 {\cal T}_{F\, \tau}\Big)(\tau ,\sigma^u ).
 \label{5.31}
 \eea

\medskip

2) Then we add the gauge fixing ${\hat {\cal K}}^r \approx 0$
 to the rest-frame conditions ${\hat {\cal
P}}^r \approx 0$: this implies $\vec \lambda(\tau) = 0$. In this way
we get $x^{\mu}(\tau) \approx Y^{\mu}(\tau)$ and we also eliminate
the internal 3-center of mass. Having chosen the Fokker-Pryce
external 4-center of inertia $Y^{\mu}(\tau)$ as origin of the
3-coordinates the constraints ${\hat {\cal K}}^r \approx 0$
correspond to the requirement $S^{\tau r} \approx 0$.

\medskip

In conclusion the effective Hamiltonian ${\cal M}\, c$ (modulo
electro-magnetic gauge transformations) of the non-inertial
rest-frame instant form is not the internal mass $Mc$, since $Mc$
describes the evolution from the point of view of the asymptotic
inertial observers. There is an additional term interpretable as an
inertial potential producing relativistic inertial effects (see
Eqs.(\ref{5.16}) for $1 + n_F(\tau ,\sigma^u)$ and Eqs.(\ref{5.15})
for $n_{F\, r}(\tau ,\sigma^u)$)

\bea
 {\cal M}\, c&=& Mc + \int d^3\sigma\, \Big({{\partial\, g^r}\over
{\partial\, \tau}}\,
 {\cal T}_{F\, r} + \frac{\partial g}{\partial\tau}\,
 {\cal T}_{F\, \tau}\Big)(\tau ,\sigma^u )=\nonumber\\
&&\nonumber\\
&=&\int d^3\sigma\,\sgn\,\Big(\Big[\,h_\mu\Big(1+\frac{\partial
g}{\partial\tau}\Big)+ \epsilon_{\mu\,r}\,\frac{\partial
g^r}{\partial\tau}\,\Big]{\cal T}_F^\mu
\Big)(\tau ,\sigma^u) =\nonumber\\
&&\nonumber\\
&=&\int d^3\sigma\, \sqrt{\gamma (\tau, \sigma^u)}\, \Big((1 +
n_F)\, T_{F\, \perp\perp} + n_F^r\, T_{F\, \perp\,r}\Big)(\tau
,\sigma^u )
 \label{5.32}
  \eea

\noindent where

\bea
 &&\sqrt{\gamma (\tau, \sigma^u)}\, T_{F\, \perp\perp}(\tau ,\sigma^u)=
 \sqrt{\gamma (\tau, \sigma^u)}\, T'_{F\, \perp\perp}(\tau ,\sigma^u) +\nonumber \\
 &&\qquad + \sum_i\,\delta(\sigma^u - \eta^u_i)\,\sqrt{m_i^2\, c^2 +
\h^{rs}_F(\tau ,\sigma^u)\,(\kappa_{ir}(\tau ) - Q_i\,A_r(\tau
,\sigma^u))\, (\kappa_{is}(\tau ) - Q_i\,A_s(\tau ,\sigma^u))},\nonumber \\
 &&\sqrt{\gamma (\tau, \sigma^u)}\, T_{F\, \perp\,r}(\tau ,\sigma^u) =  F_{rs}(\tau
,\sigma^u)\,\pi^s(\tau ,\sigma^u) - \sum_i\,\delta(\sigma^u -
\eta^u_i)\, (\kappa_{ir}(\tau )
 - Q_i\,A_r(\tau ,\sigma^u)),
 \label{5.33}
 \eea

\noindent with $T'_{\perp\perp}$ given in Eq. (\ref{4.28}).

\bigskip

Let us remark that a similar procedure should be applied also to the
gauge fixing (\ref{4.1}) if we want to reproduce the results of
Subsection B for arbitrary non-inertial frames. We do not add these
calculations, because they agrees substantially with the results of
this Subsection and do not alter the conclusions of Section IV.

\vfill\eject

\section{Conclusion}

In this paper we have defined the general theory of {\it
non-inertial frames} in Minkowski space-time. It is based on
M$\o$ller-admissible 3+1 splittings of Minkowski space-time (they
give conventions for clock synchronization, i.e. for the
identification of instantaneous 3-spaces) and on parametrized
Minkowski theories for isolated systems admitting a Lagrangian
description. The transition from a non-inertial frame to every other
one is formalized as a gauge transformation, so that physical
results do not depend on how the clock are synchronized.\medskip

The M$\o$ller conditions, implying the absence of rotational
velocities higher than the velocity of light $c$ and requiring that
the three eigenvalues of the non-inertial 3-metric inside the
instantaneous Riemannian 3-spaces has three non-null positive
eigenvalues, have to be implemented with the following two extra
conditions:

a) the lapse function must be positive definite in each point of the
instantaneous 3-space, so to avoid the intersection of 3-spaces at
different times;

b) the space-like hyper-surfaces corresponding to the Riemannian
3-spaces must become space-like hyper-planes (Euclidean 3-spaces) at
spatial infinity with a direction-independent unit normal
$l^{\mu}_{(\infty )}$ (asymptotic inertial observers to be
identified with the fixed stars).\bigskip

Among the admissible non-inertial frames we identified the {\it
non-inertial rest frames}, generalizing the inertial rest frames and
relevant for canonical gravity \cite{5,11,12}.\bigskip

All the properties of the inertial rest-frame instant form of
dynamics, studied in details in Refs.\cite{8}, have been extended to
non-inertial frames. Again every isolated system may be described as
a decoupled non-covariant external center of mass carrying a
pole-dipole structure: the internal mass of the system and an
effective spin (becoming the rest spin in the inertial rest frame).
In particular we have found the non-inertial generalization of the
second class constraints eliminating the internal 3-center of mass
inside the instantaneous 3-spaces.\bigskip

This theory of non-inertial frames is free by construction from the
coordinate singularities of all the approaches to accelerated frames
based on the 1+3 point of view, in which the instantaneous 3-spaces
are identified with the local rest frames of the observer. The
pathologies of this approach are either the horizon problem of the
rotating disk (rotational velocities higher than $c$), which is
still present in all the calculations of pulsar magnetosphere in the
form of the light cylinder, or the intersection of the local rest
3-spaces. The main difference between the 3+1 and 1+3 points of view
is that the M$\o$ller conditions forbid {\it rigid rotations} in
relativistic theories.

\bigskip

We have done a detailed study of the isolated system of
positive-energy scalar particles with Grassmann-valued electric
charges plus the electro-magnetic field extending to non-inertial
frames its Hamiltonian description given in the inertial rest frame
in Ref.\cite{8}.\medskip

By using a non-covariant (i.e. coordinate-dependent) decomposition
of the electro-magnetic potential we obtained the {\it non-inertial
radiation gauge}, in which the electro-magnetic field is described
by means of transverse quantities (the Dirac observables). This
allowed us to find the non-inertial expression of the Coulomb
potential, which is now dependent also on the field strengths and
the inertial potentials. The non-covariance of the description is
natural due to the presence in the Hamiltonian of the {\it
relativistic inertial potentials}, namely the components
$g_{AB}(\tau ,\sigma^r)$ of the 4-metric induced by the 3+1
splitting, which are {\it intrinsically coordinate dependent}. The
non-relativistic limit of the inertial potentials reproduces the
standard (again coordinate-dependent) Newtonian ones. The
Hamiltonian in non-inertial frames turns out to be the sum of the
invariant mass (now coordinate-dependent due to its dependence on
the 4-metric) of the system plus terms in the inertial potentials
disappearing in the inertial rest frame.
\bigskip

In the second paper we will give the simplest example of 3+1
splitting with {\it differential rotations}  and we will develop the
3+1 point of view for the rotating disk and the Sagnac effect. Then
we will study properties of Maxwell equations in admissible nearly
rigidly rotating frames like the wrap-up effect, the Faraday
rotation in astrophysics and the pulsar magnetosphere.

\vfill\eject

\appendix

\section{The Landau-Lifschitz Non-Inertial Electro-Magnetic Fields}

Sometimes, see for instance Ref.\cite{17}, the following {\it
generalized non-inertial electric and magnetic fields} are
introduced

 \bea
  {\cal E}^{s}_{(F)}(\tau, \sigma^u)&=&
- \left[\frac{\sqrt{\gamma_{F}}}{\sqrt{1 + n_F}}\, \h_{F}^{sr}\left(
F_{\tau r} - n_{F}^v\, F_{vr}\right)\right](\tau, \sigma^u)
\,\cir\,\pi^s(\tau, \sigma^u)),\nonumber\\
 &&\nonumber\\
  {\cal B}_{(F)}^{w}(\tau, \sigma^u)&=& \frac{1}{2}\,
  \delta^{wt}\, \epsilon_{tsr}\,
\left[(1 + n_F)\,\sqrt{\gamma_{F}}\, \h_{F}^{sv}\,\h_{F}^{ru}\, F_
{vu}- (n_F^s\,\pi^r - n_F^r\,\pi^s) \right](\tau, \sigma^u),
 \label{b1}
  \eea

\medskip

They allow us to rewrite the Hamilton-Dirac Eqs.(\ref{4.15}) in the
following form (we use a vector notation as in the 3-dimensional
Euclidean case)

 \bea
  \partial_r\, {\cal E}^r{}_{(F)}(\tau, \sigma^u)&=&
\sqrt{\gamma_{F}(\tau, \sigma^u)}\,
\overline{\rho}(\tau, \sigma^u),\nonumber\\
 &&\nonumber\\
  \epsilon_{ruv}\, \partial_u\, {\cal B}^v{}_{(F)}(\tau, \sigma^u)
  - \frac{\partial{\cal E}_{(F)}^{r}(\tau, \sigma^u)} {\partial\tau}
 &=& \sqrt{\gamma_{F}(\tau, \sigma^u)}\,
\overline{J}^r(\tau, \sigma^u),
 \label{b2}
  \eea

\noindent namely {\it in the same form of  the usual source-
dependent Maxwell equations in an inertial frame}.
\bigskip

Since Eqs.(\ref{b1}) can be rewritten in the form

 \bea
  {\cal E}^{s}_{(F)}(\tau, \sigma^u)&=&
\left[+\frac{\sqrt{\gamma_{F}}}{\sqrt{(1 + n_F)}}\, \h_{F}^{sr}
\,E_r -\frac{\sqrt{\gamma_{F}}}{\sqrt{(1 + n_F)}}\, \h_{F}^{sr}\,
 \epsilon_{ruv}\, n_F^u\, B_v \right](\tau, \sigma^u),
\nonumber\\
 &&\nonumber\\
  {\cal B}_{(F)}^{w}(\tau, \sigma^u)&=& \delta^{wt}\,
  \epsilon_{tsr}\, \left[\frac{1}{2}\,
 (1 +n_F)\,\sqrt{\gamma_{F}}\, \h_{F}^{sv}\,\h_{F}^{ru}\,
\epsilon_{vu\ell}\,B_\ell + n^s_F\, E_r \right](\tau, \sigma^u),
 \label{b3}
  \eea
\medskip

\noindent we get the following form of the Maxwell equations for the
field strengths $E_r$ and $B_r$

  \bea
\partial_r\, E_r(\tau, \sigma^u)&=&
\sqrt{\gamma_{F}(\tau, \sigma^u)}\,\Big[\, \overline{\rho}(\tau,
\sigma^u) - \overline{\rho}_R(\tau, \sigma^u)\,\Big],\nonumber\\
 &&\nonumber\\
 \epsilon_{suv}\, \partial_u\, B_v(\tau , \sigma^u )
 - \frac{\partial\, E_s(\tau, \sigma^u)}{\partial\tau}
 &=& \delta_{sr}\, \sqrt{\gamma_{F}(\tau, \sigma^u)}\,
\Big[\,\overline{J}^r(\tau, \sigma^u)- \overline{J}^r_R(\tau,
\sigma^u)\,\Big],
 \label{b4}
  \eea

\noindent where the new charge and current densities are the
following functions only of the metric tensor and of the fields
$E_r$, $B_r$

\bea
 \overline{\rho}_R(\tau, \sigma^u)&=&
\frac{1}{\sqrt{\gamma_{F}(\tau, \sigma^u)}}\, \partial_r\,
\left({\cal E}^r_{(F)}(\tau, \sigma^u) - \delta^{rs}\, E_s(\tau,
\sigma^u)\right),\nonumber\\
 &&\nonumber\\
  \overline{J}^r_R(\tau, \sigma^u)&=&
\frac{1}{\sqrt{\gamma_{F}(\tau, \sigma^u)}}\, \Big[ -
\frac{\partial}{\partial \tau}\, \left({\cal E}^r_{(F)}(\tau,
\sigma^u) - \delta^{rs}\,
E_s(\tau, \sigma^u)\right) +\nonumber \\
 &+& \delta^{rs}\, \epsilon_{suv}\, \partial_u\, \left({\cal
B}^v{}_{(F)} - \delta^{vk}\, B_k\right)(\tau ,\sigma^u)\Big].
 \label{b5}
  \eea

\medskip

Instead, as a consequence of Eqs.(\ref{4.10}), the homogeneous
equations take the form

\beq
 \epsilon_{ruv}\, \partial_u\, E_v(\tau ,\sigma^s) = - {{\partial\,
 B_r(\tau ,\sigma^s)}\over {\partial\, \tau}},\qquad \epsilon_{ruv}\,
 \partial_u\, B_v(\tau ,\sigma^s) = 0.
 \label{b5a}
 \eeq

\bigskip

By using Eq.(\ref{3.2}) of the second paper we find the results of
the Appendix A of Ref.\cite{28}

 \bea
  \vec{\cal E}_{(F)}(\tau, \sigma^u)&=&\vec{E}(\tau, \sigma^u)
+ ({{\vec{\Omega}(\tau)}\over c} \times \vec{\sigma})
\times\vec{B}(\tau, \sigma^u),\nonumber\\
 &&\nonumber\\
  \vec{\cal B}{}_{(F)}(\tau, \sigma^u)&=& \vec{B}
+ ({{\vec{\Omega}(\tau)}\over c} \times \vec{\sigma}) \times
\vec{E}(\tau, \sigma^u) + ({{\vec{\Omega}(\tau)}\over c} \times
\vec{\sigma}) \times[ ({{\vec{\Omega}(\tau)}\over c} \times
\vec{\sigma}) \times \vec{B}(\tau, \sigma^u)].
 \label{b6}
 \eea

\bigskip

In absence of sources Eqs.(\ref{4.17}) are  the generally covariant
equations $\nabla_{\nu}\, F^{\mu\nu} \cir 0$, suggested by the
equivalence principle, in the 3+1 point of view after having taken
care of the asymptotic properties at spatial infinity.
\medskip

Let us remark that in the case of the nearly rigid limit of the
foliation (\ref{2.14}) (see Section VI) and with $\vec \Omega (\tau)
= (0,0, \tilde \Omega = const.)$  Eqs.(\ref{b4}) and (\ref{b5a})
coincide with Eqs.(9) of Schiff \cite{28} if we identify ${\bar
\rho}_R$ with $\sigma$ and ${\bar j}^r_R$ with $j^r$. This is due to
the fact that Schiff's fields $\vec E$, $\vec B$, have the
components coinciding with the covariant fields $E_r$ and $B_r$ of
Eqs.(\ref{4.10}); these fields obviously differ from the fields
(\ref{b3}) defined in Ref.\cite{17}.

\medskip

Eqs. (\ref{b4}) and (\ref{b5}), with the metric associated to the
admissible notion of simultaneity (\ref{2.14}), should be the
starting point for the calculations in the magnetosphere of pulsars,
where one always assumes a rigid rotation $\omega$ with the
consequent appearance of the so-called {\it light cylinder}  for
$\omega\, R = c$ (the horizon problem of the rotating disk). See
Refs.\cite{58} based on Schiff's equations \cite{28} (\ref{b4}) and
(\ref{b6}) or the more recent literature of Refs. \cite{59}. Instead
in Refs.\cite{60} the light cylinder is avoided using the rotating
coordinates of  Refs.\cite{19}, but at the price of a bad behavior
at spatial infinity.

\bigskip

These equations also show that the non-inertial electric and
magnetic fields ${\vec {\cal E}}_{(F)}$ and ${\vec {\cal B}}_{(F)}$
are {\it not}, in general, {\it equal} to the fields obtained from
the inertial ones ${\vec E}$ and ${\vec B}$ with a Lorentz
transformations to the comoving inertial system like it is usually
assumed following Rohrlich \cite{61} and the locality hypothesis.

\vfill\eject

\section{Covariant and Non-Covariant Decompositions of the
Electro-Magnetic Field and the Radiation Gauge in Non-Inertial Rest
frames.}

In inertial frames the identification of the physical degrees of
freedom (Dirac observables) of the free electro-magnetic field was
done in Refs. \cite{26,62,63,64} by means of the Shanmugadhasan
canonical transformation adapted to the first class constraints
$\pi^{\tau}(\tau ,\sigma^u ) \approx 0$ and $\Gamma (\tau , \sigma^u
) = \partial_r\, \pi^r(\tau ,\sigma^u ) \approx 0$. The final
canonical basis identifies the {\it radiation gauge} with its
transverse fields as the natural one from the point of view of
constraint theory.
\bigskip

In the parametrized Minkowski theories of Setion III Subsection A,
due to the last two lines of Eqs.(\ref{3.15}), we see that two
successive gauge transformations, of generators $G_i(\tau ,\sigma^u
) = \lambda^{\mu}_i(\tau ,\sigma^u )\, {\cal H}_{\mu}(\tau ,\sigma^u
)$, $i=1,2$, do not commute but imply an electro-magnetic gauge
transformation. Since the effect of the $i=1,2$ gauge
transformations is to modify the notions of simultaneity, also the
definition of the Dirac observables of the electro-magnetic field
will change with the 3+1 splitting. In general, given two different
3+1 splittings, the two sets of Dirac observables associated with
them will be connected by an electro-magnetic gauge transformation.

\medskip

Since it is not clear whether it is possible to find a
quasi-Shanmugadhasan canonical transformation adapted to ${\cal
H}_r(\tau ,\sigma^u ) = {\cal H}_{\mu}(\tau ,\sigma^u )\,
z^{\mu}_r(\tau ,\sigma^u ) \approx 0$, $\pi^{\tau}(\tau ,\sigma^u )
\approx 0$, $\Gamma (\tau ,\sigma^u ) \approx 0$ \footnote{${\cal
H}_{\perp}(\tau ,\sigma^u ) = {\cal H}^{\mu}(\tau ,\sigma^u )\,
l_{\mu}(\tau ,\sigma^u ) \approx 0$, like an ordinary Hamiltonian,
can be included in the adapted Darboux-Shanmugadhasan basis only in
case of integrability of the equations of motion.}, the search of
the electro-magnetic Dirac observables must be done with the
following strategy:

i) make the choice of an admissible 3+1 splitting by adding four
gauge-fixing constraints determining the embedding $z^{\mu}(\tau
,\sigma^u )$, so that the induced 4-metric $g_{AB}(\tau ,\sigma^u )$
becomes a numerical quantity and is no more a configuration
variable;

ii) find the Dirac observables on the resulting completely fixed
simultaneity surfaces $\Sigma_{\tau}$ with a suitable Shanmugadhasan
canonical transformation adapted to the two remaining
electro-magnetic constraints.

\medskip

Let us remark that a similar scheme has to be followed also in the
canonical Einstein-Maxwell theory: only after having fixed a 3+1
splitting (a system of 4-coordinates on the solutions of Einstein's
equations) we can find the Dirac observables of the electro-magnetic
field.

\medskip

This strategy is induced by the fact that, while the Gauss law
constraint $\Gamma (\tau ,\sigma^u ) = \partial_r\, \pi^r(\tau ,
\sigma^u ) \approx 0$ is a scalar under change of admissible 3+1
splittings \footnote{$\pi^r(\tau ,\sigma^u )$ is a vector density
like in canonical metric gravity.},  the gauge vector potential
$A_r(\tau ,\sigma^u )$ is the pull-back to the base of a connection
one-form and can be considered as a tensor only with topologically
trivial surfaces $\Sigma_{\tau}$ (like in the case we are
considering). Since a Shanmugadhasan canonical transformation
adapted to the Gauss law constraint transforms $\Gamma (\tau ,
\sigma^u )$  in one of the new momenta, it is not clear how to
define a conjugate gauge variable $\eta_{em}(\tau ,\sigma^u )$ such
that $\{ \eta_{em}(\tau ,\sigma^u ), \Gamma (\tau ,{\sigma}^u_1) \}
= \delta^3(\sigma^u , {\sigma}^u_1)$ and two conjugate pairs of
Dirac observables having vanishing Poisson brackets with both
$\eta_{em}(\tau ,\sigma^u )$ and  $\Gamma (\tau ,\sigma^u )$ when
the 3-metric on $\Sigma_{\tau}$ is not Euclidean ($g_{rs}(\tau ,
\sigma^u ) \not= -\sgn\, \delta_{rs}$).
\medskip

With every fixed type of instantaneous 3-space $\Sigma_{\tau}$ with
non-trivial 3-metric, $g_{rs}(\tau ,\sigma^u ) \not= -\sgn\,
\delta_{rs}$, we have to find suitable gauge variable
$\eta_{em}(\tau ,\sigma^u )$ and the Dirac observables replacing
$A^r_{\perp}(\tau ,\sigma^u )$ and $\pi^r_{\perp}(\tau ,\sigma^u )$.
\bigskip

Let us consider an arbitrary admissible non-inertial frame
identified by the embedding $z^{\mu}_F(\tau ,\sigma^u) =
x^{\mu}(\tau ) + F^{\mu}(\tau ,\sigma^u)$ of Eq.(\ref{4.1}). In it
the fields $A_r(\tau ,\sigma^u)$ and $\pi^r(\tau ,\sigma^u)$ admit
both a {\it covariant} and a {\it non-covariant} decomposition.
\bigskip

The {\it covariant decomposition} \cite{65} is

\bea
 &&\pi^r(\tau,\sigma^u)={\hat \pi}^r_{\perp}(\tau,\sigma^u) + {\hat \pi}^r_L(\tau,\sigma^u)\nonumber \\
 &&{}\nonumber \\
 &&{\hat \pi}^r_{\perp}(\tau ,\sigma^u) = \Big(\delta^r_s - \nabla^r_{F}\, {1\over
 {\Delta_{F}}}\, \nabla_{F\,s}\Big)\, \pi^s(\tau ,\sigma^u) = \Big(\delta^r_s - \nabla^r_{F}\, {1\over
 {\Delta_{F}}}\, \partial_s\Big)\, \pi^s(\tau ,\sigma^u),\nonumber \\
  \qquad&& \Rightarrow\,\nabla_{F\,r}{\hat \pi}^r_{\perp}(\tau ,\sigma^u) = 0,\nonumber \\
&&\nonumber\\
 &&{\hat \pi}^r_L(\tau ,\sigma^u) =  \nabla^r_{F}\, {1\over
 {\Delta_{F}}}\, \nabla_{F\,s}\, \pi^s(\tau ,\sigma^u) = \nabla^r_{F}\, {1\over
 {\Delta_{F}}}\, \partial_s\, \pi^s(\tau ,\sigma^u),\nonumber \\
 &&{}\nonumber \\
 &&{}\nonumber \\
&& A_r(\tau,\sigma^u) = {\hat A}_{\perp r}(\tau,\sigma^u) + {\hat A}_{L\, r}(\tau,\sigma^u),\nonumber\\
&&\nonumber\\
 &&{\hat A}_{\perp r}(\tau ,\sigma^u) = \Big(\delta^s_r - \nabla_{F\, r}\, {1\over {\Delta_{F}}}\,
 \nabla^r_{F})\, A_r(\tau ,\sigma^u)\,\Rightarrow\,\nabla^r_{F}{\hat A}_{\perp\,r}(\tau
 ,\sigma^u) = 0,\nonumber \\
&&\nonumber\\
 &&{\hat A}_{L\, r}(\tau ,\sigma^u) = \nabla_{F\, r}\, {1\over {\Delta_{F}}}\,
 \nabla^s_{F}\, A_s(\tau ,\sigma^u).
 \label{c1}
  \eea

Here $\nabla^r_{F}$ and $\triangle_{F} = \nabla^r_{F}\, \nabla_{F\,
r} = {1\over { \sqrt{\gamma_{F}(\tau ,\sigma^u )}}}\,
\partial_r\, \Big( \sqrt{ \gamma_{F}(\tau ,\sigma^u ) }\,
\gamma_{F}^{rs}(\tau ,\sigma^u )\, \partial_s\Big)$ are the
covariant derivative and the Laplace-Beltrami operator associated to
the positive 3-metric $h_{F\, rs}(\tau ,\vec \sigma^u)$,
respectively. The inverse of Laplace-Beltrami operator
$(1/\Delta_F)$ is defined by the {\em fundamental solution} of the
Laplace-Beltrami operator $G(\sigma^u,\sigma^{\,\prime\,u})$
\footnote{His existence is assured by existence's theorem (see for
example Ref.\cite{66},  but a closed analytic form is not known. A
general property of these fundamental solutions is a {\em
singularity} when the geodesic distance
$s(\sigma^u,\sigma^{\,\prime\,u})$ between $P=\{\sigma^u\}$ and
$Q=\{\sigma^{\,\prime\,u}\}$ goes to zero $\lim_{s\mapsto 0}\,
G(\sigma^u, \sigma^{\,\prime\,u}) \mapsto\, - \frac{1}{4\pi}\,
\frac{1}{s(\sigma^u,\sigma^{\,\prime\,u})} $.}
$f(\sigma^u)=\frac{1}{\Delta_F}\,g(\sigma^u)\byd\int
d^3\sigma^{\,\prime}\, \sqrt{\gamma(\sigma^{\,\prime\,u})}\,
G(\sigma^u,\sigma^{\,\prime\,u})\, g(\sigma^{\,\prime\,u})$, such
that $\Delta_F\,f(\sigma^u) = g(\sigma^u)$.

\bigskip

Since $\pi^r(\tau ,\sigma^u)$ is a vector density, we have
$\partial_r\, \pi^r(\tau ,\sigma^u) = \nabla_{F\, r}\, \pi^r(\tau
,\sigma^u)$: this quantity is a {\it 3-scalar density} on
$\Sigma_{\tau}$.

\bigskip

Instead the {\it non-covariant decomposition} \cite{1,5,9,59} in a
transverse and a longitudinal part (${\hat \partial}^r\, {\buildrel
{def}\over =}\,  \delta^{rs}\, \partial_r$, $\triangle =
\partial_r\, {\hat \partial}^r = {\vec \partial}^2$)  is

\bea
 &&\pi^r(\tau,\sigma^u) = \pi^r_{\perp}(\tau,\sigma^u) + \pi^r_L(\tau,\sigma^u),\nonumber \\
 &&{}\nonumber \\
 &&\pi^r_{\perp}(\tau ,\sigma^u) =
 \Big(\delta^r_s - {\hat \partial}^r\, {1\over {\Delta}}\,
 \partial_s \Big)\, \pi^s(\tau ,\sigma^u)\,\Rightarrow\,\partial_r\,
 \pi^r_{\perp}(\tau ,\sigma^u) = 0,\nonumber\\
 &&\nonumber\\
 &&\pi^r_L(\tau ,\sigma^u) = {\hat \partial}^r\, {1\over {\Delta}}\,
 \partial_s\,\pi^s(\tau ,\sigma^u),\nonumber\\
 &&{}\nonumber \\
 &&{}\nonumber \\
 &&A_r(\tau,\sigma^u) = A_{\perp\,r}(\tau,\sigma^u) + A_{L\,r}(\tau,\sigma^u),\nonumber \\
 &&{}\nonumber \\
 &&A_{\perp\,r}(\tau ,\sigma^u) =
 \Big(\delta^s_r - \partial_r\, {1\over {\Delta}}\,
  {\hat \partial}^s\Big)\, A_s(\tau ,\sigma^u)\,\Rightarrow\,
  {\hat\partial}^r\, A_{\perp\,r}(\tau ,\sigma^u) = 0,\nonumber\\
 &&\nonumber\\
 &&A_{L\,r}(\tau ,\sigma^u) = \,\partial_r {1\over {\Delta}}\,
 \,{\hat \partial}^s A_s(\tau ,\sigma^u).
 \label{c2}
 \eea

In Eq.(\ref{c2}) ${\hat \partial}^r\, A_r = \triangle\, \eta_{em}$
is a {\it non-covariant quantity}.\medskip

Here the inverse of Laplacian is defined using the standard
(Euclidean-like) fundamental solution: $c(\sigma^u -
\sigma^{\,\prime\,u}) = - \frac{1}{4\pi}\,
\frac{1}{\sqrt{\sum_{u=1}^3\,(\sigma^u - \sigma^{\,\prime\,u})^2}}$,
so that $f(\sigma^u) = \frac{1}{\Delta}\, g(\sigma^u)\byd \int
d^3\sigma^{\,\prime}\, c(\sigma^u-\sigma^{\,\prime\,u})
g(\sigma^{\,\prime\,u})$ and $\Delta\, f(\sigma^u) =
\Big(\sum_{r=1}^3\, {\hat \partial}^r\, \partial_r\Big)\,
f(\sigma^u) = g(\sigma^u)$.

\bigskip

Eq.(\ref{c2}) allow us to define the following non-covariant
Shanmugadhasan canonical transformation

\bea
  &&\begin{minipage}[t]{1cm}
\begin{tabular}{|l|} \hline
$A_A$ \\  \hline
 $\pi^A$ \\ \hline
\end{tabular}
\end{minipage} \ {\longrightarrow \hspace{.2cm}} \
\begin{minipage}[t]{2 cm}
\begin{tabular}{|l|l|l|} \hline
$A_{\tau}$ & $\eta_{em}$   & $A_{\perp\, r}$   \\ \hline
$\pi^{\tau}\approx 0$& $\Gamma \approx 0$ &$\pi^r_{\perp}$ \\
\hline
\end{tabular}
\end{minipage}\nonumber \\
 &&{}\nonumber \\
 &&{}\nonumber \\
 A_r(\tau ,\sigma^u )&=& - {
{\partial}\over {\partial \sigma^r} }\, \eta_{em} (\tau ,\sigma^u
) + A_{\perp\,r}(\tau ,\sigma^u ),\nonumber \\
 &&{}\nonumber \\
 \pi^r(\tau ,\sigma^u )&=&\pi^r_{\perp}(\tau
,\sigma^u ) + {1\over {\Delta}}\, {\hat \partial}^r\,
\Gamma (\tau ,\sigma^u ),\nonumber \\
 &&{}\nonumber \\
  \eta_{em}(\tau ,\sigma^u )&=&- {\hat \partial}^r\, A_r(\tau,\sigma^u),\nonumber \\
 &&{}\nonumber \\
A_{\perp\,r}(\tau ,\sigma^u) &=&
 \Big(\delta^s_r - \partial_r\, {1\over {\Delta}}\,
  {\hat \partial}^s\Big)\, A_s(\tau ,\sigma^u),\nonumber \\
&&\nonumber\\
  \pi^r_{\perp}(\tau ,\sigma^u) &=&
 \Big(\delta^r_s - {\hat \partial}^r\, {1\over {\Delta}}\,
 \partial_s \Big)\, \pi^s(\tau ,\sigma^u),\nonumber \\
 &&{}\nonumber \\
  &&\lbrace \eta_{em}(\tau ,\sigma^u ),\Gamma
(\tau ,{\sigma}^{\,\prime\,u} ) \rbrace = \delta^3(
\sigma^u, {\sigma}^{\,\prime\,u}),\nonumber \\
 &&{}\nonumber \\
 &&\lbrace A_{\perp\,r}(\tau ,
\sigma^u ),\pi^s_{\perp}(\tau ,{\sigma} ^{\,\prime\,u})\rbrace =
c\,(\delta^{rs} - {{\partial_r\, {\hat \partial}^s} \over
{\Delta}})\, \delta^3(\sigma^u, { \sigma}^{\,\prime\,u}).
 \label{c3}
 \eea
\medskip

If we add the gauge fixing $\eta_{em}(\tau ,\sigma^u) \approx 0$,
then its $\tau$-constancy implies $A_{\tau}(\tau ,\sigma^u) \approx
0$ and we get a non-inertial realization of the {\it non-covariant
radiation gauge}.

\bigskip

While with the non-covariant decomposition we can easily find a
Shanmugadhasan canonical transformation adapted to the Gauss law
constraint with the standard canonically conjugate (but
non-covariant) Dirac observables ${\vec A}_{\perp}$ and ${\vec
\pi}_{\perp}$ of the radiation gauge, it is not clear whether the
covariant decomposition can produce such a canonical basis. In any
case, as shown in Ref.\cite{65}, the radiation gauge formalism is
well defined in both cases if we add suitable gauge fixings.

\bigskip

In the inertial rest-frame instant form reviewed in Section III
Subsection B the 3-metric inside the Wigner 3-spaces is $g_{rs}(\tau
,\sigma^u ) = - \sgn\, h_{rs}(\tau ,\sigma^u ) = - \sgn\,
\delta_{rs}$ and the two decompositions coincide.

\bigskip

In Subsection B of Section IV there is the non-covariant
Shanmugadhasan canonical transformation in non-inertial frames in
presence of charged particles.
\bigskip

Let us remark that in the non-Euclidean 3-space we are using a delta
function $\delta^3(\sigma^u, \sigma^{\,\prime\,u})$, with the
properties $\delta^3(\sigma^u, \sigma^{\,\prime\,u}) =
\delta^3(\sigma^{\,\prime\,u}, \sigma^u)$ and
$\frac{\partial}{\partial\sigma^r}\, \delta^3(\sigma^u,
\sigma^{\,\prime\,u}) = - \frac{\partial}{\partial
\sigma^{\,\prime\,r}}\, \delta^3(\sigma^u, \sigma^{\,\prime\,u})$,
such that $d^3\sigma^\prime\, \delta^3(a^u,
\sigma^{\,\prime\,u})\,f(\sigma^{\,\prime\,u}) = f(a^u)$, and not a
covariant one $D^3(\sigma^u, \sigma^{\,\prime\,u}) =
\frac{1}{\sqrt{\gamma(\tau,\sigma^{\,\prime\,u})}}\,
\delta^3(\sigma^u, \sigma^{\,\prime\,u}) =
\frac{1}{\sqrt{\gamma(\tau,\sigma^{u})}}\, \delta^3(\sigma^u,
\sigma^{\,\prime\,u})$ such that $\int d^3\sigma^\prime\,
\sqrt{\gamma(\tau,\sigma^{\,\prime\,u})}\, D^3(a^u,
\sigma^{\,\prime\,u})\, f(\sigma^{\,\prime\,u}) = f(a^u)$.

\vfill\eject

\end{document}